\pdfoutput=1
\newcommand*{\ATLASLATEXPATH}{}
\documentclass[PAPER, atlasdraft=false, texlive=2016, UKenglish, cernpreprint]{\ATLASLATEXPATH atlasdoc}
\LEcontact{Lily Asquith lily.asquith@cern.ch}

\usepackage{multirow}
\usepackage{\ATLASLATEXPATH atlaspackage}
\usepackage{\ATLASLATEXPATH atlasbiblatex}
 
\usepackage{\ATLASLATEXPATH atlasphysics}
 
\usepackage{graphicx}

\graphicspath{{logos/}{figures/}}
 
\usepackage{ANA-TRIG-2018-05-PAPER-defs}

\AtlasTitle{Performance of electron and photon triggers in ATLAS during LHC Run 2}
 
\AtlasAbstract{
Electron and photon triggers covering transverse energies from 5~\GeV{} to
several \TeV{} are essential for the ATLAS experiment to record signals for a wide variety of physics: from
Standard Model processes to searches for new phenomena in both proton--proton and
heavy-ion collisions. To cope with a fourfold increase of peak LHC luminosity
from 2015 to 2018 (Run 2), to 2.1$\times$\highL{}, and a similar increase in the number
of interactions per beam-crossing to about 60, trigger algorithms and
selections were optimised to control the rates while retaining a high
efficiency for physics analyses.
For proton--proton collisions, the single-electron trigger efficiency relative to a single-electron offline selection
is at least 75\%
for an offline electron of 31~\GeV{}, and rises to 96\% at 60~\GeV;
the trigger efficiency of a 25~\gev{} leg of the primary diphoton trigger
relative to a tight offline photon selection is more than 96\% for an offline photon of 30~\GeV{}.
For heavy-ion collisions, the primary electron and photon trigger efficiencies relative to
the corresponding standard offline selections are at least 84\% and 95\%, respectively,
at 5~\gev{} above the corresponding trigger threshold.
}
 
\author{The ATLAS Collaboration}
 
\AtlasRefCode{TRIG-2018-05}

\PreprintIdNumber{CERN-EP-2019-169}

\AtlasCoverSupportingNote{Electron supporting note}{https://cds.cern.ch/record/2651056}
\AtlasCoverSupportingNote{Photon supporting note}{https://cds.cern.ch/record/2652154}
\AtlasCoverSupportingNote{Ringer supporting note}{https://cds.cern.ch/record/2652142}
 
\AtlasCoverCommentsDeadline{11 August 2019}
 
\AtlasCoverAnalysisTeam{M.~Backes, T.~Bold, K.~Burka, J.~Da~Fonseca~Pinto, D.~Damazio, M.~Feng, L.~Flores, W.~S.~Freund, A.~Giannini, I.~Grabowska-Bold, K.~Hill, J.~Hoya, T.~Hryn'ova, Q.~Hu, P.~A.~Janus, S.~Jones, D.~Koeck, J.~Kremer, J.~Kroll, G.~Lerner, M.~Levy, D.~Millar, F.~Monticelli, A.~Morley, G.~Orellana, R.~Owen, D.~Perepelitsa, J.~Reichert, A.~Ruiz~Martinez, F.~Salvatore, B.~Sefarzadeh~Samani, J.~M.~Seixas, Ph.~Sommer, M.~Spina, E.~Thomson, R.~Vallance, M.~Verissimo~de~Araujo, H.~Wahlberg, R.~White}
 
\AtlasCoverEdBoardMember{Christos~Anastopoulos~(chair), Yanping~Huang, Takanori~Kono}

\AtlasCoverEgroupEditors{atlas-TRIG-2018-05-editors@cern.ch}
 
\AtlasCoverEgroupEdBoard{atlas-TRIG-2018-05-editorial-board@cern.ch}
 
\AtlasJournal{EPJC}
 
\AtlasJournalRef{Eur. Phys. J. C 80 (2020) 47}
 
\AtlasDOI{10.1140/epjc/s10052-019-7500-2}
\hypersetup{pdftitle={ATLAS document},pdfauthor={The ATLAS Collaboration}}
 
\begin{document}
 
\maketitle

\section{Introduction}
\label{sec:intro}
 
Electrons and photons are present in many Standard Model processes
as well as in searches for phenomena beyond the Standard Model.
The ATLAS physics programme relies on an efficient trigger system to record a highly signal-rich
subset of all collision events produced by the Large Hadron Collider (LHC) at CERN.
 
The ATLAS Collaboration has published several electron and photon trigger performance results since
the start of data-taking: the early 2010 data are covered in Ref.~\cite{PERF-2011-02}, the 2011
data in Ref.~\cite{ATLAS-CONF-2012-048} and the 2015 data in Ref.~\cite{TRIG-2016-01}. This paper
addresses the evolution of performance of the electron and photon triggers from 2015 to 2018 (Run 2).
The major challenge for the trigger in this period was the need to maintain excellent performance for
the ATLAS physics programme while adapting to a nearly fourfold increase in the LHC peak luminosity and
in the  number of interactions per beam-crossing.
 
This paper is organised as follows. The ATLAS detector is described in Section~\ref{sec:detector}.
The trigger system is introduced in Section~\ref{sec:menu}.
Section~\ref{sec:datasets} introduces the data sets used in this publication. The following two sections detail electron and photon reconstruction
and identification at the analysis level (offline) and the trigger level (online). Section~\ref{sec:tools} presents the techniques used to measure
the trigger performance. The performance of the photon and electron triggers from 2015 to 2018 is described in Sections~\ref{sec:trig_l1}--\ref{sec:trig_hi}.
The data quality monitoring is described in Section~\ref{sec:ops}, and the conclusions are presented in Section~\ref{sec:conclusion}.
 
\section{The ATLAS detector}
\label{sec:detector}
 
The ATLAS detector~\cite{PERF-2007-01,ATLAS-TDR-19,PIX-2018-001}
is a multipurpose detector designed to observe particles
produced in high-energy proton--proton ($pp$) and heavy-ion (HI) collisions.
It is composed of a tracking detector in the innermost region
around the interaction point, surrounded by
calorimeters and muon chambers.
 
The inner tracking detector (ID) is immersed in a 2$\,$T magnetic field produced by a
thin superconducting solenoid, and provides precise reconstruction of
charged-particle tracks in a pseudorapidity range\footnote{ATLAS uses
a right-handed coordinate system with its origin at the nominal interaction
point (IP) in the centre of the detector and the $z$-axis along the beam-pipe.
The $x$-axis points from the IP to the centre of the LHC ring, and the $y$-axis
points upward. Cylindrical coordinates $(r, \phi)$ are used in the transverse
plane, $\phi$ being the azimuthal angle around the beam-pipe.
The pseudorapidity is defined in terms of the polar angle $\theta$ as
$\eta=-\ln\tan(\theta/2)$. The angular distance $\Delta R$ is defined as
$\Delta R \equiv \sqrt{(\Delta\eta)^2 +(\Delta\phi)^2}$. Transverse momenta and energies are defined as
$\pt=p\sin\theta$ and $\et=E\sin\theta$, respectively.} $|\eta|<2.5$.
The innermost part consists of a silicon pixel detector with four layers.
The layer closest to the beam-pipe, the insertable B-layer, was installed before Run~2 and
provides high-resolution hits in three-dimensions with pixels at a radius of 3.3~cm to improve the tracking performance.
A silicon microstrip tracker surrounds the pixel detector with typically four layers of sensor modules.
Each module is composed of multiple pairs of sensors with a stereo-angle to measure
three-dimensional hit positions.
The outermost region of the tracker in the range $|\eta|<2.0$ is covered by a transition radiation
tracker (TRT). 
It consists of
straw drift tubes filled with a gas mixture of 70\% Xe, 27\% CO$_2$ and 3\% O$_2$
interleaved with polypropylene/polyethylene transition radiators creating
transition radiation for particles with a large Lorentz factor. This radiation
is absorbed by the Xe-based gas mixture, discriminating electrons from hadrons over a wide energy range.
Some of the TRT modules instead contain a gas mixture of 70\% Ar, 28.5\% CO$_2$ and 1.5\% O$_2$ as a mitigation for gas leaks that
cannot be repaired without an invasive opening of the inner detector.
At the end of Run 2 data-taking the two innermost TRT
barrel layers, i.e.\ about half of the modules in $|\eta|<0.6$,
and 3 (2) out of 14 endcap wheels in $-2<\eta<-1$ ($1<\eta<2$) were running with the argon-based gas mixture.
The presence of this gas mixture is taken into account in the simulation and the corresponding loss in identification power is
partially mitigated by a dedicated TRT particle-identification algorithm~\cite{ATLAS-CONF-2011-128}.
For charged particles with transverse momenta $>0.5$~\GeV{} the TRT provides typically 35 hits per track.
 
The calorimeter system has both electromagnetic (EM) and hadronic components.
It is designed to provide a full $\phi$ coverage and covers the pseudorapidity range $|\eta|<4.9$, with finer granularity
over the region matched to the inner detector. The EM calorimeter is a lead/liquid-argon (LAr) sampling calorimeter
with an accordion-geometry. It is divided into two half-barrels ($-1.475<\eta<0$ and $0<\eta<1.475$) and
two endcap components ($1.375<|\eta|<3.2$). The transition region
between the barrel and endcaps ($1.37<|\eta|<1.52$) contains significant additional
inactive material~\cite{PERF-2007-01}. Over the region devoted to
precision measurements ($|\eta|<2.5$, excluding the transition region),
the EM calorimeter is segmented into three layers longitudinal in shower depth.
The first layer consists of strips finely grained in the $\eta$ direction,
offering excellent discrimination between isolated photons and pairs of closely spaced
photons coming from  $\pi^{0}\rightarrow \gamma\gamma$ decay. For electrons
and photons with high transverse energy, most of the energy is collected in the second layer, which has a
lateral granularity of $0.025\times 0.025$ in $(\eta, \phi)$ space.
The third layer provides measurements of energy deposited in the tails of
the shower. In front of the accordion calorimeter, a thin presampler layer, covering the pseudorapidity
interval  $|\eta|<1.8$, is used to correct for energy loss upstream of the calorimeter.
 
Three hadronic calorimeter layers surround the EM calorimeter. For electrons and photons, they
provide additional background discrimination through measurements of hadronic energy.
The barrel hadronic calorimeter ($|\eta| < 1.7$) is an iron/scintillator tile sampling calorimeter
with wavelength-shifting fibers. For the hadronic endcaps, copper/LAr calorimeters are used. The forward
regions (FCal) are instrumented with copper--tungsten/LAr calorimeters for both the EM and hadronic energy measurements
up to $|\eta| = 4.9$. The LAr-based detectors are housed in one barrel and two endcap cryostats.
 
The outermost layers of ATLAS consist of an external
muon spectrometer (MS) in the pseudorapidity range $|\eta|<2.7$, incorporating three large toroidal magnet assemblies with eight coils each.
The field integral of the toroids ranges between 2.0 and 6.0$\,$Tm for most of the acceptance.
The MS includes precision tracking chambers and fast detectors for triggering.
 
 
\section{ATLAS trigger system}
\label{sec:menu}
 
A two-level trigger system~\cite{TRIG-2016-01} is used to select events of interest. The first-level (L1) trigger,
implemented in custom hardware, utilises coarser-granularity signals from the calorimeters and the muon chambers to
reduce the event rate from the \SI{40}{\MHz} bunch crossing rate to below \SI{100}{\kHz}; it has 2.5$\,\mu$s to decide which events to keep to satisfy this factor 400 reduction.
L1 also defines regions-of-interest (RoIs) which have calorimeter clusters with high transverse energy, \et{}, or muon tracks in the muon chambers.
 
Events accepted by L1 are processed by the high-level trigger (HLT),
based on algorithms implemented in software which must further reduce the number of events recorded to
disk to an average rate of about \SI{1}{\kHz} within a few seconds. The HLT uses fine-granularity calorimeter information,
precision measurements from the muon spectrometer and tracking information from the ID, which are not available at L1.
HLT reconstruction can be executed either within the RoIs identified at L1 or for the full detector (full-scan).
The selection of particle candidates by the HLT is performed at each step, so that if it
fails at a certain step, subsequent steps are not executed. This is essential to
reduce the time needed by the HLT to reconstruct the event and make a decision.
 
A sequence of L1 and HLT trigger algorithms is called a `trigger' and is meant to identify
one or more particles of a given type and a given threshold of
transverse energy or momentum. For example,
electron and photon triggers are meant to select events with one or more electrons or photons in the detector.
The configuration of the trigger is controlled by the `trigger menu', which defines a full list of the
L1 and HLT triggers and their configurations. Menu composition and trigger thresholds are
optimised for the LHC running conditions (beam type, luminosity, etc.) to fit within
the event acceptance rate and the bandwidth constraints of the data acquisition system of the ATLAS detector
as well as the offline storage constraints.
 
In addition to the triggers described above, there are `rerun' triggers which never accept an event on their own, but are configured to run
only on the events accepted by other triggers, and their decision is recorded for offline use.
This information is used for studies of the trigger efficiency, which is calculated separately
for each object (leg) of the multi-object triggers.
 
Trigger thresholds and identification criteria have to be modified sometimes to maintain a stable output rate.
To ensure an optimal trigger menu within the rate constraints of a given LHC luminosity,
prescale factors can be applied to both the L1 and HLT triggers independently and configured during data-taking.
They allow the experiment to either disable triggers completely or to set the fraction of events that may be accepted by them.
 
 
\section{Data sets and simulation samples }
\label{sec:datasets}
 
The results described in this paper use the full $pp$ collision data set recorded by ATLAS
between 2015 and 2018 with the LHC operating at a centre-of-mass energy of $\sqrt{s}=13\,$\tev.
The maximum instantaneous luminosities increased by a factor of four during the four years of Run 2, resulting in
an increase in the average number of interactions per bunch crossing, $\langle\mu\rangle$, also referred to as `pile-up'.
In addition to $pp$ data, the heavy-ion (HI) physics programme is realised for one month per year, typically starting in November.
During it, the LHC provides either lead--lead ($PbPb$) ion collisions, or
special reference runs with either low-pile-up $pp$ or proton--lead ($pPb$) ion collisions.
The per-year values of maximum instantaneous luminosity, pile-up and integrated luminosity after requiring
stable beam conditions and a functional detector are summarised in Table~\ref{tab-lumi} for the
standard $pp$ collisions and in Table~\ref{tab-lumihi} for the HI programme.
 
\begin{table}[ht!]
\small
\centering
\caption{The per-year values of maximum instantaneous luminosity
(\textbf{L}), peak and average pile-up $(\langle\mu\rangle)$, and integrated luminosity for $pp$ data-taking.
The uncertainty in the combined 2015--2018 integrated luminosity is 1.7\%~\cite{ATLAS-CONF-2019-021},
obtained using the LUCID-2 detector \cite{LUCID2} for the primary luminosity measurements.
It should be noted that in 2017 the peak $\langle\mu\rangle=80$ was reached only in a few dedicated
runs, so a maximum value of $\langle\mu\rangle=60$ is used for the results shown in this paper.
}
\begin{tabular}{@{\extracolsep{\fill}} l c c c c}
\toprule
\multirow{2}{*}{ \textbf{Year} } & \multirow{2}{*}{ \textbf{Peak L [cm$^{-2}$s$^{-1}$]} } &
\multirow{2}{*}{ \textbf{Peak $\langle\mu\rangle$} } &
\multirow{2}{*}{ \textbf{Average $\langle\mu\rangle$} } & \multirow{2}{*}{ \textbf{$\int$Ldt [fb$^{-1}$]}}  \\
&&&\\
\midrule
2015 &  0.5 $\times 10^{34}$   & 15 & 13.4 & 3.2 \\ 
2016 &  1.4 $\times 10^{34}$   & 45 & 25.1 & 32.9 \\ 
2017 &  2.1 $\times 10^{34}$  & 80 & 37.8 & 43.9 \\ 
2018 &  2.1 $\times 10^{34}$   & 60 & 36.1 & 58.5 \\
\bottomrule
\end{tabular}
\label{tab-lumi}
\end{table}
 
\begin{table}[ht!]
\small
\centering
\caption{The per-year values of centre-of-mass energy per nucleon pair, maximum instantaneous luminosity, and integrated luminosity
for the heavy-ion data-taking.
A $\langle\mu\rangle$ value related to pile-up is not listed as it is negligible compared to the nominal $pp$ data-taking
(below 0.04 for the 2015 $PbPb$ data set).}
\begin{tabular}{@{\extracolsep{\fill}} l c c c r }
\toprule
\multirow{2}{*}{ \textbf{Beam type}} & \multirow{2}{*}{ \textbf{$\sqrt{s_{NN}}$ }}
& \multirow{2}{*}{ \textbf{Year} } & \multirow{2}{*}{ \textbf{Peak L [cm$^{-2}$s$^{-1}$]} }
& \multirow{2}{*}{ \textbf{$\int$Ldt}} \\
&&&\\ \midrule
$PbPb$ & 5.02~\tev & 2015 & 2.7 $\times 10^{27}$ & $0.48$~\inb \\
& 5.02~\tev & 2018 &  6.2 $\times 10^{27}$ & $1.73$~\inb\\
$pp$ & 5.02~\tev & 2015 & 3.8 $\times 10^{32}$ & $25$~\ipb \\
$pPb$ & 8.16~\tev & 2016 & 8.6 $\times 10^{29}$ & $165$~\inb{} \\
\bottomrule
\end{tabular}
\label{tab-lumihi}
\end{table}
 
Samples of simulated $Z\rightarrow ee$ and $W\rightarrow e\nu$ decays are used to benchmark the
expected electron trigger efficiencies and to optimise the electron identification criteria.
\POWHEGBOX v1 Monte Carlo (MC) generator~\cite{Nason:2004rx,Frixione:2007vw,Alioli:2010xd,Alioli:2008gx}
is used for the simulation of the hard-scattering in these samples.
It is interfaced to \PYTHIAV{8.186}~\cite{pythia8}
for the modelling of the parton shower, hadronisation, and underlying
event (UE), with parameters set according to the AZNLO tune~\cite{STDM-2012-23}. The CT10 PDF set~\cite{ct10} is used
for the hard-scattering processes, whereas the CTEQ6L1 PDF
set~\cite{Pumplin:2002vw} is used for the parton shower. The effect of
QED final-state radiation is simulated with \textsc{Photos++}
(v3.52)~\cite{Golonka:2005pn,photospp}. The \textsc{EvtGen}~v1.2.0
program~\cite{Lange:2001uf} is used to decay bottom and charm hadrons.
For optimisation of the low-\et{} electron selection, $J/\psi\rightarrow ee$ samples are used.
These were generated with \PYTHIAV{8.186}, the A14 set of tuned parameters~\cite{ATL-PHYS-PUB-2014-021}, and the CTEQ6L1 PDF set for both
the hard-scattering processes and the parton shower.
For high-\et{} electron trigger studies, a MC event sample for the
$gg\rightarrow\mathrm{radion\,(3~\TeV)}\rightarrow VV\rightarrow ee qq$
process was produced with \textsc{MadGraph5}-2.6.0~\cite{Alwall:2014hca} interfaced to \PYTHIAV{8.212}.
 
Background samples for electron processes were simulated with two-to-two processes in \PYTHIAV{8.186}
with the A14 set of tuned parameters and NNPDF23LO~\cite{Ball:2012cx}. These processes include multijet production,
$qg\rightarrow q\gamma$, $q\bar{q}\rightarrow q\gamma$, $W/Z$ boson production (plus other electroweak processes)
and top-quark production. A filter is applied to enrich the sample in electron backgrounds: selected events have
particles (excluding muons and neutrinos) produced in the hard scatter
with a summed transverse energy exceeding 17~\GeV{} in a region of
$\Delta\eta\times\Delta\phi=0.1\times 0.1$. For the background studies, electrons from
$W/Z$ boson production are excluded using generator-level information.
 
For low-\et photon trigger studies, samples of \Zlly\ ($\ell = e,\mu$) events with transverse energy of the photon above 10~\GeV{}
were generated with \SHERPA 2.1.1~\cite{Gleisberg:2008ta} and the CT10 PDF set.
For high-\et photon trigger studies, MC samples of prompt-photon production generated with \PYTHIAV{8.186} are used.
These samples include the leading-order $\gamma$+jet events from $qg\to q\gamma$ and $q\overline{q} \to g\gamma$
hard-scattering processes, as well as prompt photons from quark fragmentation in QCD dijet events.
In addition to the samples detailed above, samples of a Standard Model Higgs boson produced via gluon--gluon fusion
decaying into two photons were generated using \POWHEGBOX, NNLOPS implementation~\cite{Hamilton:2013fea,Hamilton:2015nsa},
with the PDF4LHC15 PDF set~\cite{Butterworth:2015oua}, and interfaced to \PYTHIAV{8.186} for parton showering,
hadronisation and the UE using the AZNLO set of tuned parameters.
 
Simulation of collision events includes the effect of multiple $pp$ interactions in the same or neighbouring bunch crossings. The simulation of pile-up collisions was performed with \PYTHIAV{8.186} using the ATLAS A3 set of tuned parameters~\cite{ATL-PHYS-PUB-2016-017} and the NNPDF23LO PDF set, and weighted to reproduce the average number of pile-up interactions per bunch crossing observed in data.
The generated events were passed through a full detector simulation~\cite{SOFT-2010-01} based on\ \GEANT~4~\cite{geant}.
 
\section{Offline object reconstruction and identification}
\label{sec:egselectionoff}
 
The offline electron and photon reconstruction~\cite{ATLAS-EGAM-2018-01}
uses dynamic, 
variable-size clusters of energy deposits measured in
topologically connected EM and hadronic calorimeter cells~\cite{PERF-2014-07}, called topo-clusters,
to recover energy from bremsstrahlung photons or from electrons from photon conversions.
After applying initial position corrections and energy calibrations to the topo-clusters, they are
matched to ID tracks re-fitted to account for bremsstrahlung, following the procedure described in
Ref.~\cite{ATLAS-CONF-2012-047}, to reconstruct electron candidates.
Topo-clusters not matched to any track or matched to conversion vertices are reconstructed as photon candidates.
The electron and photon candidates to be used for analyses then have their energies recalibrated.
 
Identification of photon candidates in ATLAS relies on rectangular selection requirements based on
calorimetric variables~\cite{ATLAS-EGAM-2018-01} which deliver good separation between prompt photons and fake signatures.
Fake photon signatures can result either from non-prompt photons originating from the decay of
neutral hadrons in jets, or from jets depositing a large energy fraction in the EM calorimeter.
Two identification working points (WPs), \loose\ and \tight, are defined for photons.
Photon identification WPs are strictly inclusive, i.e.\ photons satisfying the \tight\ selection
are a subset of those satisfying the \loose\ selection.
The \loose\ selection is based on shower shapes in the second layer of the EM calorimeter and
on the energy deposited in the hadronic calorimeter.
In addition to the \loose\ selection criteria, the \tight\ selection uses information from
the finely segmented first layer of the calorimeter. 
For a collection of photons radiated from leptons in $Z$ decays and with \et $>$25~\GeV,
the efficiency integrated over 2015--2017 data sets of the \loose\ (\tight) selection is 98.9\% (87.5\%)
for photons not matched to any track and 96.3\% (87.6\%) for photons matched to conversion vertices~\cite{ATLAS-EGAM-2018-01}.
 
Prompt electrons entering the central region of the detector ($|\eta| < 2.47$) are selected
using a likelihood-based (LH) identification~\cite{ATLAS-EGAM-2018-01}, which exploits the characteristic
features of energy deposits in the EM calorimeters (longitudinal and lateral shower shapes),
track quality, track--cluster matching, and particle identification by the TRT.
The LH probability density functions (pdfs) for the \et{} range of 4.5 to 15~\GeV{} are derived
from $J/\psi\rightarrow ee$ and for \et$\,>15$~\GeV{} from $Z\rightarrow ee$
events as described in Ref.~\cite{ATLAS-EGAM-2018-01}.
Different pdfs are obtained for each identification quantity
in separate bins in electron-candidate \et{} and $\eta$. To ensure a smooth variation of the electron
identification efficiency with the electron \et, the discriminant requirements are varied in
finer bins than those defined for the pdfs and, at the \et{} bin boundaries, a linear
interpolation between the neighbouring bins in \et{} is used to determine both the pdf
values and the discriminant requirements at the bin boundaries. This procedure is referred
to as `smoothing'. The discriminant threshold is also adjusted linearly as a function of the number of
reconstructed vertices to yield a stable rejection of background electrons.
Three operating points, corresponding to increasing threshold values for the LH
discriminant, are defined (identification efficiencies quoted are averages for electroweak processes integrated over 2015--2017 data sets):
`loose' (93$\%$), `medium' (88$\%$) and `tight' ($80\%$)~\cite{ATLAS-EGAM-2018-01}.
 
Muon candidates, used in photon performance studies, are identified by matching ID tracks
to tracks reconstructed in the muon spectrometer~\cite{PERF-2015-10}.
 
To reduce backgrounds from misidentified jets and from light- and heavy-flavour hadron
decays inside jets, photon and lepton candidates are often required to be isolated. This
isolation selection is specific to the analysis topology.
The calorimeter isolation \EtIsol\ is computed as the sum of transverse energies of topo-clusters in the calorimeters,
in a cone around the candidate.
The energy deposited by the photon or lepton candidate and the contributions from the UE and
pile-up are subtracted on an event-by-event basis~\cite{Cacciari:2007fd}.
The track isolation variable, \ptIsol\, is obtained by summing the scalar \pt{} of good-quality tracks in
a cone around the candidate; good tracks are defined here as having \pt$>1$~\GeV\ and a distance
of closest approach to the primary vertex along the
beam axis $|z_0\sin\theta|<3$~mm, and exclude the tracks associated with the photon conversion or the
lepton  candidate. The exact definitions of a few WPs used in this paper are provided in Table~\ref{tab-iso}
and described in detail in Refs.~\cite{ATLAS-EGAM-2018-01,PERF-2015-10}.
 
\begin{table}[ht!]
\small
\centering
\caption{Definition of isolation working points.
For electron (muon) track isolation the cone size $\Delta R^\mathrm{var}$
has a maximum value of 0.2 (0.3) and decreases as a function of $\pt$ as 10~\GeV/\pt{}[\GeV].
Prefix `FC' highlights a fixed requirement applied to the calorimeter and track isolation variables.
}
\begin{tabular}{@{\extracolsep{\fill}} l l c c}
\toprule
\textbf{Object} & \textbf{WP} & \textbf{Calorimeter isolation} & \textbf{Track isolation} \\
\midrule
Photon & Calorimeter-only tight & \EtIsol$(\Delta R<0.4) < 0.022 \cdot \et+2.45${\GeV} & - \\
Electron & FCTight & \EtIsol$(\Delta R<0.2)/\et < 0.06$ & \ptIsol$(\Delta R^\mathrm{var}<0.2)/\pt < 0.06$\\
& FCLoose & \EtIsol$(\Delta R<0.2)/\et < 0.2$ & \ptIsol$(\Delta R^\mathrm{var}<0.2)/\pt < 0.15$\\
Muon & FCLoose & \EtIsol$(\Delta R<0.2)/\pt < 0.3$ & \ptIsol$(\Delta R^\mathrm{var}<0.3)/\pt < 0.15$\\
\bottomrule
\end{tabular}
\label{tab-iso}
\end{table}
 
\section{Trigger reconstruction and identification of photons and electrons}
\label{sec:egselectiontrig}
 
\begin{figure}[h!]
\centering
\includegraphics[width=0.75\textwidth]{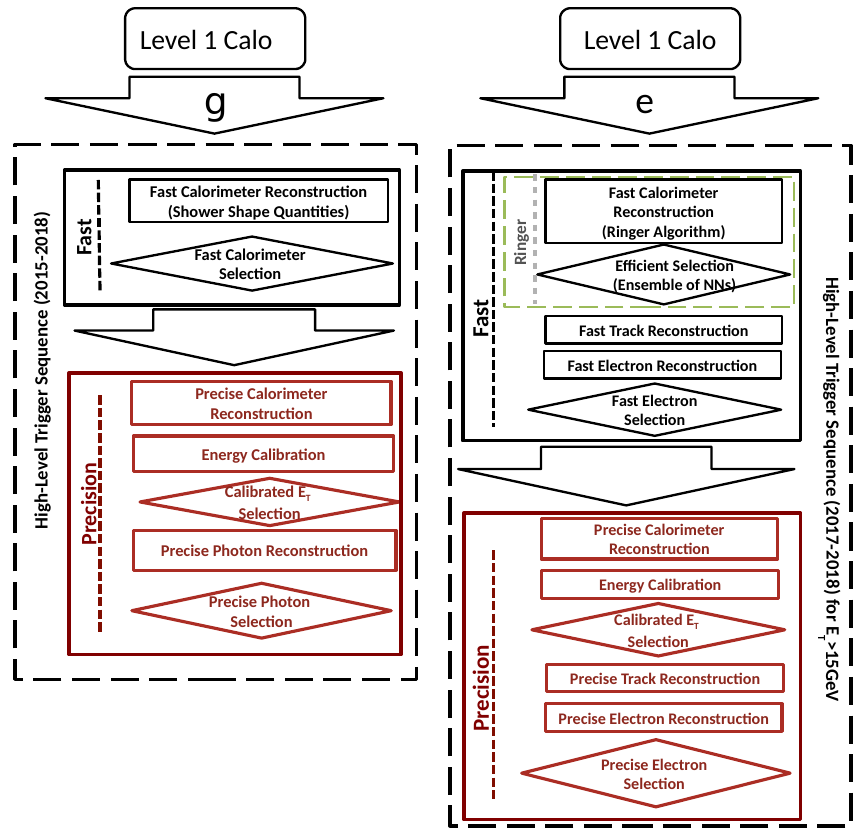}
\caption{Simplified non-isolated photon (g) and electron (e)
trigger sequences for $pp$ data-taking.}
\label{fig-seq}
\end{figure}

Photon and electron reconstruction at the HLT stage is performed on each EM RoI
provided by L1, which satisfies \et{} and isolation requirements as specified by
the trigger menu. It proceeds in a series of sequential steps as shown in Figure~\ref{fig-seq}.
In the HLT, fast algorithms are executed first, allowing precision algorithms
(which closely reproduce the offline algorithms and require more CPU time)
to run at a reduced rate later in the trigger sequence.
 
Fast algorithms are executed using calorimeter and ID
information within the RoI to perform the initial selection and identification
of the photon and electron candidates, and achieve early background rejection.
 
If a particle candidate satisfies the criteria defined for the fast selection,
the precision algorithms are executed in the HLT, where access to
detector information outside the RoI is possible. These precision
online algorithms are similar to their offline counterparts, with the
following exceptions: the bremsstrahlung-aware re-fit of electron tracks~\cite{ATLAS-CONF-2012-047}
and electron and photon dynamic, variable-size topo-clusters~\cite{ATLAS-EGAM-2018-01} are not used online;
photon candidates are identified using only the calorimeter information online;
the online algorithms use $\langle\mu\rangle$ to assess pile-up, while
the number of primary vertices is used offline. In addition to the above,
some cell-energy-level corrections are not available online,
such as the correction for transient changes in LAr high-voltage~\cite{LARG-2009-01}, or
differ in implementation, such as the bunch crossing position-dependent
pile-up correction~\cite{ATL-DAQ-PUB-2017-001, PERF-2017-03}.
 
\subsection{Photon and electron triggers at L1}
\label{sec:L1vh}
 
The details of the Run 2 L1 trigger algorithms can be found in Ref.~\cite{TRIG-2016-01},
and only the basic elements are described here.
The L1 trigger for photons and electrons uses calorimeter information in
the central ($|\eta|<2.5$) region to build an EM RoI. A sliding window
algorithm is used, with a window consisting of $4\times 4$ trigger
towers with granularity $0.1\times 0.1$ in $\eta$ and $\phi$,
longitudinally segmented into electromagnetic and hadronic towers.
Within the window the algorithm uses the maximum \et~from the four
possible pairs of nearest-neighbour electromagnetic towers in a $2\times 2$
central region; this is used for EM transverse energy reconstruction.
The energy of the trigger towers is calibrated at the electromagnetic energy scale (EM scale).
This EM scale is not the same as the one used in the offline reconstruction,
which can lead to trigger inefficiencies relative to offline reconstruction
as discussed in Section~\ref{sec:trig_l1}.
 
A nominal transverse energy threshold is applied (e.g.\ \et $>$ 22~\GeV,
denoted by the trigger name EM22). The threshold can also be $\eta$-dependent,
due to the energy scale depending on $\eta$,
within the granularity of 0.1. Typical variations of the threshold are $-2$ to $+3\,$\GeV{}
relative to the nominal value.
Optionally, a selection to reject hadronic activity can be applied:
candidate electrons and photons are rejected if the sum of transverse energies
in hadronic towers matched to the $2\times 2$ central region
is at least 1~\GeV{} and exceeds $\et/23.0-0.2$~\gev.
Finally, an EM isolation requirement
can be applied: candidate photons and electrons are rejected if the
sum of transverse energies in the 12 towers
surrounding the $2\times 2$ central region
in the EM layer is at least 2~\GeV{} and exceeds $\et/8.0-1.8$~\gev.
No requirements based on hadronic activity or EM isolation
are applied above 50$\,$\GeV{} of \et{} reconstructed at L1.
These additional selections were optimised to maintain a
fixed L1 efficiency at the lowest possible rate.
The effect of these additional selections on the rate and
efficiency is discussed in Section~\ref{sec:trig_l1}.
 
\subsection{HLT photon reconstruction and identification for $pp$ data-taking}
 
The HLT fast algorithm reconstructs clusters from the calorimeter
cells within the EM RoIs identified by L1. To minimise the HLT latency, the fast algorithm
uses only the second layer of the EM calorimeter to find the cell with the largest
deposited transverse energy in the RoI. This cell is referred to as the `pre-seed'.
Nine possible $3\times7$ windows ($\Delta\eta\times\Delta\phi=0.075\times0.175$) around the pre-seed cell
are checked to ensure that the local maximum, the cluster seed, is found.
The final cluster position is obtained by calculating the energy-weighted average cell
positions inside a $3\times7$ window centred on the cluster seed.
To compute the accumulated energy in all EM calorimeter layers,
a cluster size of $3\times7$ is used in the barrel and a cluster size of $5\times5$ in the endcaps.
Several corrections, based on the offline
reconstruction algorithms, are used at the fast algorithm step in order to improve the resolution
of the cluster position and energy.
 
In this fast reconstruction step, only selections on the cluster \et{} and shower shape parameters\footnote{$R_\mathrm{had}$
is the ratio of the cluster transverse energy in the hadronic calorimeter to that in the EM calorimeter.
\reta{} is based on the cluster shape in the second layer of the EM calorimeter and defined as the ratio of the \et{} in a core region
of $3\times 7$ cells in $\eta\times\phi$ to that in a $7\times 7$ region, expanded in $\eta$ from $3\times 7$ core. $E_\mathrm{ratio}$
is based on the cluster shape in the first layer of the EM calorimeter and
defined as the ratio of the energy difference between the maximum energy deposit and the energy deposit in a secondary
maximum in the cluster to the sum of these energies.} $R_\mathrm{had}$, \reta{} and $E_\mathrm{ratio}$, which have
good discrimination power between background and signal, are applied.
 
In the precision step, offline-like algorithms are used for the reconstruction of calorimeter quantities.
After retrieving the cell information from regions slightly larger than the RoI, the precision HLT
reconstruction uses the offline sliding-window algorithm to construct clusters~\cite{ATL-LARG-PUB-2008-002}.
The energy of the clusters is calibrated using a multivariate technique such that the response of
the calorimeter layers is corrected in data and simulation~\cite{PERF-2017-03}. The online photon
identification relies on the same cluster shower shapes that are used in the offline algorithms
(details are given in Ref.~\cite{PERF-2017-02}), and three identification WPs are defined: `loose',
`medium', and `tight'. The \medium\ identification is used only in the HLT.
 
An optional requirement on calorimeter-only isolation in photon triggers uses topo-clusters,
similar to the offline isolation calculation~\cite{ATLAS-EGAM-2018-01}. Full-scan topo-cluster reconstruction is needed
to compute the energy density of the event on-the-fly in the HLT; this is then used to subtract
ambient noise in the isolation cone. An isolation cone of size $\Delta R=0.2\,(0.4)$ around the photon
candidate is used for the very-loose (tight) isolation requirement, denoted by `icalovloose'
(`icalotight'). If the ratio of the transverse energy in the topo-clusters to the
transverse energy of the photon candidate is less than 10\% (3\%, with an energy offset of 2.45~\gev),
then the photon is considered isolated for `icalovloose' (`icalotight') by the HLT.
These isolation criteria are over 98\% efficient for offline photons satisfying tight isolation.
The full-scan topo-cluster reconstruction
is executed only once per event at the end of the trigger sequence
(as it is very CPU intense) and is common to all isolated triggers and all trigger signatures.

\subsection{HLT electron reconstruction and identification for $pp$ data-taking}
\label{sec:hlt-electron}
 
HLT electron reconstruction also has fast and precision steps.
The description below corresponds to the implementation at the end of Run~2 data-taking;
modifications made to the initial Run~2 implementation are described in Section~\ref{sec:trig_e_evol}.
 
The fast calorimeter reconstruction and selection steps for electrons
have two implementations: a cut-based algorithm and a neural-network-based `Ringer' algorithm.
The former algorithm, the same as described above for photons, is used
for electron triggers with \et$<15$~\GeV{}.
The Ringer algorithm, described in detail in Section~\ref{sec:ringer},
is used for triggering electrons with \et$\ge 15$~\GeV{}.
For both fast algorithms, electron candidates are required to have tracks from the
fast track reconstruction step, performed inside the RoI only,
matching the corresponding clusters as detailed in Section~\ref{sec:trig_e_evol}.
 
In the precision calorimeter reconstruction step, the cluster reconstruction and calibration
are similar to those for photons, and to those used offline in early Run~2 analyses~\cite{PERF-2017-01}.
Precision tracks within the RoI are extrapolated to the second layer of the EM calorimeter and are required to match the
clusters within $|\Delta\eta\textrm{(track, cluster)}|<0.05$ and $|\Delta\phi\textrm{(track, cluster)}|<0.05$~rad.
The offline reconstruction uses a looser, asymmetric condition for the matching in $\phi$~\cite{ATLAS-EGAM-2018-01}
to mitigate the effects of the
energy loss due to bremsstrahlung; this leads to some inefficiency at the trigger level.
In the precision step, the electron selection relies on a multivariate technique using a
LH discriminant with four operating points: `lhvloose', `lhloose', `lhmedium', and `lhtight'.
The identification in the trigger is designed to be as close as possible to the offline version,
but there are a few necessary differences: the discriminating variables used online have different resolutions;
the momentum loss due to bremsstrahlung, $\Delta p/p$, is not accounted for in the online LH.
Triggers with `nod0' suffixed to their names do not include
the transverse impact parameter relative to the beam-line, $d_0$, and its significance, $|d_0/\sigma(d_0)|$,
in the online LH: this reduces inefficiency due to the absence of the bremsstrahlung-aware re-fit in the HLT
and preserves efficiency for electrons from exotic processes which do not originate at the primary vertex.
 
An additional, optional requirement of isolation denoted `ivarloose' is also available for electron triggers.
This tracking-only isolation is required to satisfy \ptIsol$(\Delta R^\mathrm{var}<0.2)/\pt < 0.10$ and is
calculated similarly to the offline isolation working points detailed in Table~\ref{tab-iso} in Section~\ref{sec:egselectionoff}.
 
Some triggers with non-standard electron sequences are also used. For example, triggers with only \et{}
requirements applied in the HLT (fast and precision levels) and no tracking requirements are called `etcut' and
are used both as high-\et\ unprescaled triggers described in Section~\ref{sec:se} and as
prescaled triggers for electron performance studies described in Ref.~\cite{ATLAS-EGAM-2018-01}.
 
\subsubsection{Ringer algorithm}
\label{sec:ringer}
 
\begin{figure}[ht!]
\centering
\includegraphics[width=0.9\textwidth]{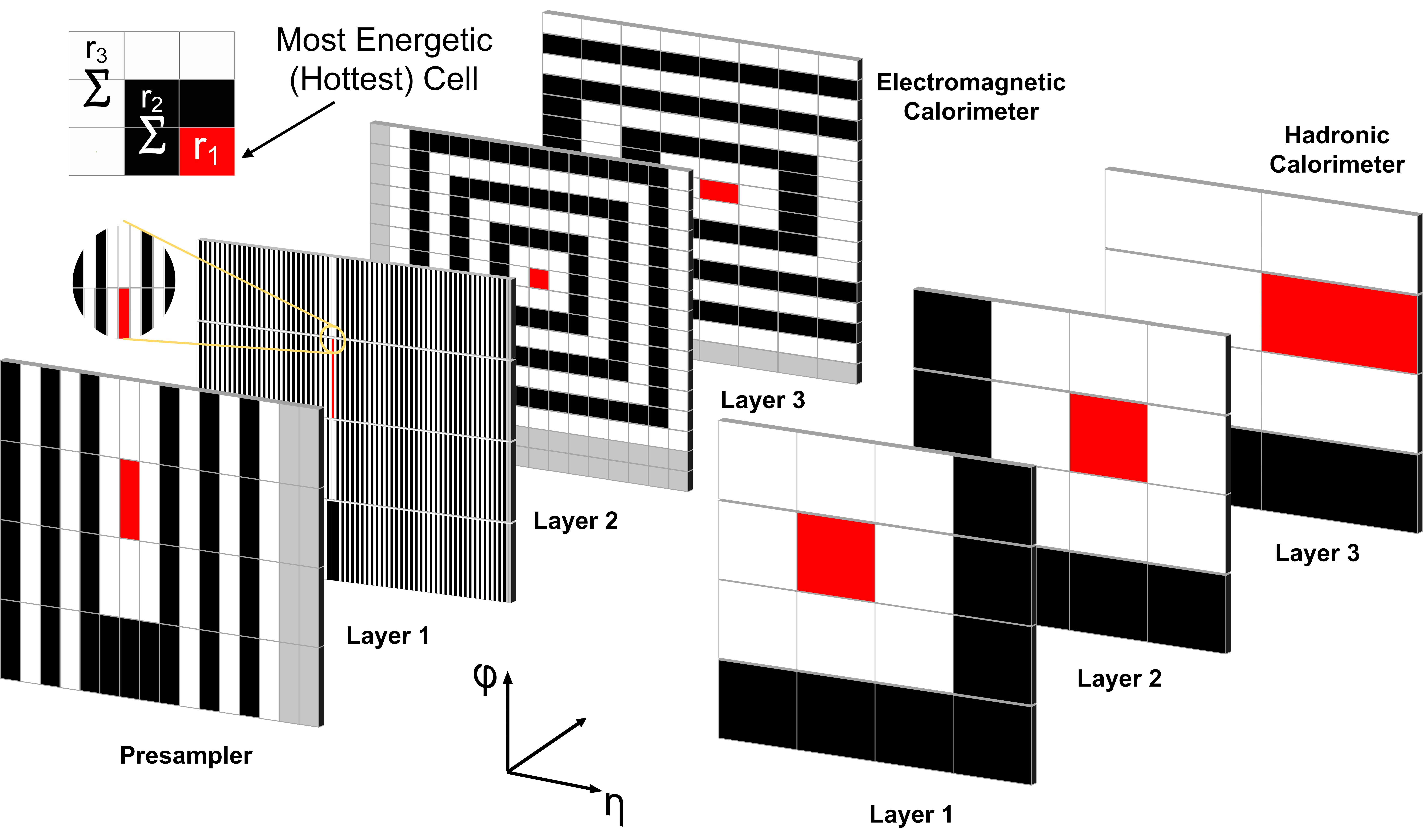}
\caption{The concept of ring-shaped energy reconstruction in a calorimeter slice corresponding to the RoI
size for $|\eta|<1.35$. The most energetic cell ($r_1$) is shown in red, while the rings around it alternate between
black and white. In this scenario there are some areas not used by the Ringer algorithm calculation:
the vertical strip at the right edge of the presampler and layer 1, and the horizontal strip(s) at the lower edge of layers 2 and 3.}
\label{fig-ringer}
\end{figure}

The Ringer algorithm exploits the property of
EM showers to develop in the lateral direction in an approximately conical structure around the initial particle.
This feature allows the relevant information from the calorimeters to be encoded into quantities describing
energy sums ($\Sigma$) of all the cells in a concentric ring~\cite{SEIXAS1996143}, referred to as `rings', in each calorimeter sampling layer.
The rings ($r_2, ... r_n$) are rectangular in shape because of the calorimeter cell structure~\cite{PERF-2007-01}
as illustrated in Figure~\ref{fig-ringer}.
In the EM calorimeter, these rings are centred around the most energetic cell at each layer,
while in the hadronic calorimeter the position of the most energetic cell in the second layer of the EM calorimeter is used as an axis.
A hundred rings are defined in total within an RoI. There are $n=8$ rings for each of the presampler,
second, and third layers of the EM calorimeter, 64 rings in the first layer of the EM calorimeter,
and 4 rings in each of the three layers of the hadronic calorimeter. The transverse energy deposited
in each ring is normalised to the total transverse energy in the RoI.
 
The concatenated vector of 100 normalised ring transverse energies feeds an ensemble of multilayer perceptron (MLP)
neural networks (NN)~\cite{mpl} for each $\et\times\eta$ region. The activation function of the hidden
layer is a hyperbolic tangent. For the 2017 data-taking, the model parameters were optimised on the
simulated $Z\rightarrow ee$ and background data sets described in Section~\ref{sec:datasets}.
In 2018, this optimisation was performed on the 2017 data. The training procedure and parameters
are the same for all specific NNs, except for the number of hidden units in a single-hidden-layer MLP,
optimised in the range of 5 to 20 units using tenfold cross-validation efficiency measurements.
 
The NNs are trained and selected for operation with a heuristic technique to avoid local
optima and achieve an optimal fake rejection, while maintaining the trigger electron efficiency unchanged relative to the cut-based fast-electron selection.
The ensemble composition comprises 20 (25) NNs for 2017 (2018) data-taking, with typically no more than 10 neurons in the hidden layer.
 
Each discriminant requirement is computed as a linear function of $\langle\mu\rangle$ to ensure pile-up
independence of the signal efficiency. To account for mismodelling in the MC simulations, the parameters
of the linear threshold correction were derived using 2016 (2017) collision data for the 2017 (2018) data-taking period.
 
The Ringer algorithm increases the time taken by the fast calorimeter reconstruction step to 1--2$\,$ms
per event, approximately 45$\%$ slower than the cut-based algorithm. However, it reduces the number
of input candidates for the more CPU-demanding fast tracking step (which takes about 64$\,$ms per event)
by a factor of 1.5--6. This factor depends on the detailed trigger configuration.
Overall, the use of the Ringer algorithm enabled at least a $50\%$ reduction in the CPU demand
for the lowest-threshold unprescaled single-electron trigger.
In Run 2, the Ringer algorithm was used only in the electron triggers with \et{} thresholds above 15~\GeV,
because of the availability of a large $Z\rightarrow ee$ event sample in data, useful for the Ringer algorithm validation.
Its implementation for the electron triggers with \et{} thresholds below 15~\GeV{}
was finalized only during 2018, too late to be used for the data-taking.
The Ringer algorithm was not used in the photon triggers during Run 2,
as they do not have any CPU-demanding track reconstruction steps and gains from its implementation are smaller than for electrons.
 
\subsection{HLT photon and electron reconstruction and identification for heavy-ion data-taking}
 
Triggering on both reference $pp$ and $pPb$ collisions relies on
strategies developed for high-pile-up $pp$ data-taking, while for $PbPb$
a dedicated set of triggers is required.
 
One of the main characteristics of HI collisions is event centrality, which is determined
by the total transverse energy measured in the forward calorimeter, FCal $\Sigma\et$.
Small (large) values correspond to events with a small (large) overlap region of two colliding nuclei.
In a $PbPb$ collision, the average background originating from the UE in the calorimeter and
ID can affect the performance of the online reconstruction and identification of photons and electrons.
Unlike pile-up in $pp$ collisions, the UE background cannot be assumed to change slowly with time;
on the contrary, it can be dramatically different event-by-event due to the varying HI
collision centrality.  The tracking performance is approximately centrality-independent, so the
track-related requirements are identical to those for $pp$ collisions.
 
To maintain centrality-independence of the photon and electron trigger performance,
a UE correction is performed in a two-stage approach. First, at the fast calorimeter stage of the HLT,
no shower shapes are used for the online selection and only an \et{} requirement is imposed.
This allows the rate with which the UE correction is applied to be reduced, while consuming
only the resources required for reconstruction in RoIs. Next, the UE correction is evaluated per cell
as an average in $\phi$ for each slice of width $\Delta\eta = 0.1$ in each calorimeter layer.
This calculated average energy
is then subtracted from the cluster constituent cells. As a result, the cells in the RoI
(which are used as inputs to clustering and downstream identification algorithms) contain no systematic
bias due to the UE. Residual fluctuations due to the stochastic nature of the UE remain present.
 
Due to the small size of the EM cluster, this first-order UE correction is found to be sufficient;
the azimuthal modulation originating from the flow phenomenon~\cite{HION-2016-06} in the UE can be neglected.
 
Both the cut-based (photon and electron) and LH (electron) identifications are used for the HI triggers.
Their performance is presented in Section \ref{sec:trig_hi}. The cut-based electron identification
for HI collisions is similar to that used for the $pp$ data-taking in Run 1 (2010--2012)~\cite{PERF-2011-02}.
Two WPs are defined: `loose\_ion' and `medium\_ion', based on a subset of variables used in the
standard electron LH selection. These working points are strictly inclusive. The `medium\_ion'
working point uses more variables than `loose\_ion' to increase background rejection.
 
\subsection{Trigger naming convention}
 
\begin{table}[h!]
\small
\centering
\caption{Selected optional requirements which can be suffixed to photon and electron trigger names.
}
\label{tab:name}
\begin{tabular}{@{\extracolsep{\fill}}l c c}
\toprule
& \textbf{Photon (g)} & \textbf{Electron (e)} \\
\midrule
Identification & loose, medium, tight & lhvloose, lhloose, lhmedium, lhtight \\
Modified identification & & \verb|nod0| -- transverse impact parameter not used \\
Isolation & icalovloose, icalotight & ivarloose \\
Special reconstruction & \multicolumn{2}{c}{\texttt{etcut} -- \et-only requirement applied in the HLT}  \\
& \multicolumn{2}{c}{\texttt{ion} -- triggers for heavy-ion data-taking}  \\
\bottomrule
\end{tabular}
\end{table}
 
The ATLAS Run 2 trigger naming convention used is \texttt{[Trigger level]\_[object multiplicity][object type][minimum \et{} value in \GeV{}]}
and, optionally, an additional string detailing other requirements listed in Table~\ref{tab:name}. Trigger level is L1 or HLT, often omitted for brevity.
Possible object types are `EM' for L1 EM cluster, `g' for HLT photon and `e' for HLT electron.
Additional options at L1, detailed in Section~\ref{sec:L1vh}, are $\eta$-dependence
of the \et{} threshold (denoted by the letter
`V' appended to the trigger name), the hadronic activity veto
(denoted by `H') and the EM isolation requirement (denoted by `I').
Although each HLT trigger is configured with a L1 trigger as its `seed',
the latter is not always mentioned explicitly
as part of the trigger name. 
For example, a trigger with name 2g20\_tight\_icalovloose\_L12EM15VHI is designed to identify at least
two photons at the HLT with \et{}~$>20\,$~\GeV{} each satisfying  `tight' photon identification criteria
and calorimeter-only very loose isolation requirements; here the name explicitly mentions the `seed'
L1 trigger, which requires two isolated L1 EM clusters with $\eta$-dependent
threshold centred on \et{} of 15~\GeV{}.

 
\section{Performance measurement techniques}
\label{sec:tools}
 
\subsection{Rate measurements}
 
The ATLAS data-taking conditions are archived with a time interval of the order of a minute, which defines a luminosity block.
In order to obtain the rate of a given trigger as a function of the instantaneous luminosity~\cite{DAPR-2013-01, LUCID2},
individual rate measurements on different luminosity blocks from all data collected in a given year are used.
If, for a given rate measurement, the ratio of trigger rate to instantaneous luminosity varies by more than 20\% from the average
of other measurements, that measurement is not taken into account as an estimator of the rate for that trigger.
This avoids averaging rate measurements that fluctuate because of unpredictable and temporary changes of LHC collisions.
 
\subsection{Measurement of the electron trigger efficiency}
\label{sec:tnp}
 
The electron trigger efficiency, denoted by $\epsilon_\text{trig}$, can be measured either
for electrons at the HLT (including L1) or for EM clusters at L1. It is estimated
directly from data using the tag-and-probe method described in detail in Ref.~\cite{PERF-2016-01}.
This method selects, from a known resonance such as $Z\rightarrow ee$, an unbiased sample
of `probe' electrons by using strict selection requirements on the second `tag' object.
The efficiency of any given requirement can then be determined by applying
it to the probe sample, after accounting for residual background contamination.
 
The total efficiency, $\epsilon_\mathrm{total}$, may be factorised as a product of two efficiency terms:
\begin{equation*}
\epsilon_\mathrm{total}=\epsilon_\mathrm{offline}\times\epsilon_\mathrm{trig}=
\bigg(\frac{N_\mathrm{offline}}{N_\mathrm{all}}\bigg)\times\bigg(\frac{N_\mathrm{trig}}{N_\mathrm{offline}}\bigg),
\end{equation*}
where $N_\mathrm{all}$ is the number of produced electrons, $N_\mathrm{trig}$ is the number of triggered electron candidates,
$N_\mathrm{offline}$ is the number of isolated, identified and reconstructed offline electron candidates
and $\epsilon_\mathrm{offline}$ is the offline efficiency~\cite{ATLAS-EGAM-2018-01}.
The efficiency of a trigger is computed with respect to a specific offline isolation and identification WP.
Therefore, when presenting the results in Section~\ref{sec:trig_e}, several efficiencies per trigger are
provided and these correspond to a few representative offline electron selections.
 
Events with $Z\rightarrow ee$ decays are collected using unprescaled single-electron triggers (see Section~\ref{sec:trig_e_evol}
for details). The tag electron must be an electron identified offline with the tight selection criteria
(hereafter called `tight offline electron') associated geometrically with the object that
fired the trigger, with \et$>27\,$GeV and $|\eta|<2.47$ and outside the transition region between
the barrel and the endcaps $1.37<|\eta |<1.52$.
For the electron trigger efficiency measurement, the isolation and identification
requirements on the probe are always specified and they have to correspond to the
electron offline identification requirements used in an analysis.
The background subtraction is performed with so-called $Z_\mathrm{mass}$ method~\cite{PERF-2016-01}, in which
the invariant-mass distribution constructed from the tag--probe pair is used to discriminate
electrons from background. The signal efficiency is extracted in a window $\pm15\,$GeV around the $Z$ boson
mass~\cite{PhysRevD.98.030001} and its statistical and systematic uncertainties are derived as described in Ref.~\cite{PERF-2017-01}.
 
Simulated events need to be corrected to reproduce as closely as possible the efficiencies measured in
data. This is achieved by applying `an efficiency correction factor', defined as the ratio of the efficiency
measurement in data to that determined in simulated events, to the event weight in simulation.
The impact of the choice of $Z\rightarrow ee$ events for the efficiency measurement, and
uncertainties in the background estimation, are assessed by varying the
requirements on the selection of both the tag and probe electron candidates and by varying the details of
the background subtraction method as detailed in Ref.~\cite{PERF-2017-01}. The scaling factor and its systematic uncertainty are obtained
from the mean and standard deviation, respectively, of the results produced by the set of independent variations of all these parameters.
The statistical uncertainty is calculated as the average over the statistical uncertainties in all variations.
 
\subsection{Measurement of the photon trigger efficiency}
 
The photon trigger efficiency at the HLT (including L1), denoted by $\epsilon^{\gamma}_{\mathrm{trig}}$,
can be measured by two complementary data-driven methods.
The Bootstrap (BS) method uses photons triggered by a lower level or unbiased trigger,
while the second method uses photons from radiative $Z\rightarrow \ell\ell\gamma$ decays.
 
The BS event sample is collected by L1-only triggers or by loose, low-\ET{} photon triggers.
In the BS method, the photon trigger efficiency can be factorised as the product of two efficiency terms:
\begin{equation*}
\epsilon^{\gamma}_{\mathrm{trig}} = \epsilon_{\mathrm{HLT}|\mathrm{BS}} \times \epsilon_{\mathrm{BS}}.
\end{equation*}
The efficiency of the HLT photon trigger relative to the corresponding BS sample efficiency, $\epsilon_{\mathrm{HLT}|\mathrm{BS}}$,
is measured with offline photons on events in the BS sample.
The BS sample efficiency, $\epsilon_{\mathrm{BS}}$, is computed on collision events recorded by a special `random' trigger,
which runs at a rate of a few Hz, by comparing the number of the BS events with the
number of isolated, identified and reconstructed offline photon candidates in the sample.
The background contamination in this sample is large, which could lead to biases towards a lower efficiency estimate.
Those biases are expected to be small because the photon trigger efficiency is evaluated with respect to
\tight\ and isolated offline photons and a few \GeV{} above the trigger threshold. The trigger efficiency for
background photons fulfilling the \tight\ offline identification is also very high, close to the one of signal photons;
an additional systematic uncertainty is assigned as described below to account for any potential biases.
 
The systematic uncertainty of the trigger efficiency is computed as the discrepancy between the efficiency measured in data
and in simulated \hgg\ (high-\et) and prompt-photon (low-\et) samples. This approach to compute the systematic uncertainties is conservative,
as it also includes the discrepancies between simulation and real data (mismodelling). The main underlying assumption is that
the trigger efficiency in the MC simulation is close to the trigger efficiency in a pure sample of photons in data.
This assumption is supported by the observation of good agreement
between the trigger efficiencies in data and simulation for photons from \Zrad s, discussed in Section~\ref{sec:trig_g}.
 
The size of the data sample collected during Run 2 allows the use of photons from radiative $Z$ decays to
measure the photon trigger efficiencies.  In this method the photon trigger efficiency is measured using a
clean sample of prompt, isolated photons with relatively low $\pt$ from $Z\rightarrow \ell\ell\gamma$
($\ell=e, \mu$) decays, in which a photon is produced from the final-state radiation of one of the two
leptons from the $Z$ boson decay.
 
\begin{figure}[!htpb]
\centering
{\includegraphics[width=0.52\textwidth]{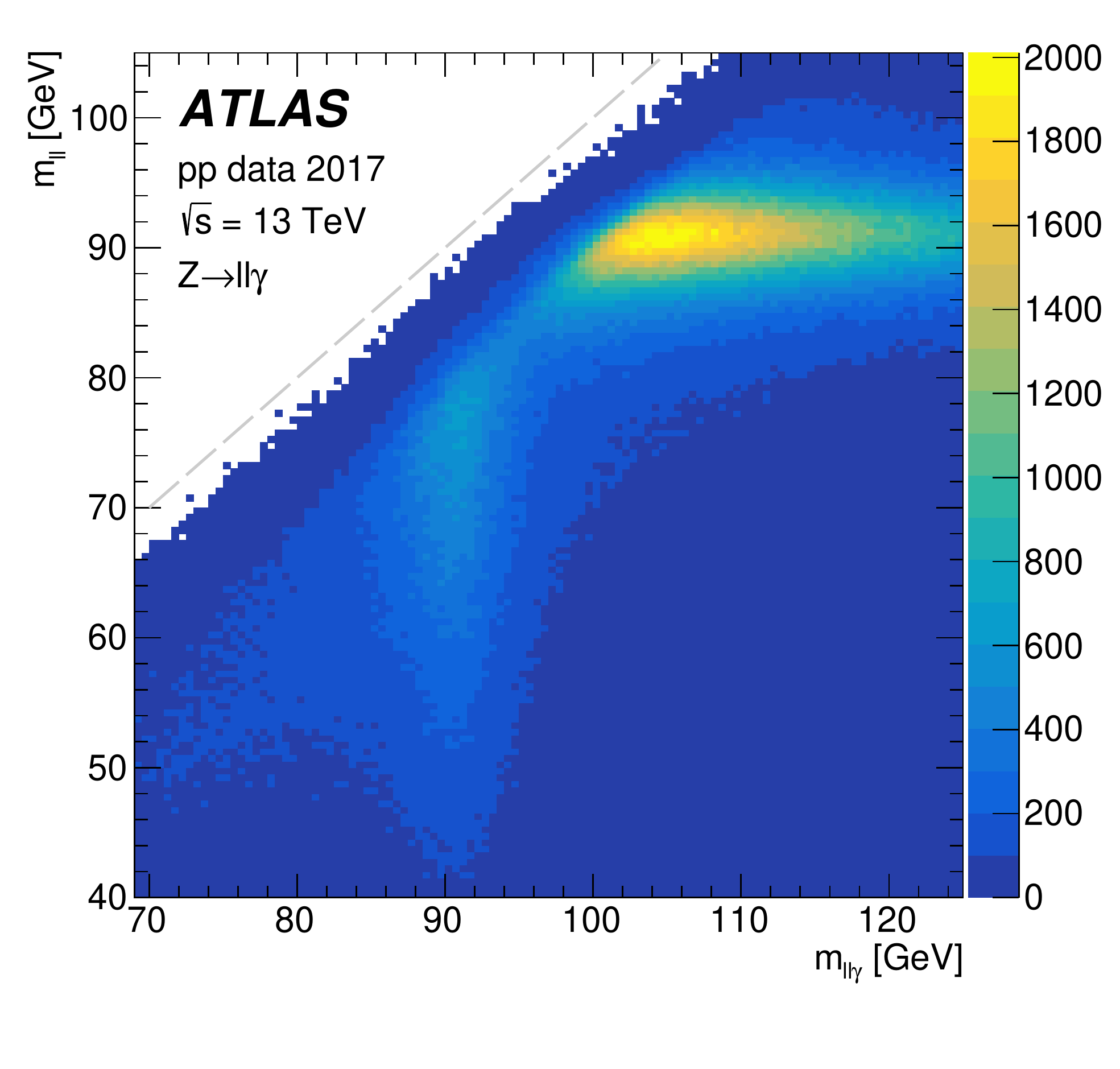}}
\caption{The invariant mass of the two leptons, $m_{\ell\ell}$, vs the invariant mass of the three-body system,
$m_{\ell\ell\gamma}$, in $Z\rightarrow ee \gamma$ and  $Z\rightarrow \mu\mu\gamma$ decays in 2017 $pp$ data.
Events with a photon and a \Zll\ decay are visible as a horizontal band, which peaks around $m_{\ell\ell} = 90$~\GeV{}
for three-body invariant masses above 96~\gev, and
can be easily separated from \Zrad\ events, which are concentrated in a vertical band around $m_{\ell\ell\gamma} = 90$~\GeV.
}
\label{fig:mllg_mll}
\end{figure}

Events triggered by the lowest-threshold unprescaled single and double electron and muon triggers are used to select $Z\rightarrow ee\gamma$ and
$Z\rightarrow \mu\mu\gamma$ event candidates. The sample is selected by requiring events with two opposite-charge
leptons ($ee$ or $\mu\mu$) with $\pt>10$~\GeV{} and a `tight' photon candidate within $|\eta|<2.37$, excluding the
calorimeter transition region, and with $\et>10$~\GeV. The photon candidate is further required to satisfy an isolation WP of interest.
Both leptons are required to satisfy the `medium' identification and `FCLoose' isolation criteria, and must have
$|\eta|<2.47$, with $|z_{0}|<10$~mm and $|d0/\sigma(d_{0})|<10$. The separation between the photon and each lepton is required to be
$\Delta R>0.2$. Figure~\ref{fig:mllg_mll} shows the distribution of $m_{\ell\ell}$  vs the three-body mass, $m_{\ell\ell\gamma}$.
The invariant mass of the two leptons must be within $40<m_{\ell\ell}<83$~\GeV{} to reject events
in which a $Z\rightarrow \ell\ell$ decay is produced in association with a photon coming from initial-state radiation.
The invariant mass of the three-body system is required to be $86<m_{\ell\ell\gamma}<96$~\GeV.
With these requirements, only photons originating from \Zrad s are selected.
If more than one $\ell\ell\gamma$ candidate is found, the one with the three-body mass closest to the $Z$ boson mass is selected.
 
The statistical uncertainty associated with the \Zrad\ method is calculated as a confidence interval of a Bayesian estimate with Jeffrey's prior~\cite{Casadei:2009ic}.
The systematic uncertainty is
estimated by following the strategy used in the $Z\rightarrow ee$ tag-and-probe method for electron measurements described in Section~\ref{sec:tnp}.
In this case, the two leptons are the tags, and the photon is the probe.
The systematic uncertainty is estimated from variations in the trigger efficiency measurement resulting from changing the requirements on the leptons and on the dilepton and three-body systems.
The requirement on the invariant mass of the dilepton system is varied from $30<m_{\ell\ell}<83$~\GeV{} to $50<m_{\ell\ell}<90$~\GeV.
The three-body system mass requirement is varied from  $65<m_{\ell\ell\gamma}<105$~\gev{} to $80<m_{\ell\ell\gamma}<95$~\GeV.
In addition, when considering the electron channel, the identification of the tags is changed from \tight\ to \medium\ for one or both electrons.
 
\begin{figure}[!htpb]
\centering
{\includegraphics[width=0.5\textwidth]{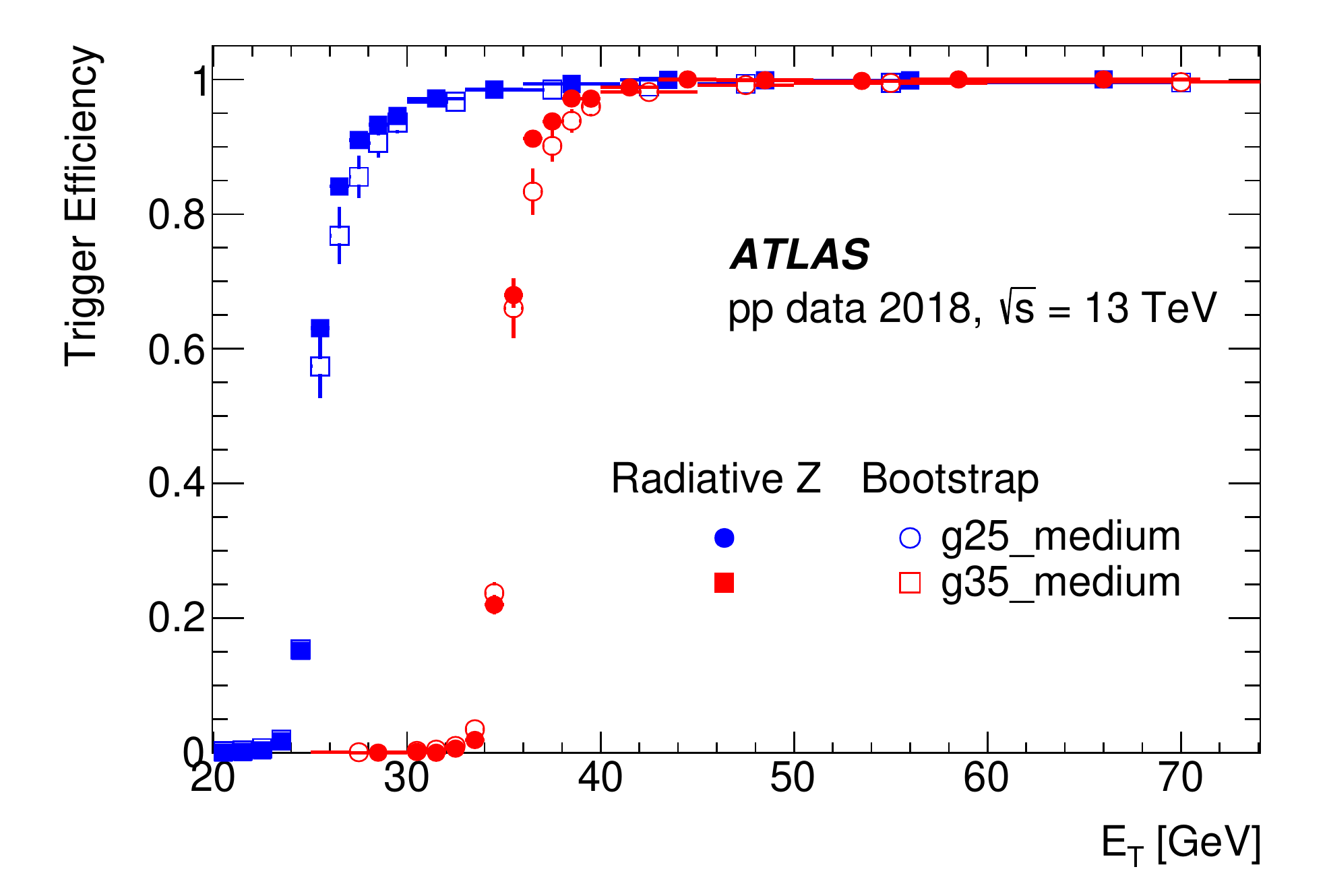}}
\caption{The photon trigger efficiency as a function of the offline photon
$\et$ for both legs of the primary diphoton trigger in the 2018 $pp$ data.
The measurement using the BS method (open markers) is compared with that
using the \Zrad\ (filled markers) method.
The efficiency is computed with respect to offline photons satisfying `tight' identification
criteria and a `calorimeter-only tight' isolation requirement.
Offline photon candidates in the calorimeter transition region
$1.37 < |\eta| < 1.52 $ are not considered.
The error bars indicate statistical uncertainties.
}
\label{fig:BS_ZRad}
\end{figure}

Figure~\ref{fig:BS_ZRad} shows a comparison between the \BS\ and \Zrad\ methods for \medium\ photon triggers in 2018.
The small difference in performance for the turn-on is due to the different purities of the samples: there are significant backgrounds
in the \BS\ sample, and almost inexistent backgrounds in the \Zrad\ measurement, leading to
slightly higher efficiency computed by the \Zrad\ method. This is expected, as the
efficiency for the trigger to select the background present in the \BS\ sample is lower than the efficiency for the trigger to select real photons.
Typically, physics analyses use the photon triggers to select objects with \et\ at least 5~\GeV{} above the trigger threshold.
In that regime, the efficiency measurements of both methods give compatible results.
The \Zrad\ method provides a data-driven sample of photons with very high purity to compute the efficiency for trigger thresholds below 60~\GeV;
above this value the \BS\ method is used.
 
\section{L1 trigger evolution and performance}
\label{sec:trig_l1}
 
\begin{table}[ht!]
\small
\centering
\caption{List of unprescaled L1 EM triggers in different data-taking periods during \RunTwo.
}
\label{tab:EM}
\begin{tabular}{@{\extracolsep{\fill}}  l | l | l | l }
\toprule
\textbf{L1 EM trigger type} & \textbf{2015} & \textbf{2016} & \textbf{2017--2018} \\
\midrule
{Single object} & L1\_EM20VH  & \multicolumn{2}{c}{L1\_EM22VHI} \\\hline
{Diobject} & L1\_2EM10VH & L1\_2EM15VH & L1\_2EM15VHI \\
\bottomrule
\end{tabular}
\end{table}
 
\begin{figure}[!ht]
\centering
\subfloat[][]{            {\includegraphics[width=0.49\textwidth]{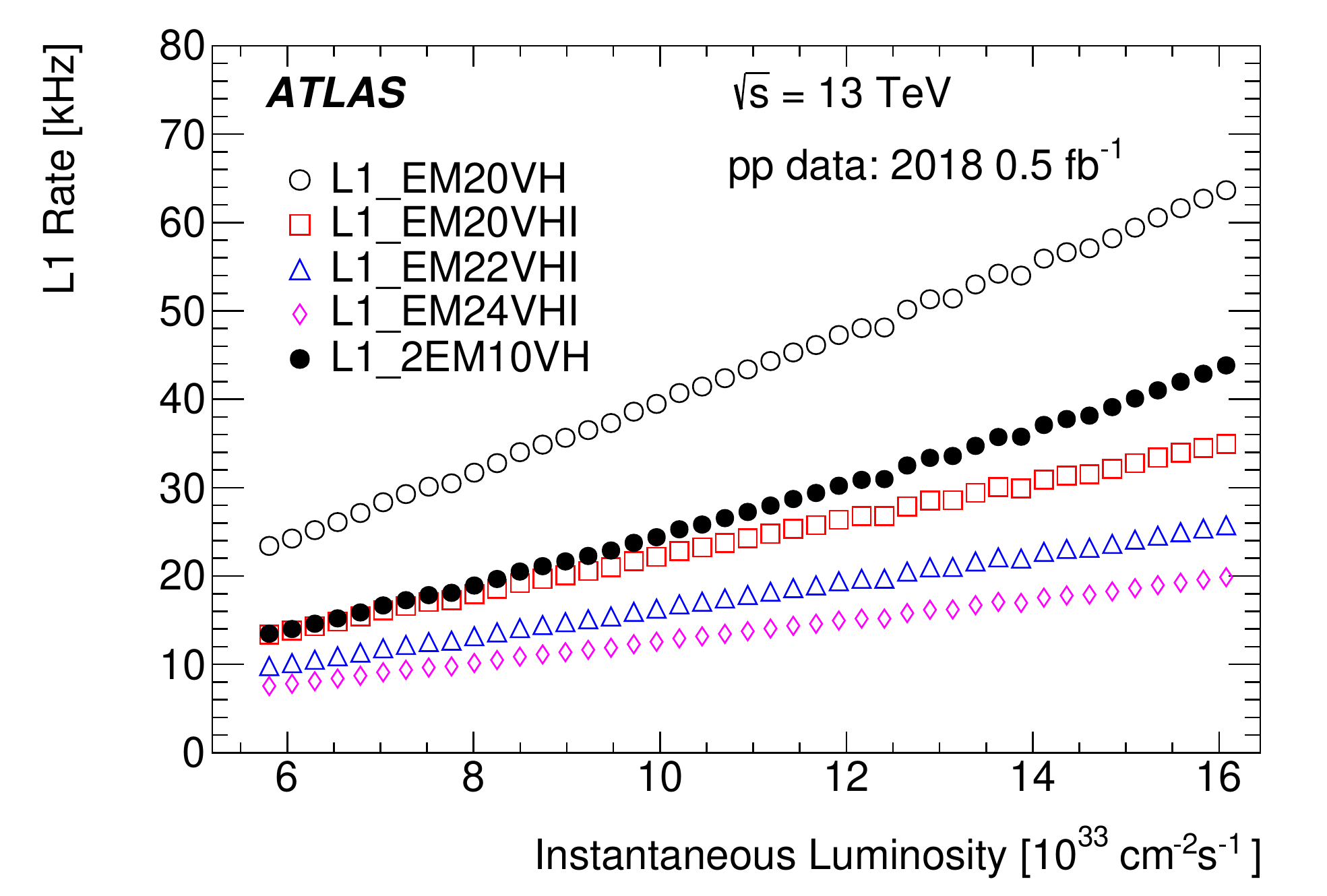}}}
\subfloat[][]{            {\includegraphics[width=0.49\textwidth]{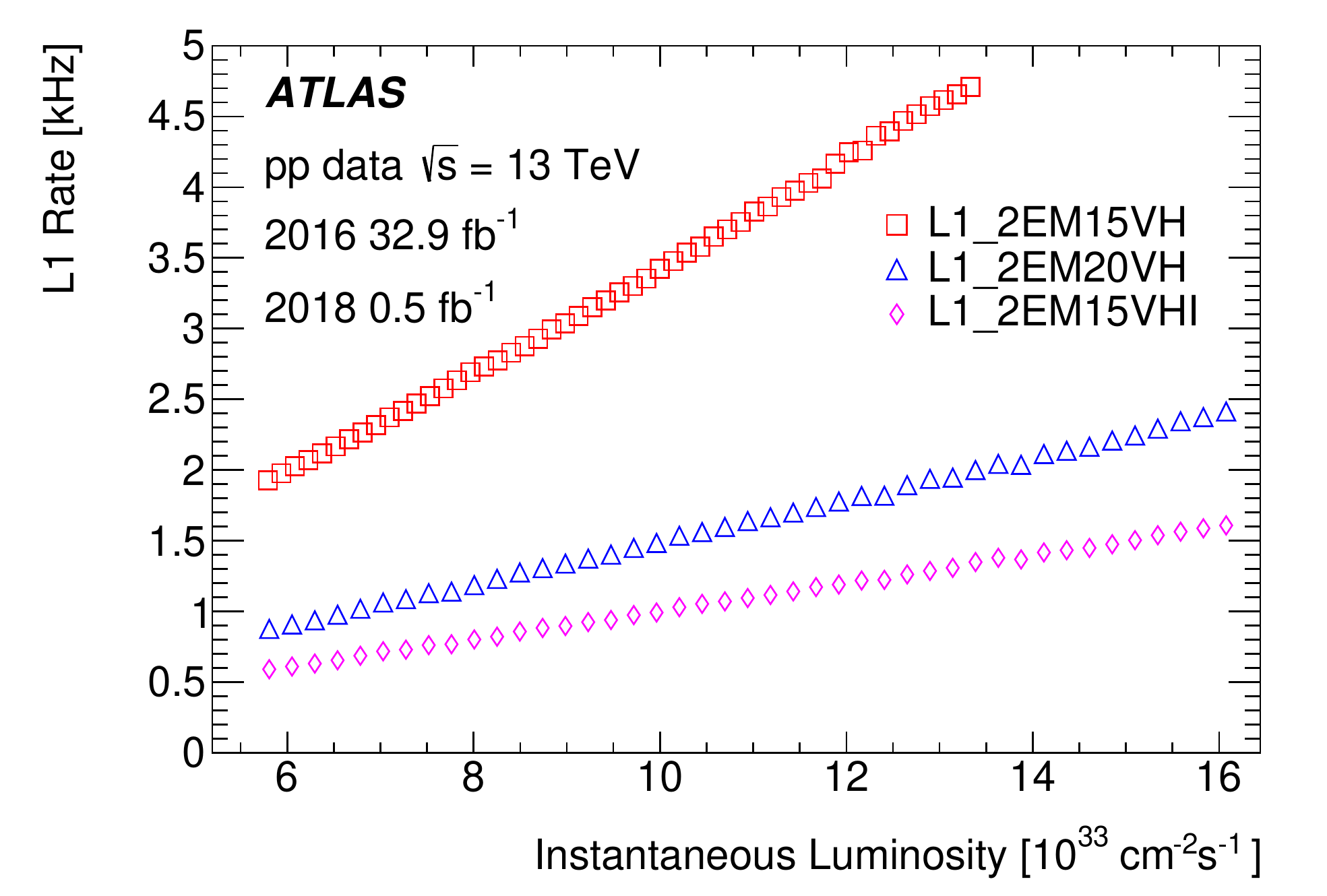}}}
\caption{Dependence of the L1 trigger rates on the luminosity
for single-object and diobject L1 EM triggers.
The 2016 $pp$ data is used for the L1\_2EM15VH trigger, as this trigger was unavailable in 2018.
}
\label{fig:l1-rate}
\end{figure}
 
\begin{figure}[!htpb]
\centering
\subfloat[][]{       {\includegraphics[width=0.50\textwidth]{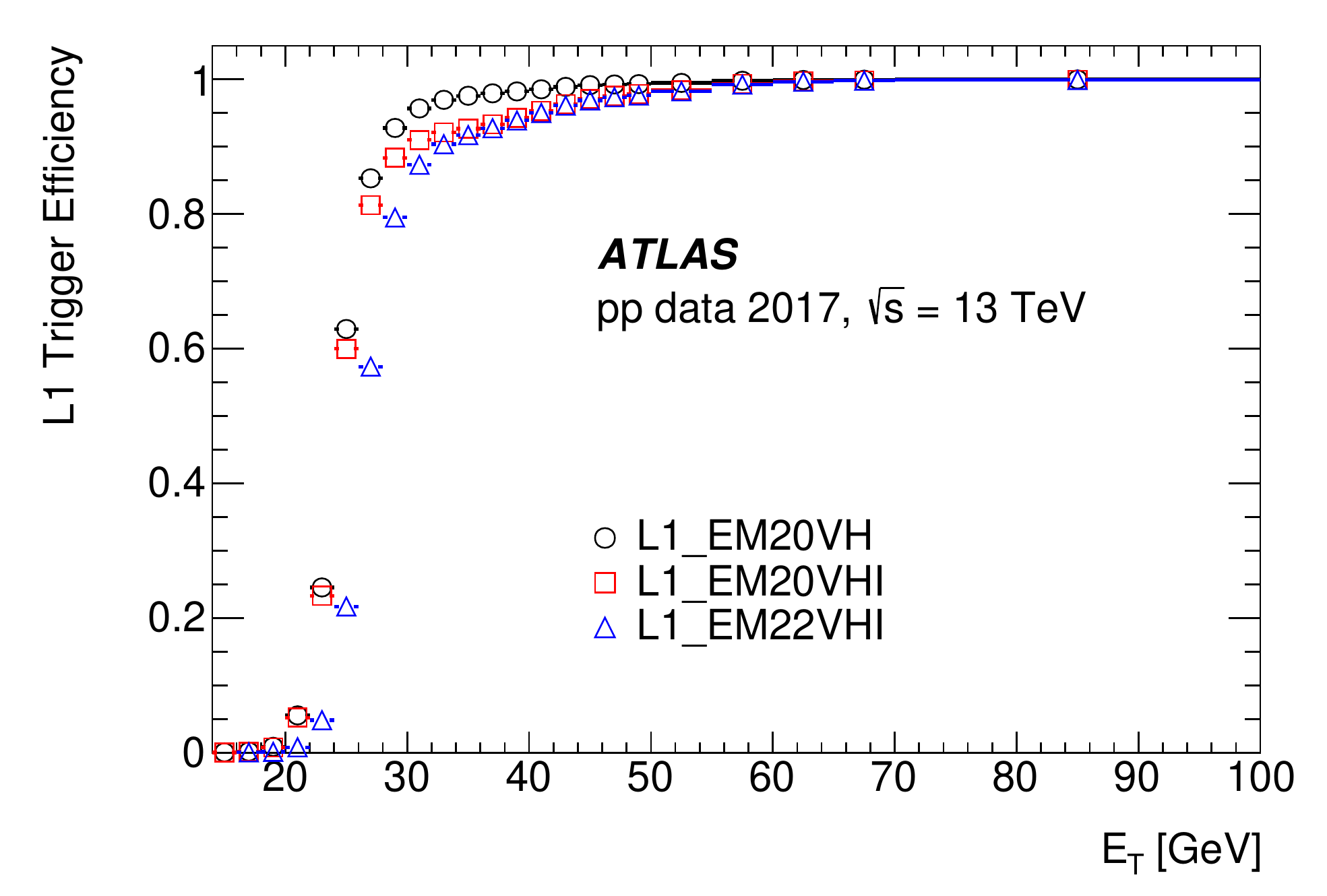}}}
\subfloat[][]{       {\includegraphics[width=0.50\textwidth]{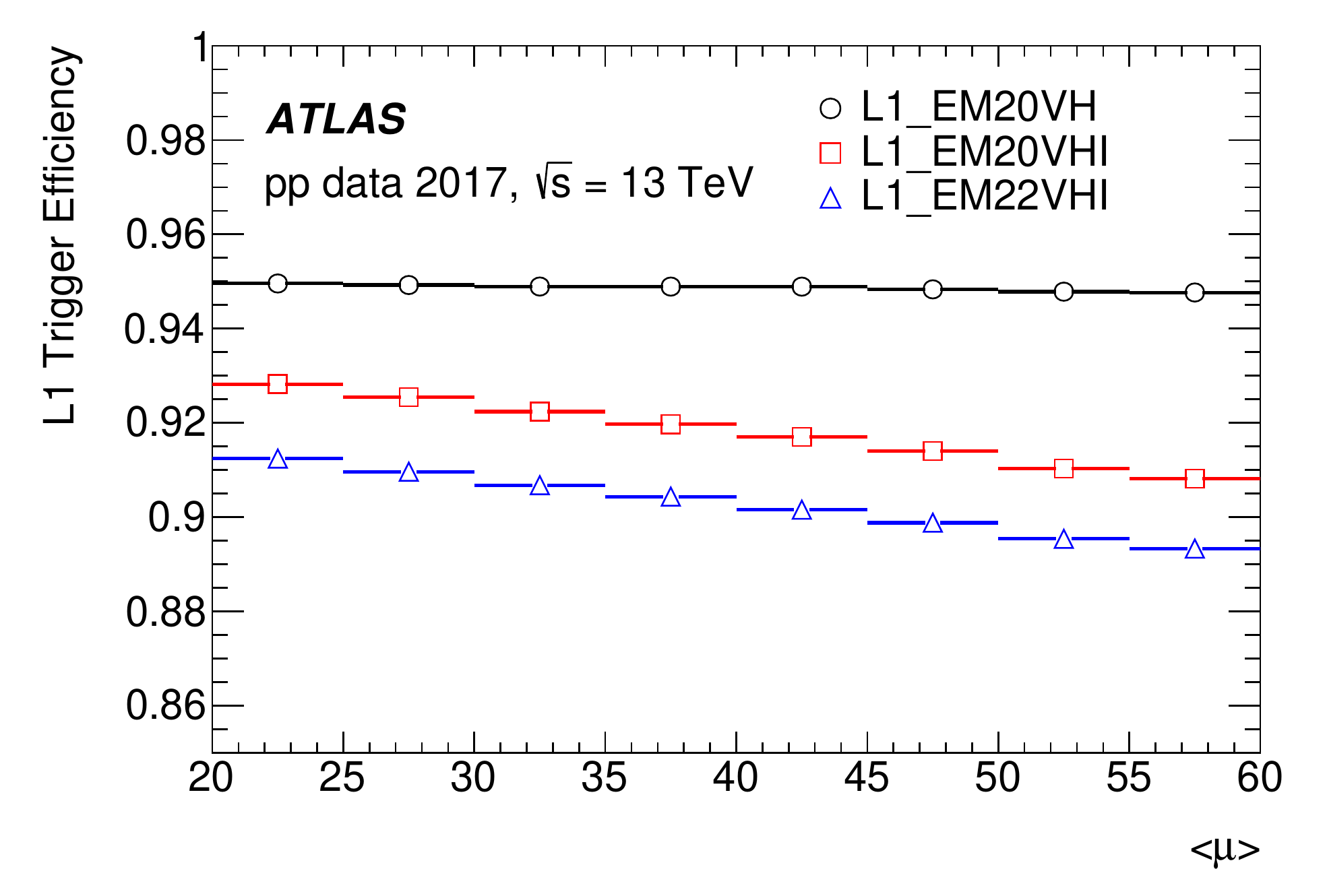}}}
\caption{Efficiency of L1 single EM object triggers
as a function of (a) the offline electron \et{} and (b) pile-up.
The offline reconstructed electron is required to pass a likelihood-based `tight' identification.
The efficiencies are measured with a tag-and-probe method in $Z\rightarrow ee$
data events, using offline monitoring tools described in Section~\ref{sec:ops}.
The error bars show the statistical uncertainties only. No background subtraction is applied,
as the effect is expected to be negligible.
For (b), only offline candidates with \et{} values at least 1~\GeV{}
above the corresponding trigger threshold are considered.
}
\label{fig:l1iso}
\end{figure}

Table \ref{tab:EM} shows the lowest-threshold unprescaled single-object and diobject L1 EM triggers
in the different data-taking periods during \RunTwo.
Figure~\ref{fig:l1-rate} shows the L1 rates for the single-object and diobject
EM calorimeter triggers during Run 2. Among the triggers shown there are two single-object triggers
not used in Run 2: the L1\_EM20VHI trigger, to highlight the additional rejection from the
EM isolation requirement, and L1\_EM24VHI, which was a `backup' trigger for single EM objects.
The single-object L1 rates in Figure~\ref{fig:l1-rate}(a) are well described
by a linear fit as a function of luminosity, with an approximately
zero intercept, indicating a negligible contribution from effects not related to $pp$ collisions,
as expected for such a narrow RoI window.
For single-object L1 EM triggers with \et{} in the range 20--24$\,$\GeV, the rate is reduced by approximately
25\% when the threshold is raised by $2\,$\GeV{}.
For diobject L1 EM triggers this reduction in rate depends on the threshold:
for a 5$\,$\gev{} increase from L1\_2EM10VH (L1\_2EM15VH) a reduction of 90\% (50\%) is achieved.
An additional EM isolation requirement (I) leads to a consistent rate reduction of about 44\% per leg
for single-object (L1\_EM20VH) and diobject (L1\_2EM15VH) triggers and
a pile-up-dependent efficiency loss of at most 5\% up to 50$\,$\GeV{}, as shown in Figure~\ref{fig:l1iso}.
No isolation requirements are applied above this \et{} value.
 
Dedicated combined-object triggers such as three EM clusters or an EM cluster with a muon, $\tau$-lepton, jet,
or missing transverse momentum are also implemented at L1,
allowing L1 EM trigger thresholds to be lowered to 7~\GeV. Some L1 triggers do not require an additional
EM object for their rate reduction. In this case, their HLT photon/electron reconstruction is seeded by
the lowest available L1 EM threshold, which is 3~\GeV.
 
Additional topological requirements (invariant mass, $\Delta R$, etc.)
can also be applied to L1 triggers to further reduce the rate.
For example, prescaled triggers used to collect $J/\psi\rightarrow ee$ events
for the low-\et~electron performance studies~\cite{PERF-2017-01,PERF-2017-03,ATLAS-EGAM-2018-01}
have at least one L1 EM threshold as low as 3~\GeV~and
a requirement on the invariant mass of the EM object pairs to lie between 1 and 5~\GeV~\cite{ATLASBook}.
The latter requirement leads to trigger rate reduction factors of 4--9, depending on the exact trigger
threshold configuration.
 
\section{Photon trigger evolution and performance in $pp$ data-taking}
\label{sec:trig_g}
 
\subsection{Evolution of photon triggers in Run 2}
\label{sec:eph}
 
Table \ref{tab:lowestunprescaledg} shows the lowest-threshold unprescaled photon triggers in
different data-taking periods during \RunTwo. The `loose' and `medium' identification
requirements remained unchanged throughout 2015--2018. An optimisation of the selection
for the online `tight' definition was performed at the end of 2017 in order to synchronise
with a reoptimised offline `tight' photon selection. The calorimeter-only isolation requirement
(icalovloose) was implemented in the HLT for tight diphoton triggers for the first time in 2017.
 
\begin{table}[ht!]
\small
\centering
\caption{List of unprescaled triggers with photons in different data-taking periods during \RunTwo.
The corresponding L1 trigger threshold is given in brackets. No L1 isolation is applied for L1 $\et>50$~\gev.}
\label{tab:lowestunprescaledg}
\begin{tabular}{@{\extracolsep{\fill}}  l | l | l | l }
\toprule
\textbf{Trigger type} & \textbf{2015} & \textbf{2016} & \textbf{2017--2018} \\
\midrule
{Primary}
& g120\_loose  & \multicolumn{2}{c}{g140\_loose } \\
{single photon}
& (EM22VHI) & \multicolumn{2}{c}{(EM22VHI)}\\[1ex]
\hline
\multirow{2}{*}{Primary diphoton}
& \multicolumn{2}{c|}{g35\_loose\_g25\_loose}  & g35\_medium\_g25\_medium  \\
& \multicolumn{2}{c|}{(2EM15VH)} & (2EM20VH)\\[1ex]
\hline
{Loose diphoton}
& \multicolumn{2}{c|}{}  & 2g50\_loose (2EM20VH)  \\[1ex]
\hline
\multirow{2}{*}{Tight diphoton}
& 2g20\_tight& 2g22\_tight & 2g20\_tight\_icalovloose \\
& (2EM15VH) & (2EM15VH)  & (2EM15VHI) \\
\bottomrule
\end{tabular}
\end{table}
 
\subsection{Primary single-photon and diphoton triggers}
 
\begin{figure}[ht!]
\centering
\subfloat[][]{\includegraphics[width=0.495\textwidth]{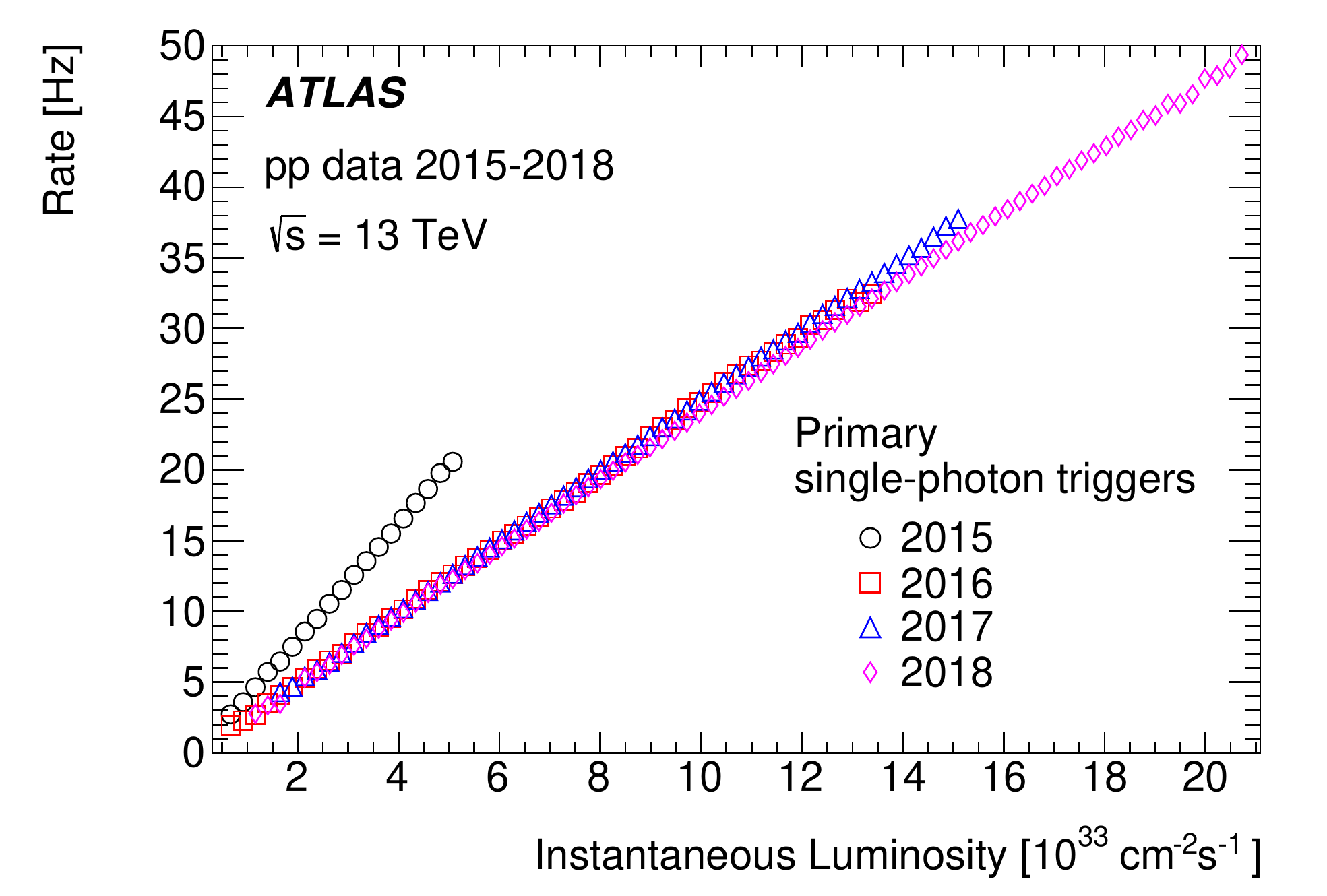}}
\subfloat[][]{\includegraphics[width=0.495\textwidth]{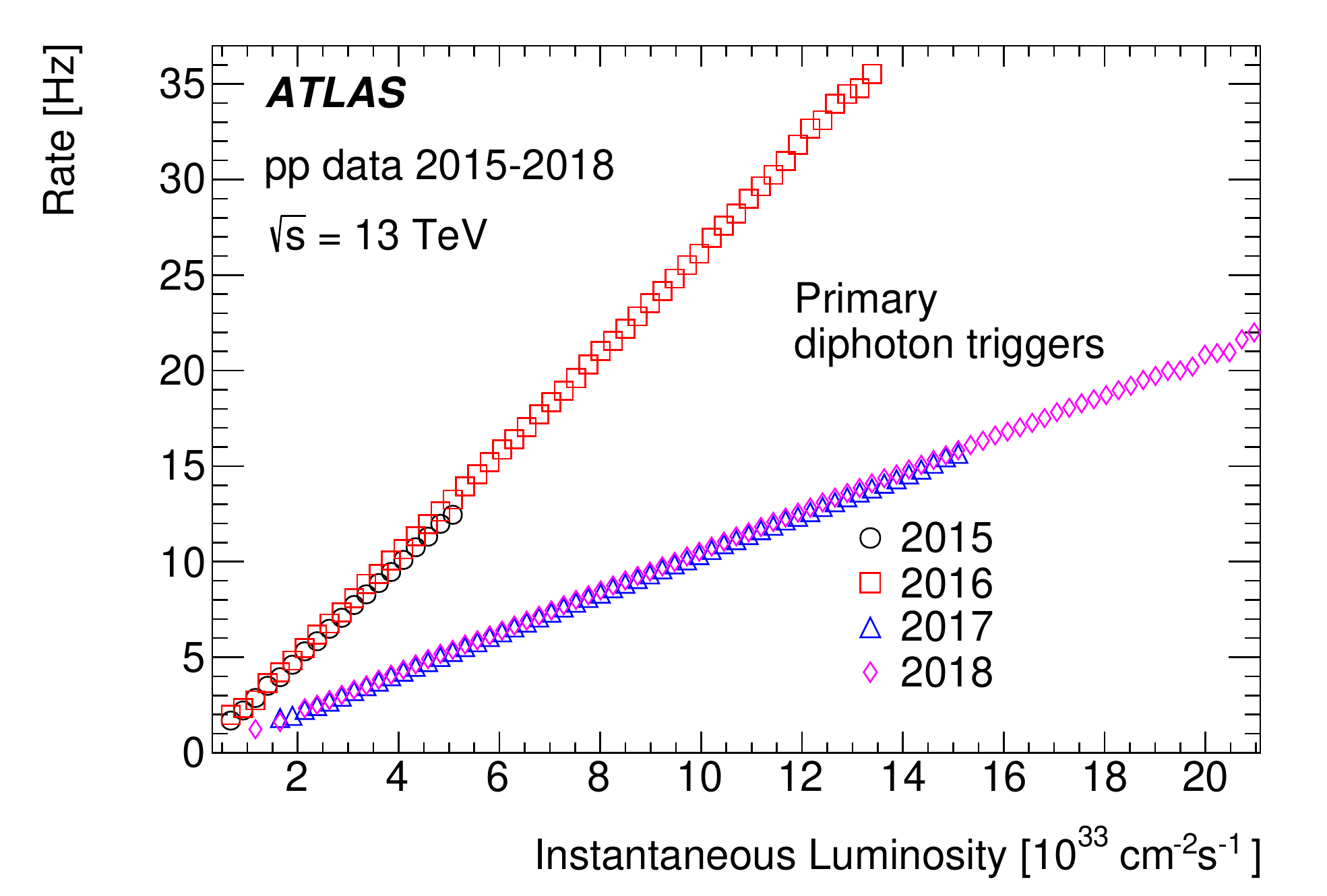}}
\caption{HLT output rate as a function of instantaneous luminosity for the primary (a) single-photon triggers
and (b) diphoton triggers in 2015--2018. The changes between years are detailed in Section~\ref{sec:eph}.}
\label{fig:photon-rates}
\end{figure}
 
The lowest-threshold unprescaled single-photon trigger is primarily designed to trigger on high-\eT{} photons in
searches for new phenomena beyond the Standard Model.
This primary single-photon trigger uses the `loose' identification requirement, with no isolation applied.
Figure~\ref{fig:photon-rates} shows the HLT trigger rates for photon triggers as a function of instantaneous luminosity.
The \et\ threshold of the single-photon trigger was increased from 120 to 140~\GeV{} in 2016 to keep its acceptance rate below 50~Hz, as shown in Figure~\ref{fig:photon-rates}(a).
 
\begin{figure}[!htpb]
\centering
\subfloat[][]{\includegraphics[width=0.495\textwidth]{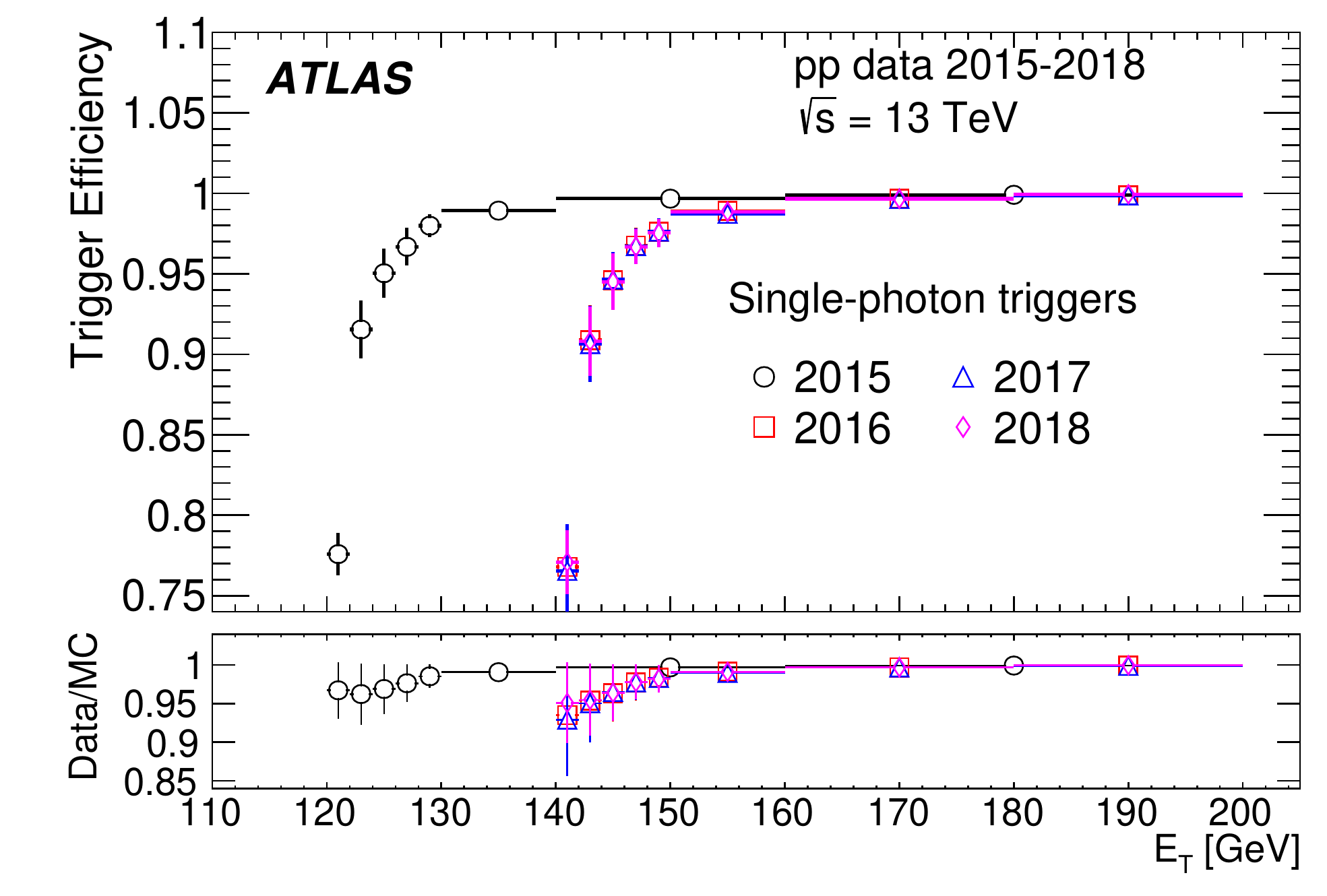}}
\subfloat[][]{\includegraphics[width=0.495\textwidth]{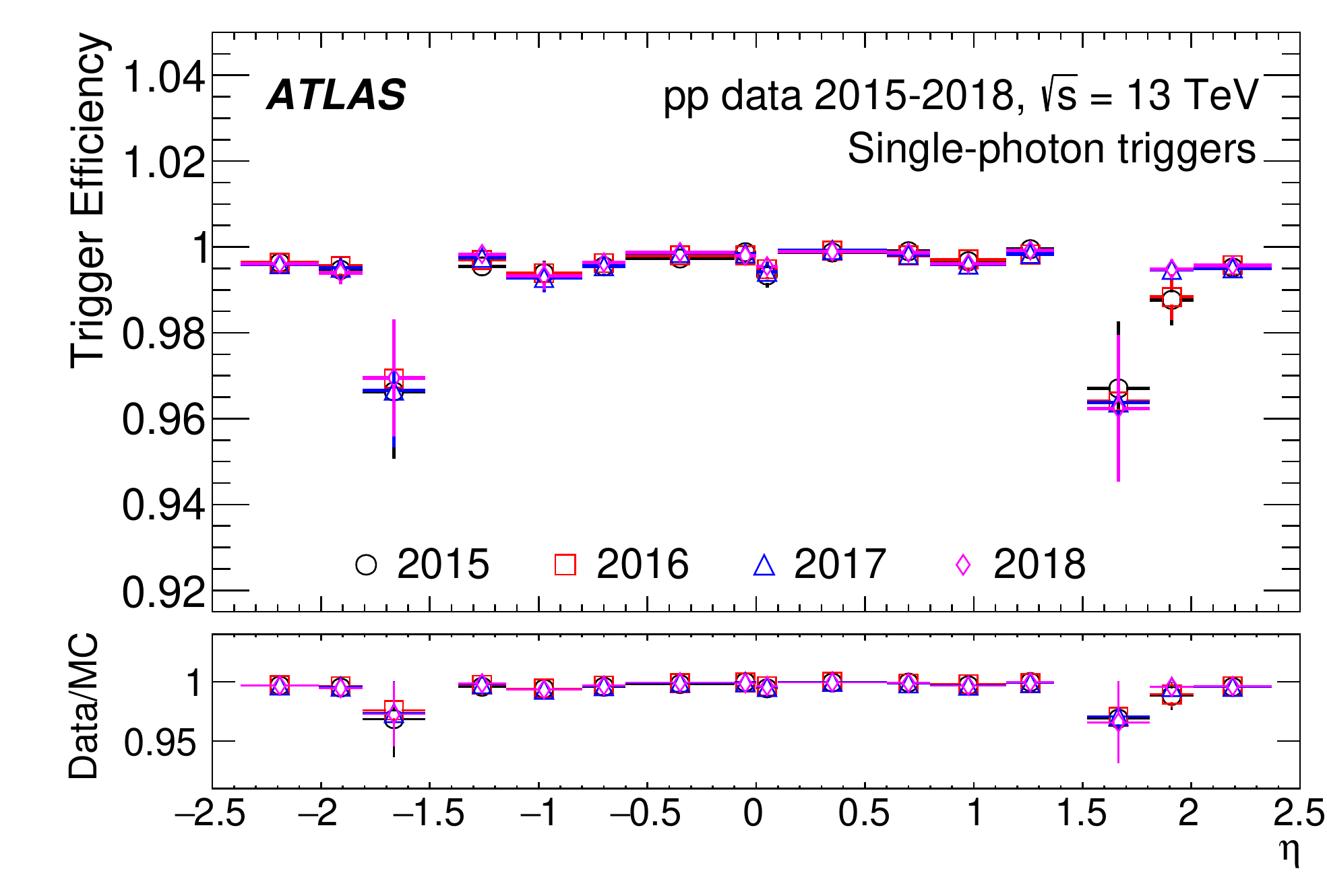}}
\caption{Efficiencies of single-photon triggers in 2015--2018 as a function of the offline photon (a) $\et$ and (b) $\eta$.
The changes between years are detailed in Section~\ref{sec:eph}. The efficiency is computed \wrt offline photons satisfying `tight' identification
criteria and a `calorimeter-only tight' isolation requirement.
The ratios of data to MC simulation efficiencies are also shown.
The total uncertainties, shown as vertical bars, are
dominated by systematic uncertainties.
Offline photon candidates in the calorimeter transition region
$1.37 < |\eta| < 1.52 $ are not considered. For (b), only offline candidates with \et{} values 5~\gev{} above the corresponding trigger threshold are used.
}
\label{fig:singlephoton}
\end{figure}

The efficiencies of the single-photon triggers in 2015--2018, measured with the  \BS\ method, are shown in Figure~\ref{fig:singlephoton} as a function of $\et$ and $\eta$.
The total uncertainties, shown as vertical bars, are dominated by systematic uncertainties, especially differences between data and Monte Carlo simulation.
The trigger efficiency measurement has a total uncertainty of the order of 1\% for photons with \et{} values 5~\GeV{}
above the trigger threshold, and an uncertainty of less than 0.1\% for photons at least 10~\GeV{} above the trigger threshold.
 
Primary diphoton triggers are mainly designed for efficient selection of events with Higgs boson candidates in the diphoton decay channel.
Trigger \et{} thresholds of 35 and 25~\GeV{} for the leading and subleading photons allow the collection of diphoton events with invariant
masses low enough for good background modelling for resonances above 120$\,$\GeV{}, such as the Higgs boson~\cite{HIGG-2012-27,CMS-HIG-12-028}.
During 2015 and 2016 `loose' identification was used at the HLT for primary diphoton triggers.
During 2017--2018, `medium' identification was used in order to keep the primary diphoton trigger rate
below 20~Hz at higher values of instantaneous luminosity, as shown in Figure~\ref{fig:photon-rates}(b).
The rate of the primary diphoton triggers shows a linear dependence on the instantaneous luminosity.
As shown in Table \ref{tab:lowestunprescaledg},
diphoton triggers with `loose' identification were maintained at higher trigger \et{} thresholds (50~\gev).
 
\begin{figure}[!htpb]
\centering
\subfloat[][]{\includegraphics[width=0.495\textwidth]{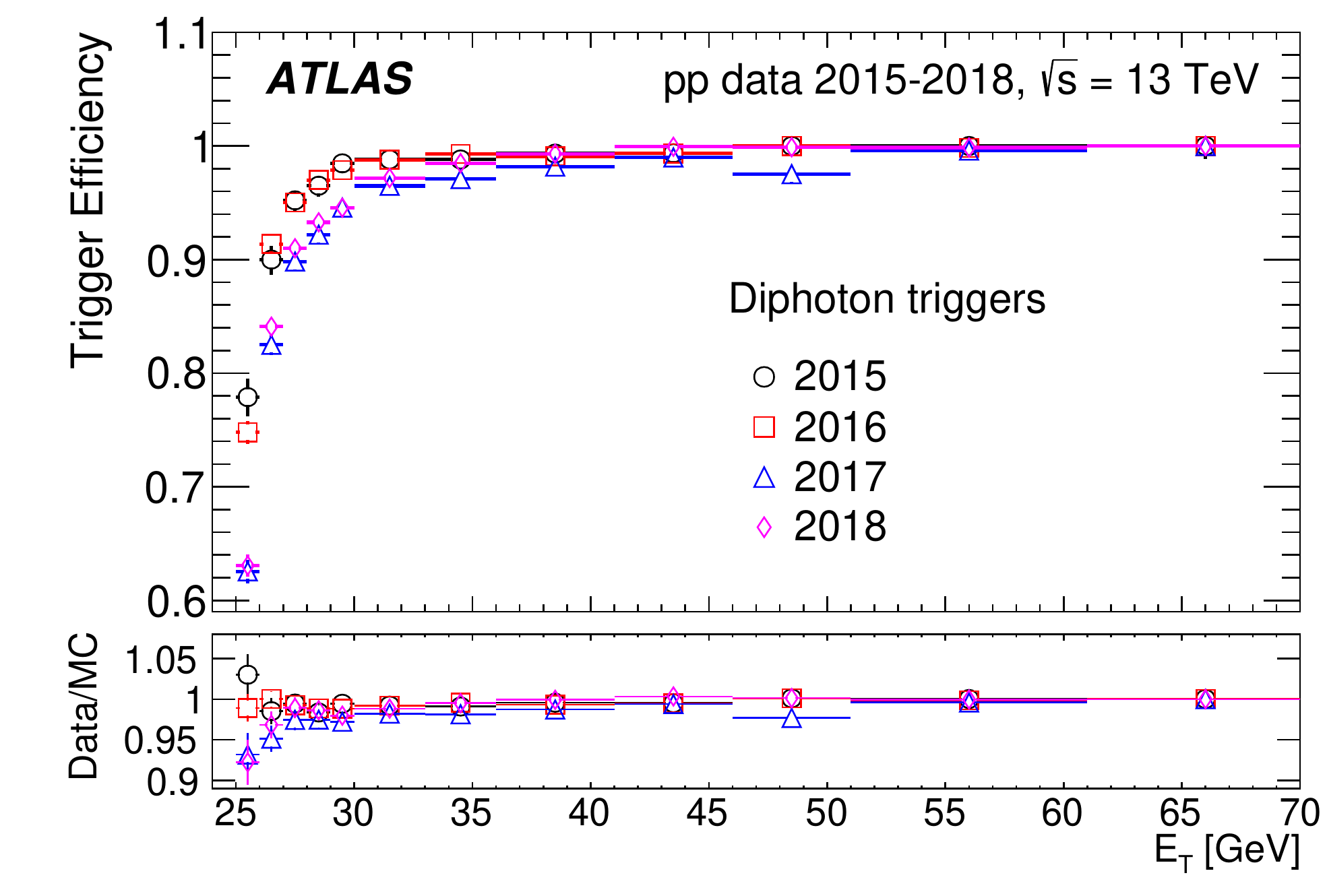}}
\subfloat[][]{\includegraphics[width=0.495\textwidth]{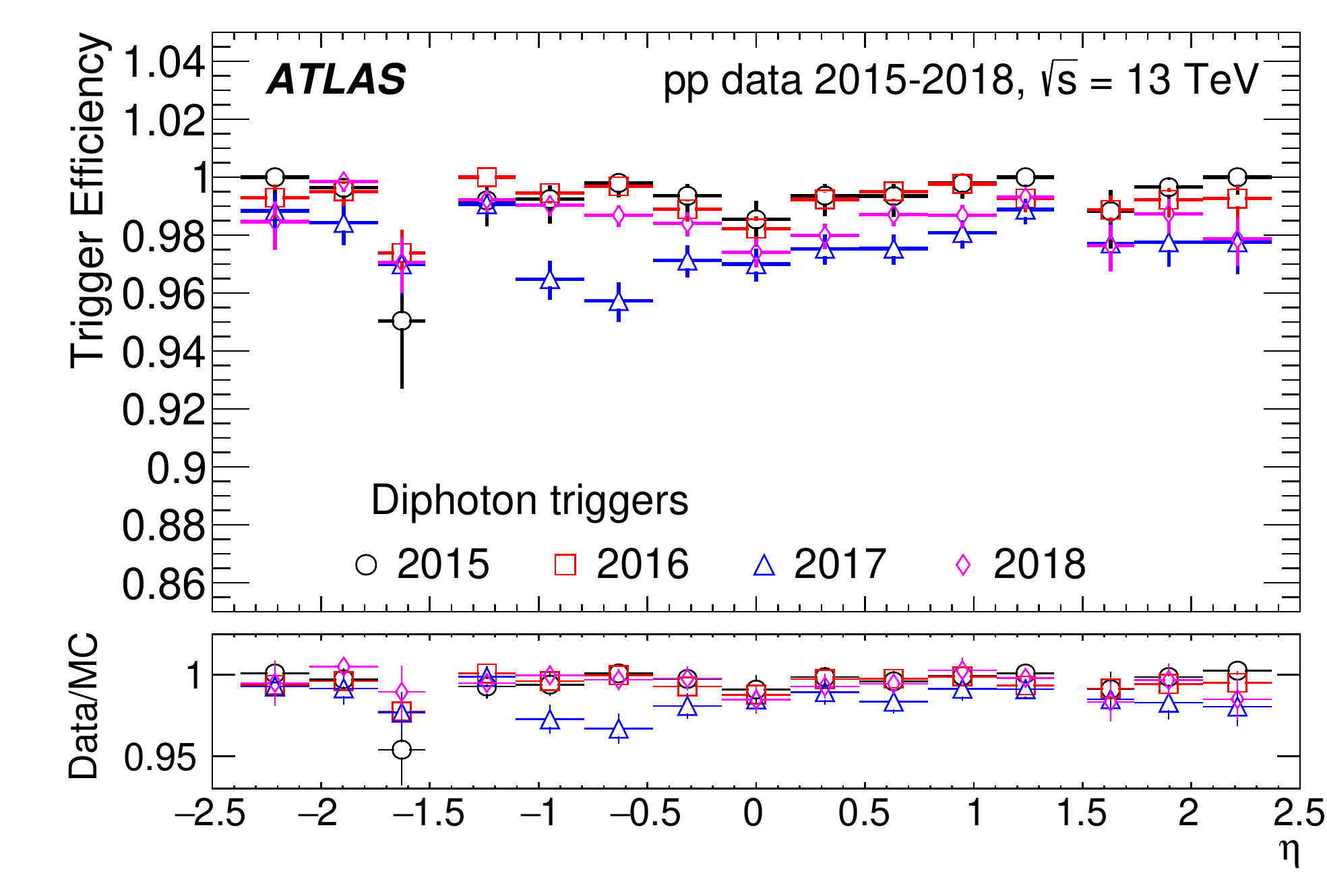}} \\
\subfloat[][]{\includegraphics[width=0.495\textwidth]{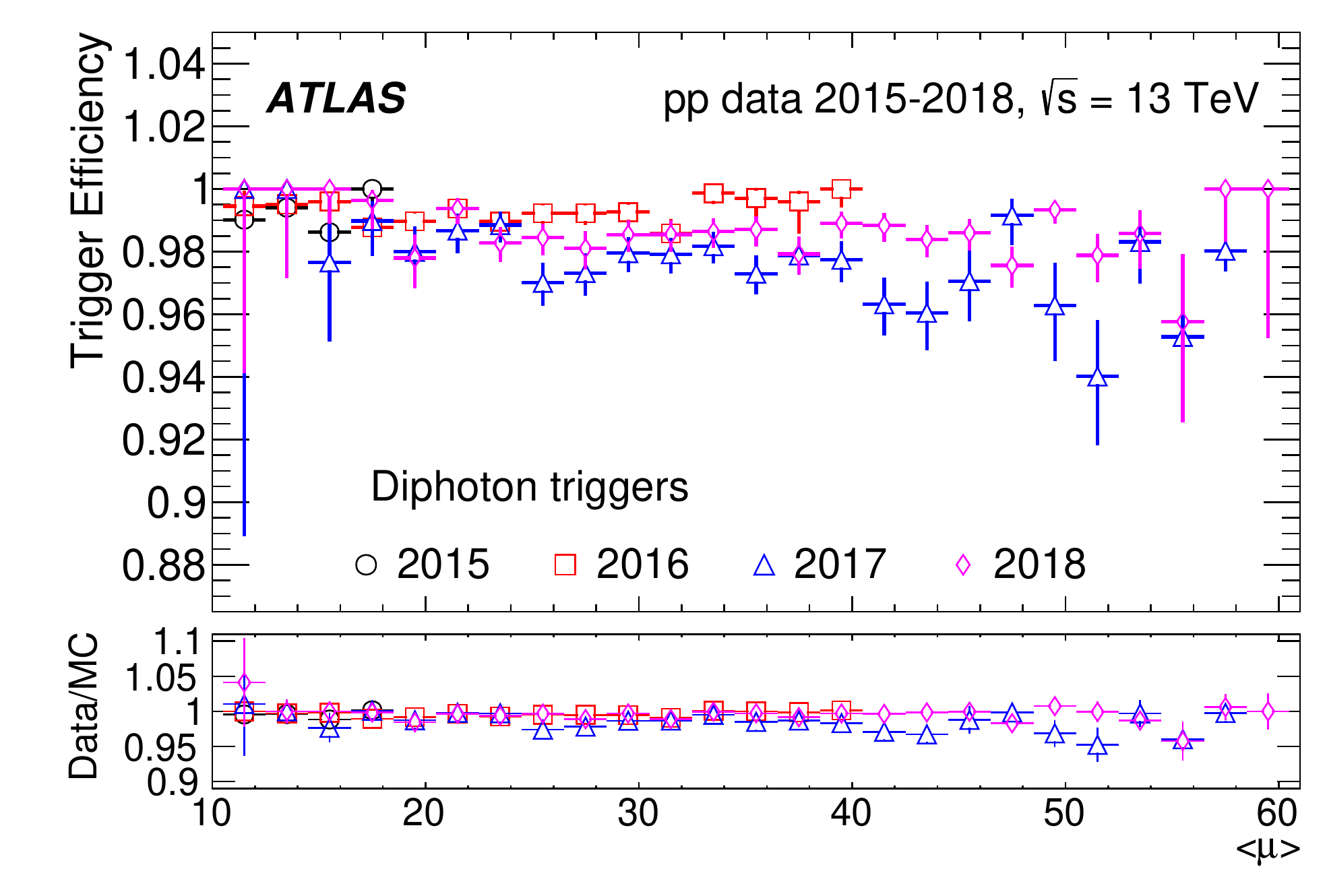}}
\caption{Efficiencies of the 25~\GeV{} leg of primary diphoton triggers in 2015--2018 as a function of
the offline photon (a) $\et$, (b) $\eta$ and (c) $\langle\mu\rangle$.
The changes between years are detailed in Section~\ref{sec:eph}.
The ratios of data to MC simulation
efficiencies are also shown. The efficiency is computed \wrt offline photons
satisfying `tight' identification criteria and a `calorimeter-only tight' isolation requirement.
Offline photon candidates in the calorimeter transition region $1.37 < |\eta| < 1.52$ are not considered.
For (b) and (c), only offline candidates with $\et>30$~\gev{} are used. The error bars indicate
statistical and systematic uncertainties combined in quadrature.
}
\label{fig:diphoton}
\end{figure}

To measure the efficiency of primary diphoton triggers in data, photons from \Zrad s are used. Trigger efficiencies for each of the legs of the
diphoton trigger are measured separately and then combined at the analysis level. This approach is used for all the multi-object and combined triggers.
The efficiencies for the 25~\GeV{} leg of the primary diphoton triggers in 2015--2018 are shown in Figure~\ref{fig:diphoton}.
Slightly lower efficiencies are observed in 2017--2018 due to the tightening of the online photon identification from the `loose' to `medium'
WP: it is $\sim 95\%$ efficient for events with offline `tight' isolated photons with \et\ at least 5$\,$\GeV{}
above the trigger threshold.
Trigger efficiencies show no significant dependence on $\eta$ or $\langle\mu\rangle$, remaining close to 100\% during most of \RunTwo;
the 2017 efficiency is the lowest of all years due to a different LHC bunch structure.
The total uncertainties, shown as vertical bars, are dominated by statistical uncertainties. The ratios of efficiency measured in data to that in
MC simulation are shown in Figure~\ref{fig:diphoton} as functions of $\et$, $\eta$ and $\langle\mu\rangle$, and are close to 1 in all cases,
confirming good data/MC simulation agreement and validating the systematic uncertainty procedure for the BS method.
 
\subsection{Tight diphoton triggers for searches for low-mass resonances}
 
\begin{figure}[ht!]
\centering
{\includegraphics[width=0.495\textwidth]{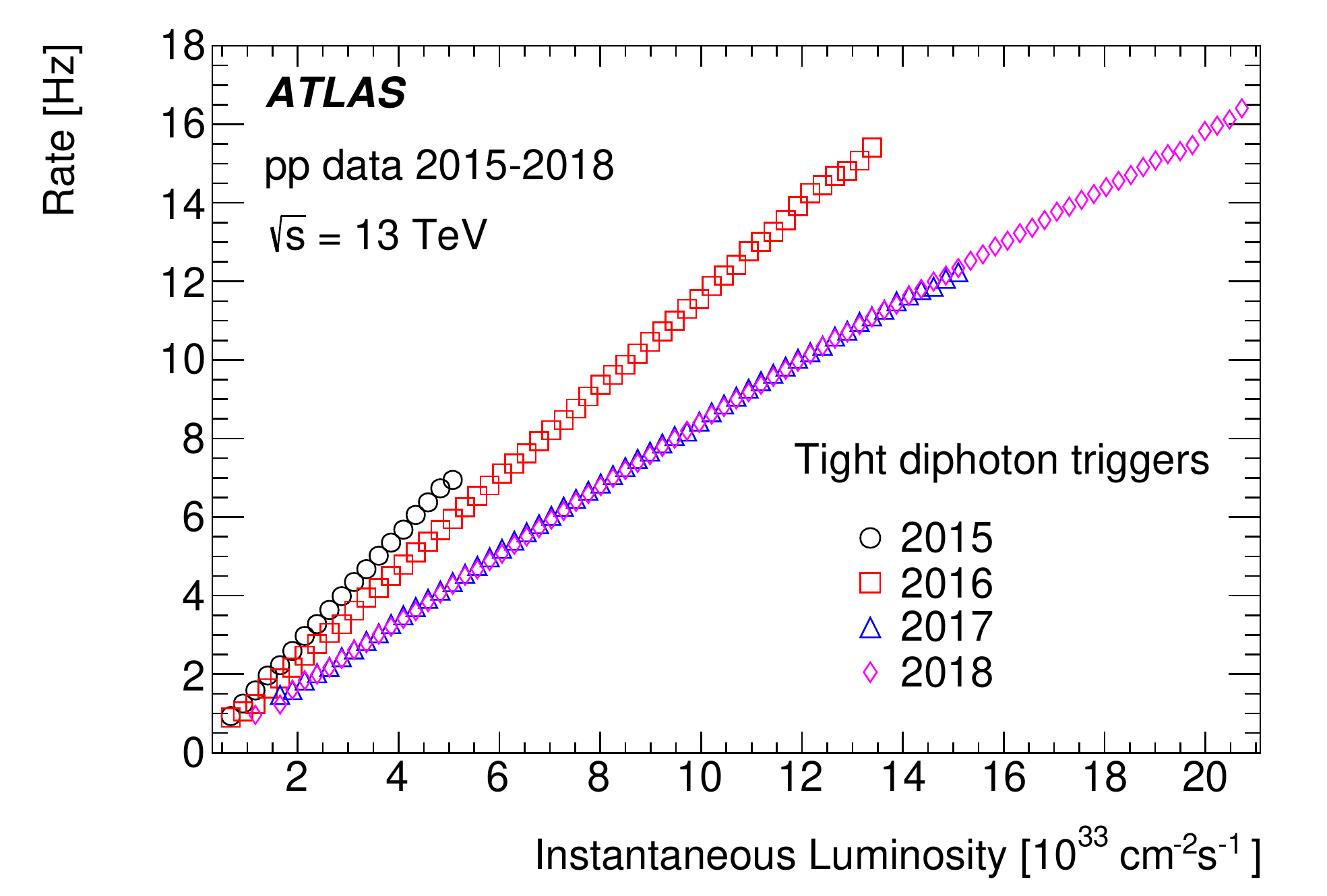}}
\caption{The HLT output rate as a function of instantaneous luminosity for tight diphoton triggers in 2015--2018.
The changes between years are detailed in Section~\ref{sec:eph}.}
\label{fig:photon-rates2}
\end{figure}

Diphoton triggers with lower \et{} thresholds and tighter identification criteria
are designed to collect events for beyond the Standard Model
low-mass diphoton resonance searches~\cite{HIGG-2014-04}.
These searches require the trigger \et{} thresholds to be kept symmetric
and as low as possible. Run 2 trigger thresholds allow searches to reach diphoton invariant masses down to $\sim$60$\,$\GeV{}.
These triggers are constrained by both the L1 and HLT rates. The L1\_2EM15VH threshold was used in 2015--2016
and L1\_2EM15VHI, which includes EM isolation at L1, was used in 2017--2018. The HLT rate for these triggers was about 16$\,$Hz
as shown in Figure~\ref{fig:photon-rates2}. The HLT thresholds were kept at 20$\,$\GeV{} in 2015, and then were increased to 22$\,$\GeV{}
as the peak luminosity rose above $1.2\times 10^{34}\,$cm$^{-2}$s$^{-1}$ in 2016. The use of the topo-cluster-based calorimeter
isolation in the HLT allowed the thresholds to be lowered back to 20$\,$\GeV{} for the 2017--2018 data-taking period, despite the
higher instantaneous luminosity and more challenging pile-up conditions.
 
\begin{figure}[!htpb]
\centering
\subfloat[][]{\includegraphics[width=0.495\textwidth]{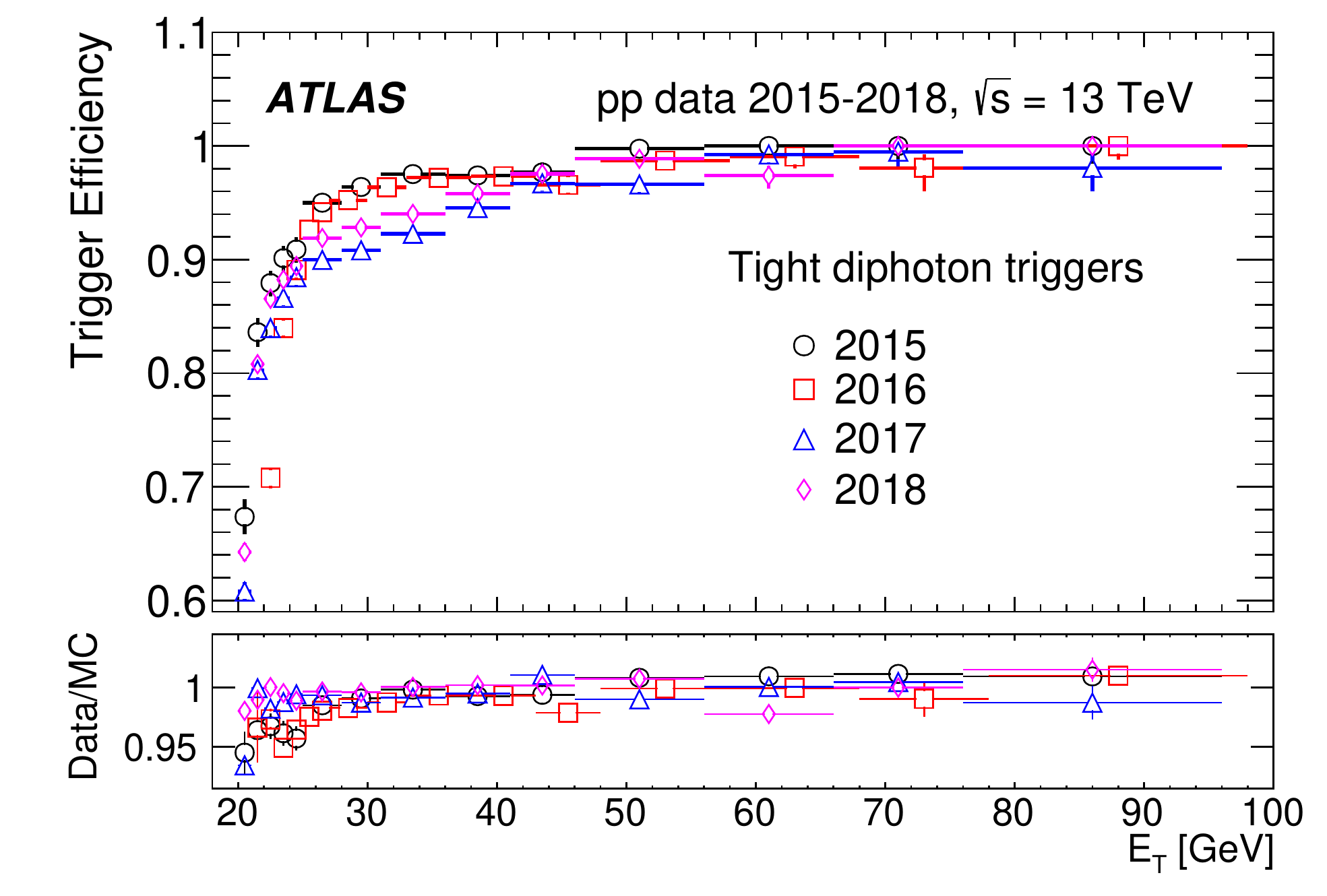}}
\subfloat[][]{\includegraphics[width=0.495\textwidth]{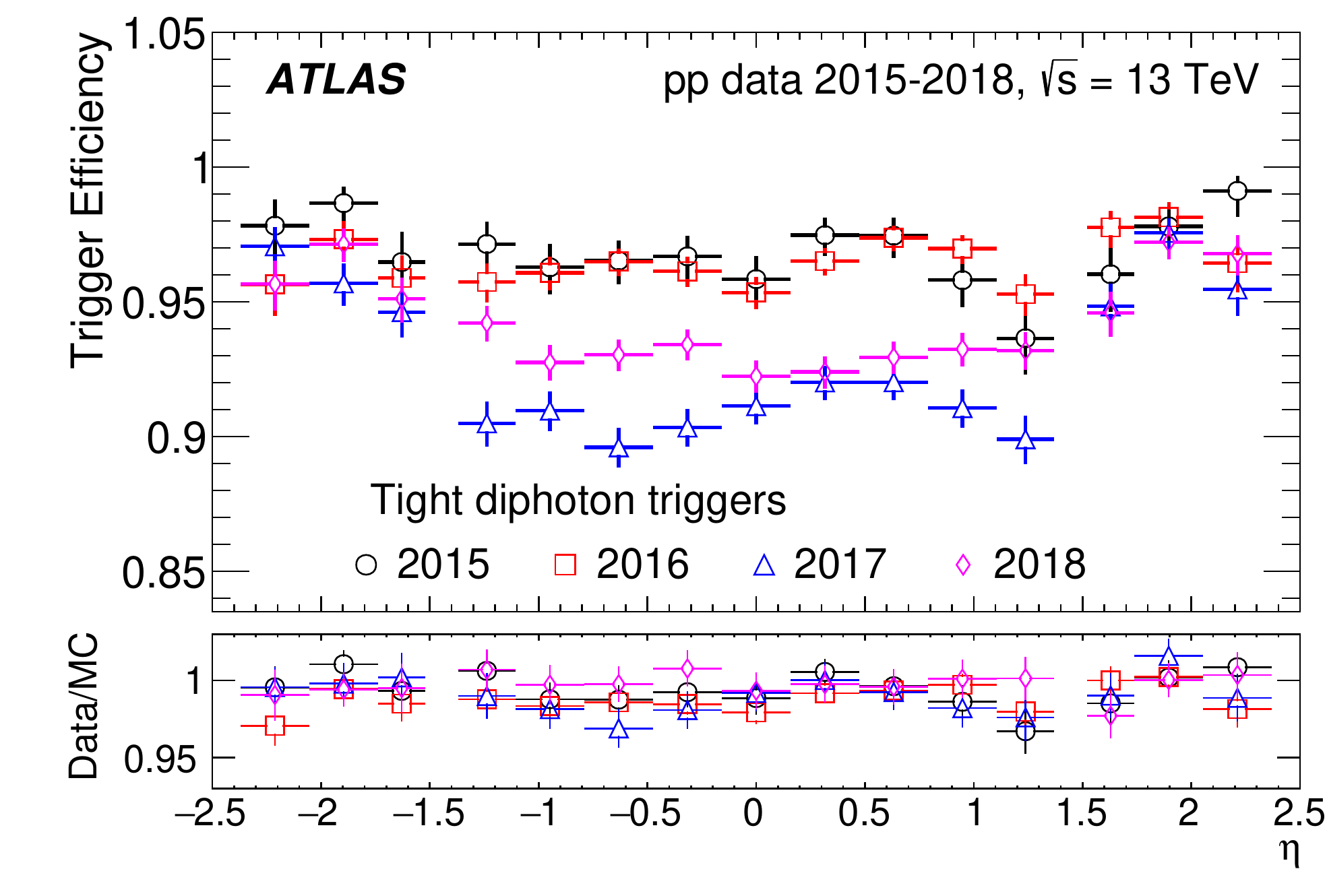}}\\
\subfloat[][]{\includegraphics[width=0.495\textwidth]{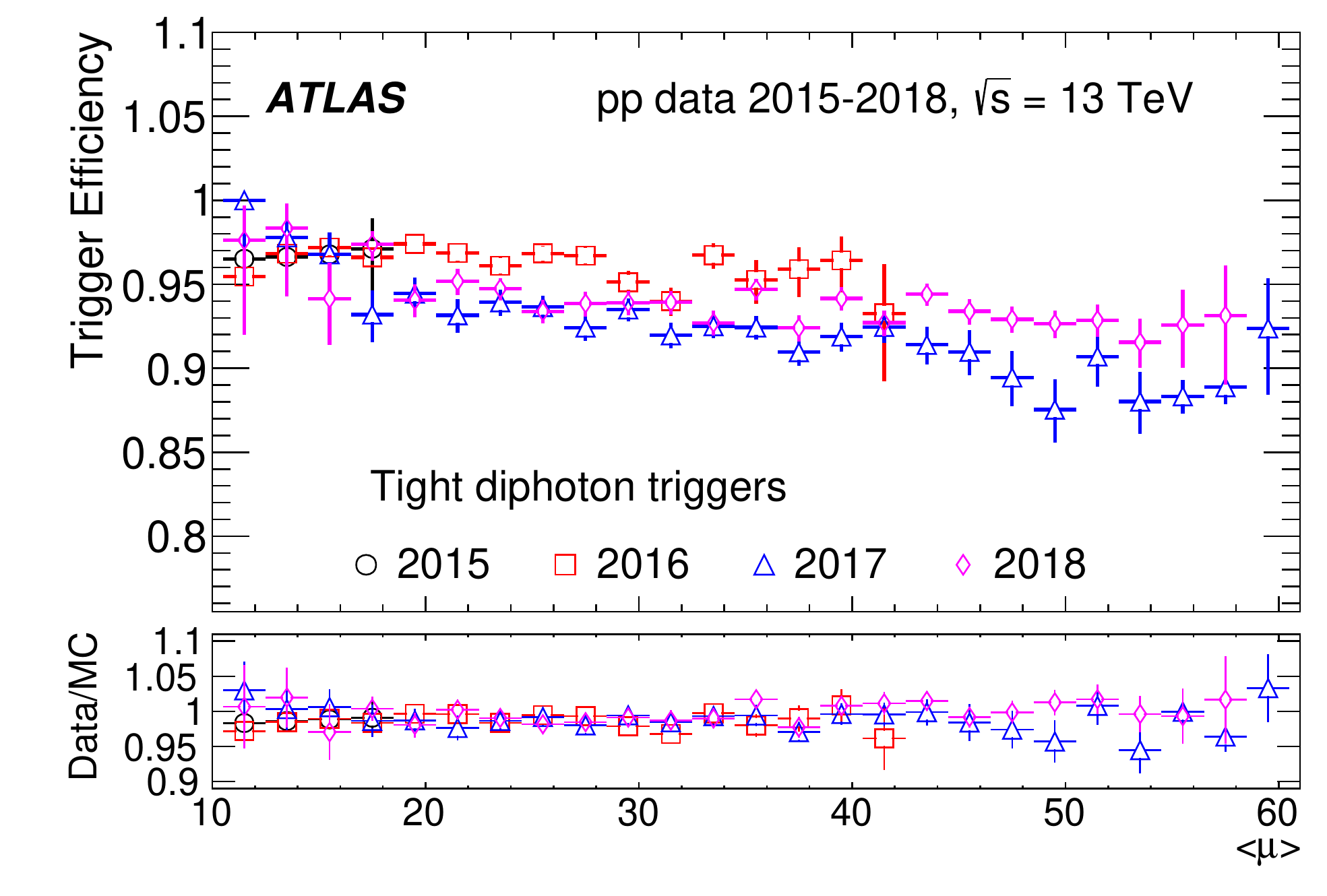}}
\caption{Evolution of efficiencies for tight diphoton trigger legs as a function of the offline photon
(a) $\et$, (b) $\eta$, and (c) $\langle\mu\rangle$ during Run 2.
The changes between years are detailed in Section~\ref{sec:eph}. The efficiency is computed
\wrt offline photons satisfying tight identification criteria and the calorimeter-only
tight isolation requirement. The ratios of data to MC simulation efficiencies are also shown.
The total uncertainties, shown as vertical bars, are
dominated by statistical uncertainties. Offline photon candidates in the calorimeter
transition region $1.37 < |\eta| < 1.52 $ are not considered. For (b) and (c), only
offline candidates with \et{} values 5~\gev{} above the corresponding trigger threshold are used.
}
\label{fig:photon_trig_lowmass}
\end{figure}

Figure~\ref{fig:photon_trig_lowmass} shows the low-mass diphoton trigger efficiencies as a function of \et, $\eta$ and pile-up.
The efficiency is computed with the \Zrad\ method for a single photon trigger leg of the diphoton trigger
with respect to offline photons satisfying the tight identification criteria and
the calorimeter-only tight isolation requirement. Slightly lower efficiency is observed for 2017 triggers due to
a different LHC bunch structure as well as loosening of
the offline tight photon selection, which was applied online only for the 2018 data-taking period. Triggers in 2017--2018 also
suffer from some inefficiency due to L1 isolation up to $\sim$50~\GeV{}, as discussed in Section~\ref{sec:trig_l1}.
The isolated trigger
g20\_tight\_icalovloose\_L1EM15VHI exhibits a degradation in efficiency of 4--5\% when $\langle\mu\rangle$ rises from 20 to 60;
this effect is visible in Figure~\ref{fig:photon_trig_lowmass}(c). Above $\langle\mu\rangle \sim$55, the trend to lower efficiency
continues and the statistical uncertainty becomes large.
The reoptimisation of the online tight identification selection criteria
improved the efficiency of these triggers in 2018 relative to 2017 at $\langle\mu\rangle$ values above $\sim$40.
 
\section{Electron trigger evolution and performance in $pp$ data-taking}
\label{sec:trig_e}
 
\subsection{Evolution of electron triggers in Run 2}
\label{sec:trig_e_evol}
 
The evolution of the Run~2 electron trigger thresholds and identification requirements for
the main unprescaled triggers is summarised in Table~\ref{tab:eltriggers}.
 
\begin{table}[h!]
\small
\centering
\caption{List of unprescaled electron triggers in different data-taking periods during \RunTwo. The corresponding L1 trigger threshold is given in brackets.
All single electron triggers start from the same L1 threshold. No L1 isolation is applied for \et$>$50~\gev.}
\label{tab:eltriggers}
\begin{tabular}{@{\extracolsep{\fill}}  l | l | l | l }
\toprule
\textbf{Trigger type} & \textbf{2015} & \textbf{2016} & \textbf{2017--2018} \\
\midrule
\multirow{4}{*}{Single electron} & e24\_lhmedium (EM20VH)          & \multicolumn{2}{l}{e26\_lhtight\_nod0\_ivarloose (EM22VHI)} \\
&   & \multicolumn{2}{l}{e60\_lhmedium\_nod0 } \\
&  e120\_lhloose & \multicolumn{2}{l}{e140\_lhloose\_nod0 } \\
& e200\_etcut                                & \multicolumn{2}{l}{e300\_etcut } \\[1ex]
\hline
\multirow{2}{*}{Dielectron}     & 2e12\_lhloose & 2e17\_lhvloose\_nod0 & 2e17\_lhvloose\_nod0 (2EM15VHI) \\
&       (2EM10VH)         &    (2EM15VH)         & 2e24\_lhvloose\_nod0 (2EM20VH) \\
\bottomrule
\end{tabular}
\end{table}
 
In addition to the threshold increases, there were also changes in the underlying electron configuration
and selection requirements as summarised in Table~\ref{tab:elconfig}. These changes are not always reflected
in the trigger names. For example, the Ringer algorithm (described in Section~\ref{sec:ringer}) was introduced in 2017.
The additional background rejection allowed looser fast electron and precision calorimeter selections. In particular,
for the latter step, in 2015--2016 there was a selection which relied on a multivariate technique using a
LH discriminant, constructed similarly to the standard offline precision selection one, but based only on calorimetric variables.
This discriminant had $\sim$4\% inefficiency relative to the offline selection.
This inefficiency was removed in 2017--2018 by moving to a simpler requirement based only on \et{}.
 
\begin{table}[h!]
\caption{Changes in the electron HLT configuration steps}
\label{tab:elconfig}
\begin{tabular}{l|l|l|l|l}
\toprule
\textbf{Step} & \textbf{2015} & \textbf{2016} & \textbf{2017} & \textbf{2018} \\
\midrule
Fast calorimeter & \multicolumn{2}{c|}{Cut-based} & \multicolumn{2}{c}{Ringer for \et$\ge 15\,$\GeV} \\
Reco and selection  & \multicolumn{2}{c|}{} & Tuned on 2016 data & Tuned on 2017 data \\\hline
Fast electron & \multicolumn{2}{c|}{track \pt$>1$~\GeV, $|\Delta\eta|<0.2$} & \multicolumn{2}{c}{track \pt$>1\,$\GeV, $|\Delta\eta|<0.3$ for \et$<20\,$\GeV} \\
Selection & \multicolumn{2}{c|}{} & \multicolumn{2}{c}{track \pt$>2\,$\GeV, $|\Delta\eta|<0.2$ for \et$\ge 20\,$\GeV} \\\hline
Precision calorimeter & \multicolumn{2}{c|}{LH calo-only selection} & \multicolumn{2}{c}{\et{} requirement} \\\hline
Precision LH& Like offline & \multicolumn{3}{c}{Same as in 2015} \\
variables, & without $\Delta p/p$, & \multicolumn{3}{c}{without $d_0$, $|d_0/\sigma(d0)|$,} \\
binning in \et{} & same \et$< 45~$\gev{} & \multicolumn{3}{c}{same full range} \\\hline
Precision & \multicolumn{2}{c|}{MC-only} & 2016 data for \et$\ge 15\,$\GeV{} & 2017 data \\
LH inputs, & \multicolumn{2}{c|}{} & MC for \et$<15\,$\GeV  & (but `lhmedium') \\
tunes & \multicolumn{2}{c|}{} & \multicolumn{2}{c}{smoothing}  \\
\bottomrule
\end{tabular}
\end{table}
 
The 2015 and 2016 pdfs for the electron LH were derived from simulation
samples described in Section~\ref{sec:datasets}. The pdfs for the trigger electrons with \et{}
below (above) 15$\,$\GeV{} were determined with $J/\psi\rightarrow ee$ ($Z\rightarrow ee$)
MC samples, and corrected for differences between data and simulation~\cite{PERF-2017-01}.
The 2017 pdfs for electrons and background above 15$\,$\GeV{} were derived from data
as detailed in Ref.~\cite{ATLAS-EGAM-2018-01};
pdf `smoothing' was also introduced online for all triggers. The 2018 electron data-driven
pdfs for all working points except the `medium' one were updated with 2017 data, maintaining the
original selection criteria and optimising for higher pile-up conditions ($\langle\mu\rangle$ up to 100).
The 2018 pdfs for electrons with \et{} below 15$\,$\GeV{} were also derived from data as detailed in Ref.~\cite{ATLAS-EGAM-2018-01}.
 
\subsection{Ringer algorithm performance}
\label{sec:eringer}
 
\begin{figure}[h!]
\centering
\subfloat[][]{\includegraphics[width=0.45\textwidth]{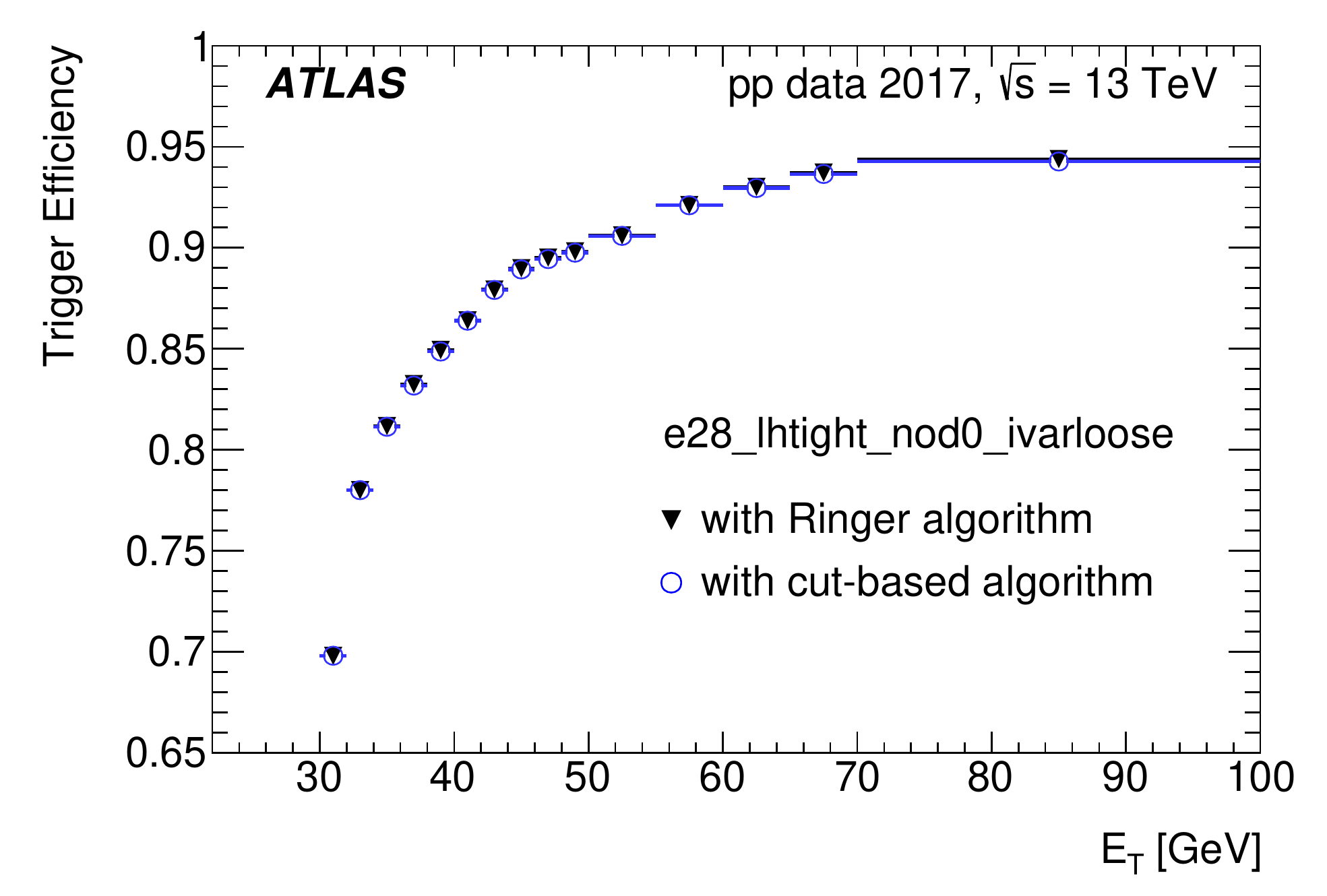}}
\subfloat[][]{\includegraphics[width=0.45\textwidth]{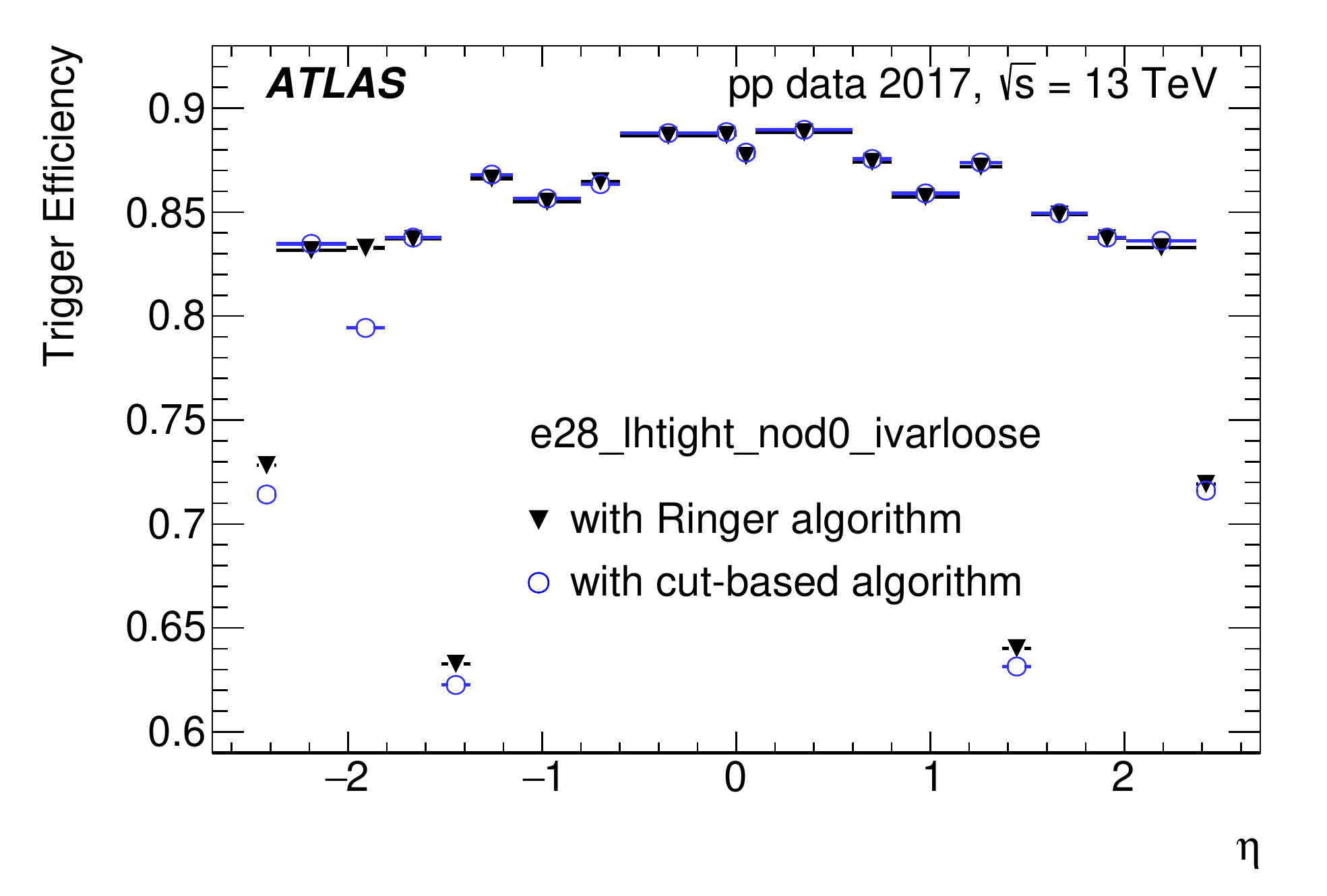}}\\
\subfloat[][]{\includegraphics[width=0.45\textwidth]{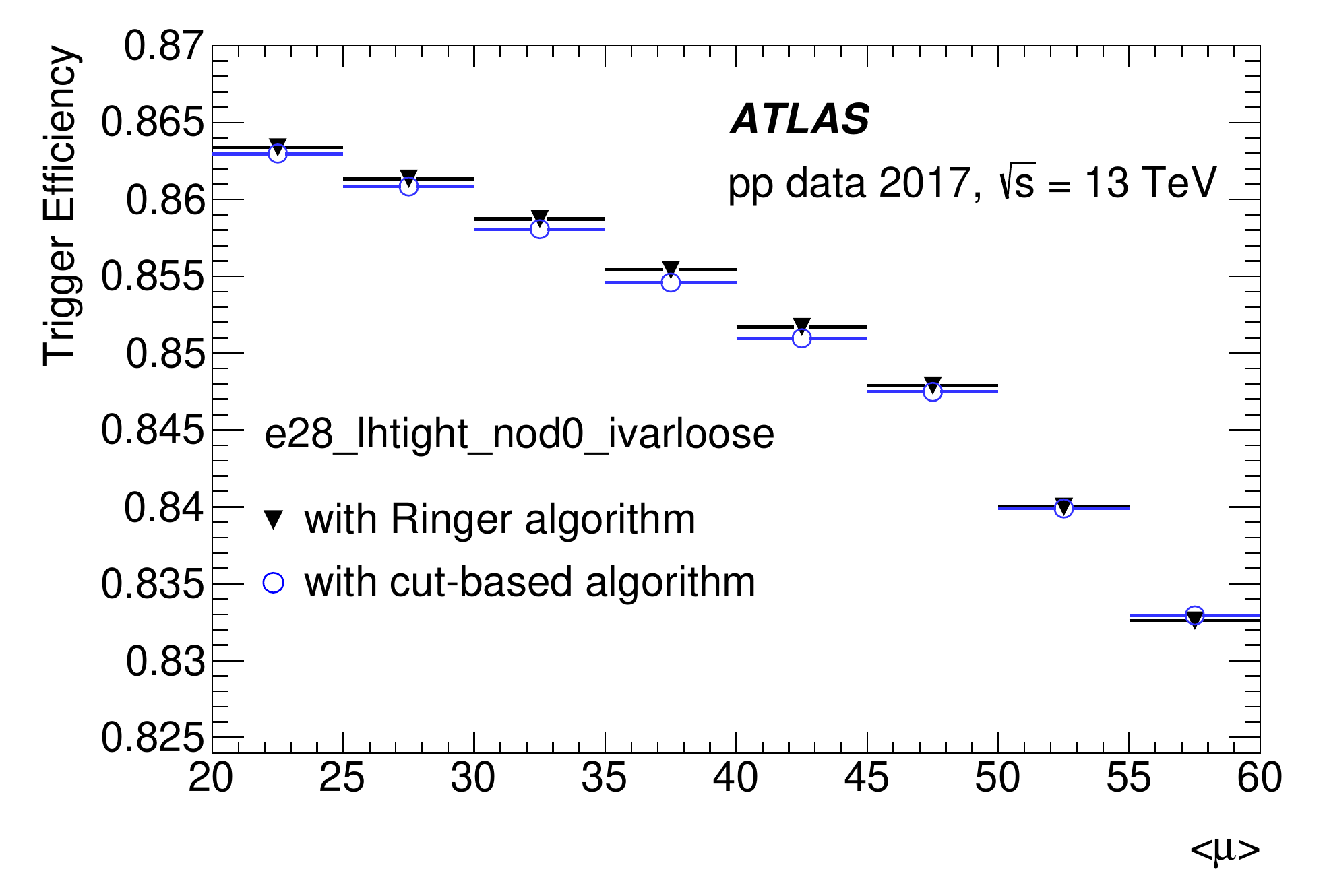}}
\caption{The efficiency of electron triggers with the Ringer algorithm
and with the cut-based algorithm
as a function of the offline electron (a) \et{}, (b) $\eta$ and (c) pile-up. Efficiency is given with respect to offline tight identification
working point. For (b) and (c), only offline candidates with $\et>29$~\gev{} are used.
}
\label{fig-ringer-eff}
\end{figure}

\begin{figure}[h!]
\centering
\subfloat[][]{
\includegraphics[width=0.45\textwidth]{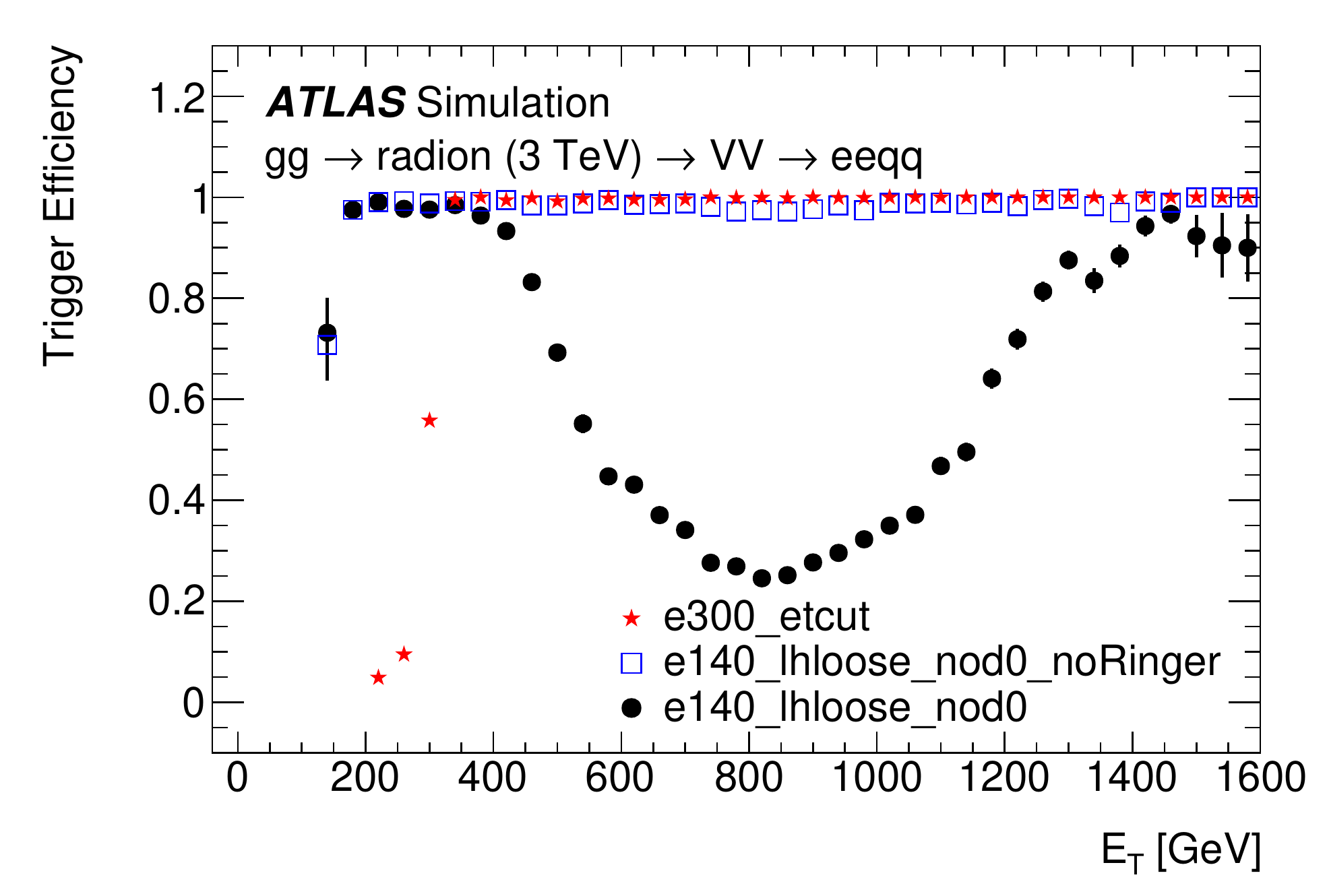}
}
\subfloat[][]{
\includegraphics[width=0.45\textwidth]{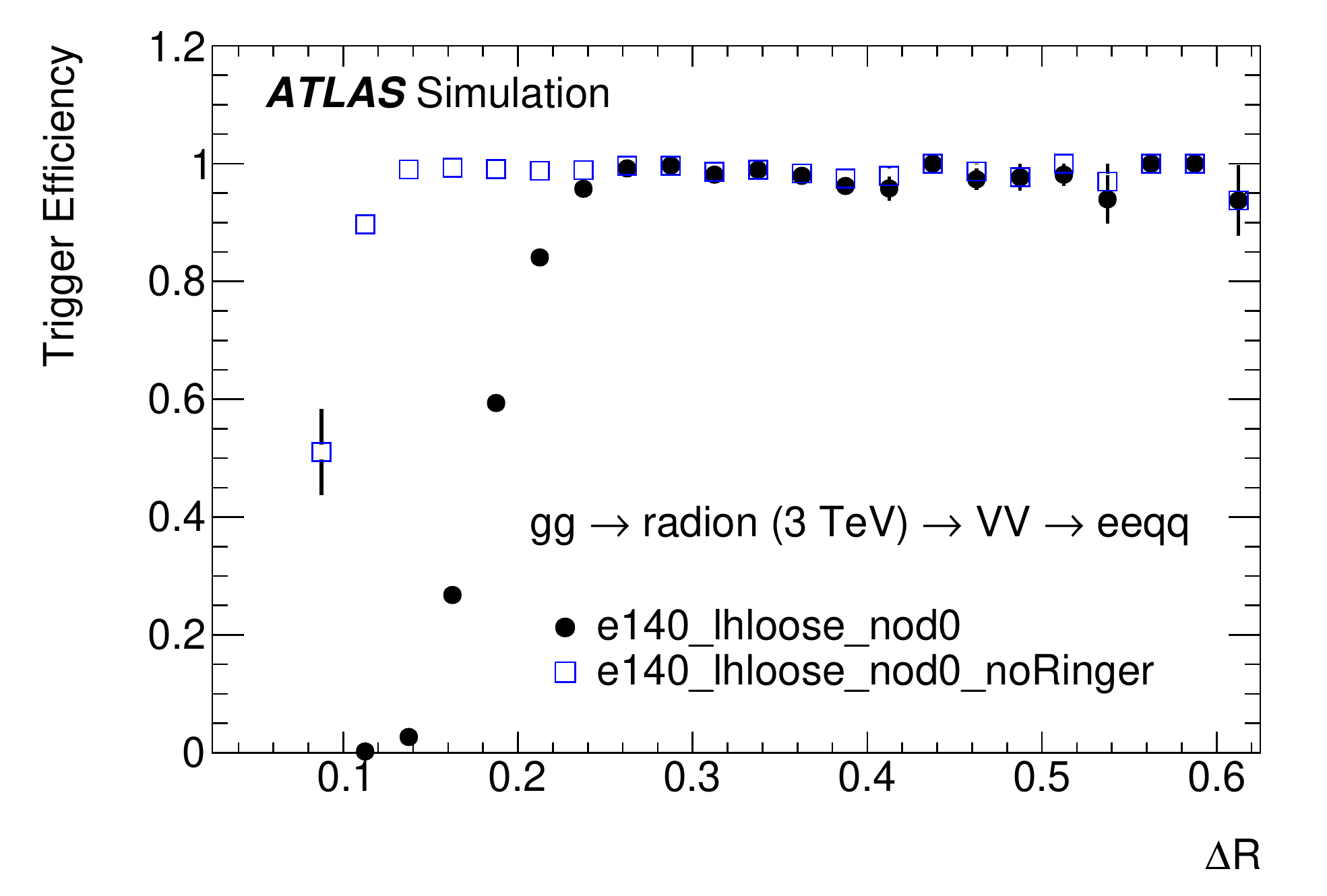}
}
\caption{The efficiency for electrons from $gg\rightarrow\mathrm{radion\,(3\,\TeV{})}\rightarrow VV\rightarrow ee qq$
as a function of (a) the offline electron \et{} and (b) $\Delta R$ between two electrons. Efficiency is given with respect to offline loose identification and the FCLoose isolation working point. For (b), only offline candidates with $\et>400$~\gev{} are used.
}
\label{fig:ringer-exot}
\end{figure}

In 2017, triggers collected data online simultaneously with and without use
of the Ringer algorithm, allowing an evaluation of its performance.
For $Z\rightarrow ee$ decays, no difference in efficiency is observed, as shown
for two triggers in Figure~\ref{fig-ringer-eff}. However, some special cases
(such as events with merged electrons coming from decays of boosted dibosons)
are found to suffer losses in efficiency as a result of using the Ringer algorithm,
as is shown in Figure~\ref{fig:ringer-exot}. The efficiency drops for $\et>400$~\gev{}
because the two clusters begin to overlap, but at very high \et{}
the two clusters become
so close that they behave as a single cluster and all triggers become efficient again.
For the \et{} range above 300~\gev{},
a trigger with only an \et{} selection in the HLT, e300\_etcut, is available.
 
\subsection{Single-electron triggers}
\label{sec:se}
 
\begin{figure}[!ht]
\centering
{\includegraphics[width=0.45\textwidth]{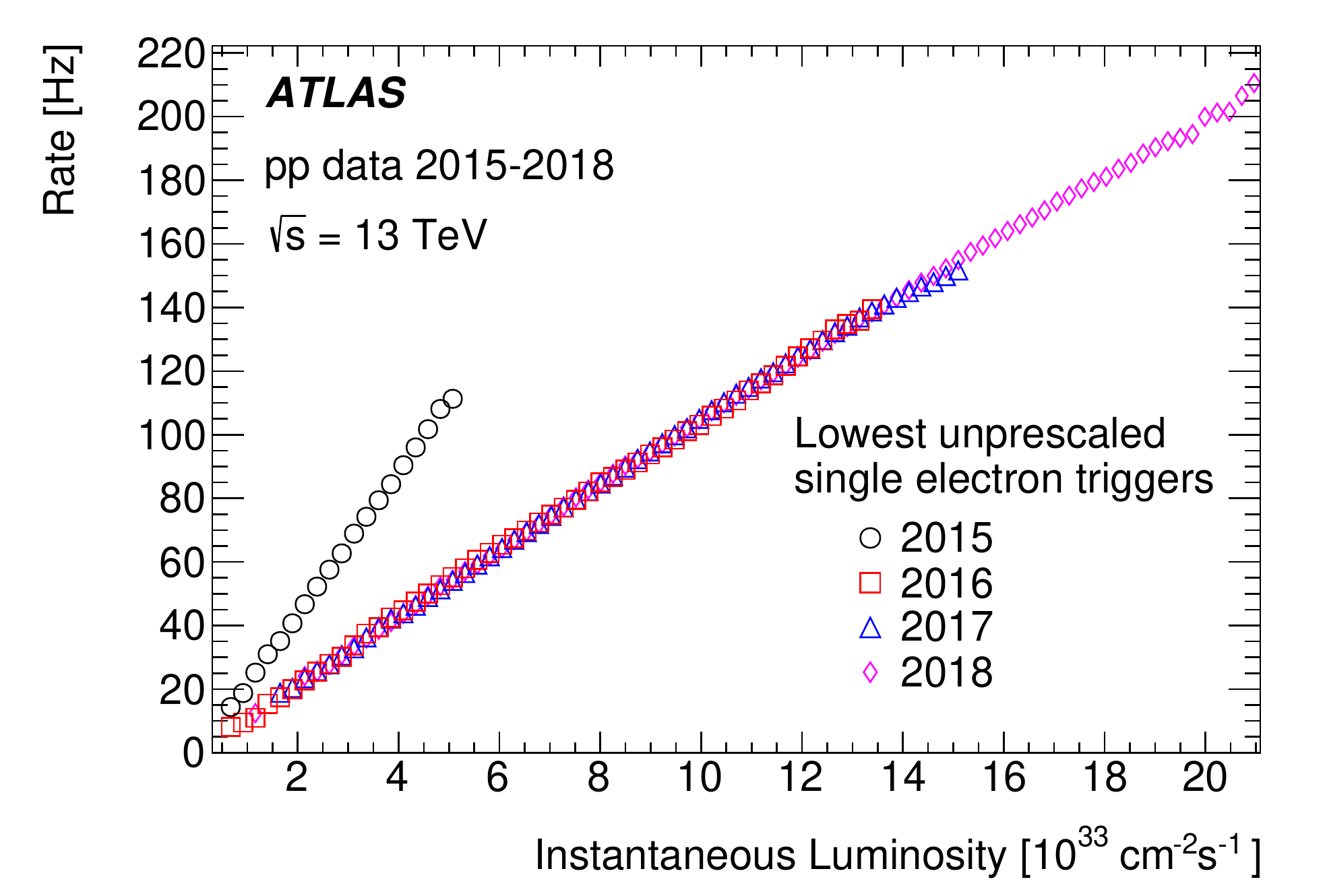}}
\caption{Dependence of the trigger rate on the luminosity for the lowest-threshold unprescaled
isolated single-electron triggers in 2015--2018. The changes between years are detailed in Section~\ref{sec:trig_e_evol}.
}
\label{fig:e-rate1}
\end{figure}
 
One of the main features of the Run~2 trigger menu is the presence of the unprescaled single-electron trigger
\et{} threshold of 24~\GeV{} for 2015 and 26~\GeV{} for 2016--2018.
This single-electron trigger ensures the collection of the majority of the events with
leptonic $W$ and $Z$ boson decays, which are present in a wide range of measurements and searches in ATLAS.
Although the threshold of this trigger is mainly constrained by the L1 bandwidth, as discussed in Section~\ref{sec:trig_l1},
the need for a low threshold and HLT rate places strong constraints on the tightness of the identification used by this trigger in the HLT.
Relying on this trigger provides a simple and inclusive strategy, widely used in the ATLAS physics programme,
at a cost of about $20\%$ of the total L1 and HLT rate.
 
\begin{figure}[!htpb]
\centering
{\includegraphics[width=0.50\textwidth]{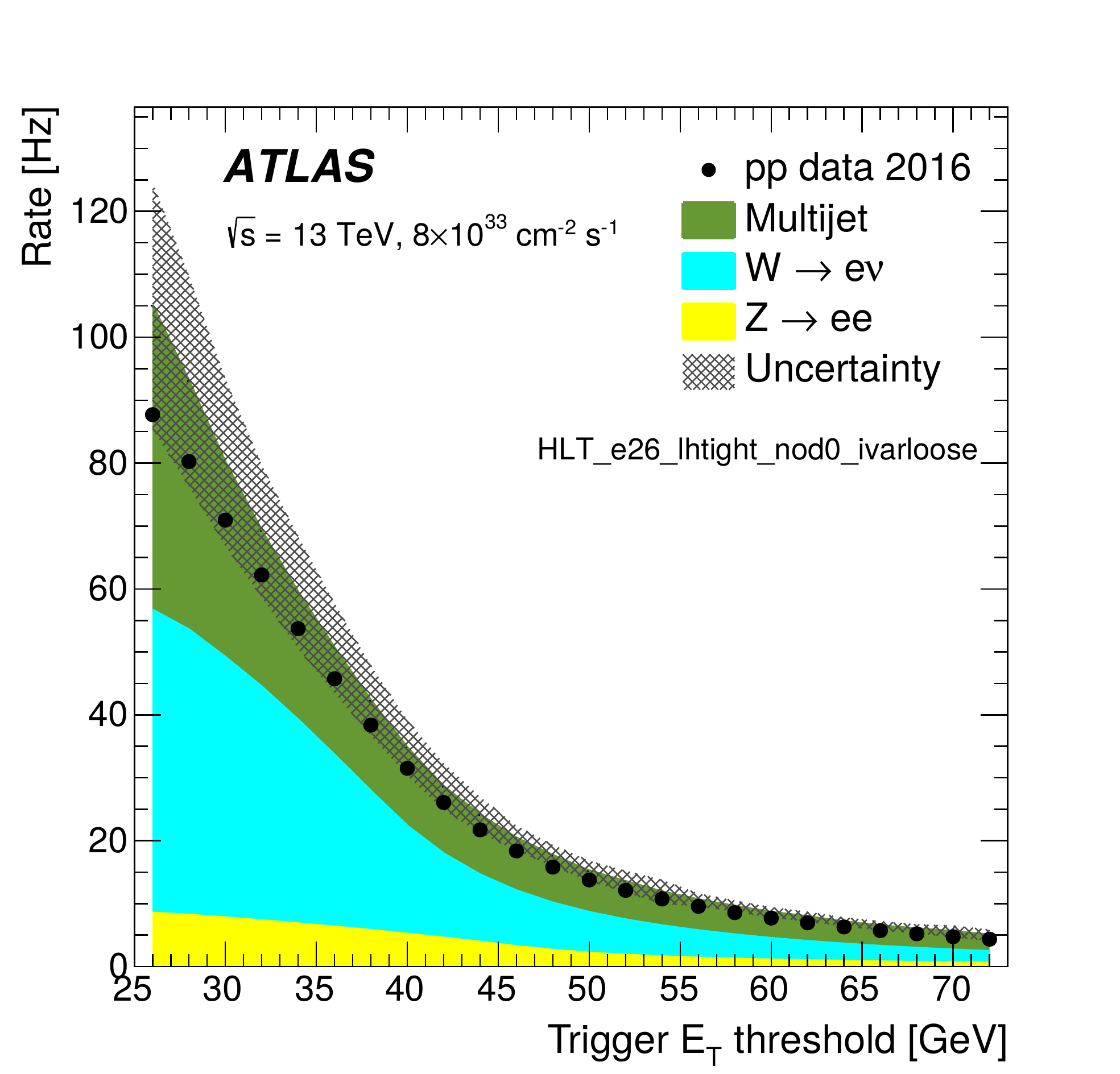}}
\caption{The lowest-threshold unprescaled isolated single-electron trigger's rate as
a function of the threshold value for the trigger electron \et{}.}
\label{fig:elcomp}
\end{figure}
 
Figure~\ref{fig:e-rate1} shows the rates for the lowest-threshold unprescaled isolated
single-electron triggers used during Run~2 as a function
of the instantaneous luminosity. The LH selection (`lhtight') of
the lowest-threshold unprescaled isolated single-electron trigger
is tuned for a given rate, which remained unchanged in 2016--2018.
 
The electron trigger candidates originate from various processes, as shown in Figure~\ref{fig:elcomp}.
This shows the lowest-threshold unprescaled isolated single-electron trigger rate as a function of the
HLT electron \et-threshold value, broken down by process.
The total rate is measured in a data set collected at a constant instantaneous luminosity of
$8\times 10^{33}\,$cm$^{-2}$s$^{-1}$ at $\sqrt{s}=13$~\TeV{}, while the individual contributions from
$W$, $Z$ and multijet production are estimated with MC simulation. The dominant uncertainty in
the multijet rate is evaluated with a data-driven technique: the rate as a function of \et{}
is obtained in a multijet-enriched region by inverting the HLT track-based electron isolation,
and the bin-by-bin disagreement between data and MC simulation is applied as a systematic uncertainty of
the multijet process.
The total expected rate is in agreement with the measured value for all
the thresholds considered. Most of the rate comes from physics processes of interest such
as $W$ and $Z$ production, while a significant but not dominant background comes from jets misidentified
as electrons.
 
\begin{figure}[!ht]
\centering
\subfloat[][]{            {\includegraphics[width=0.45\textwidth]{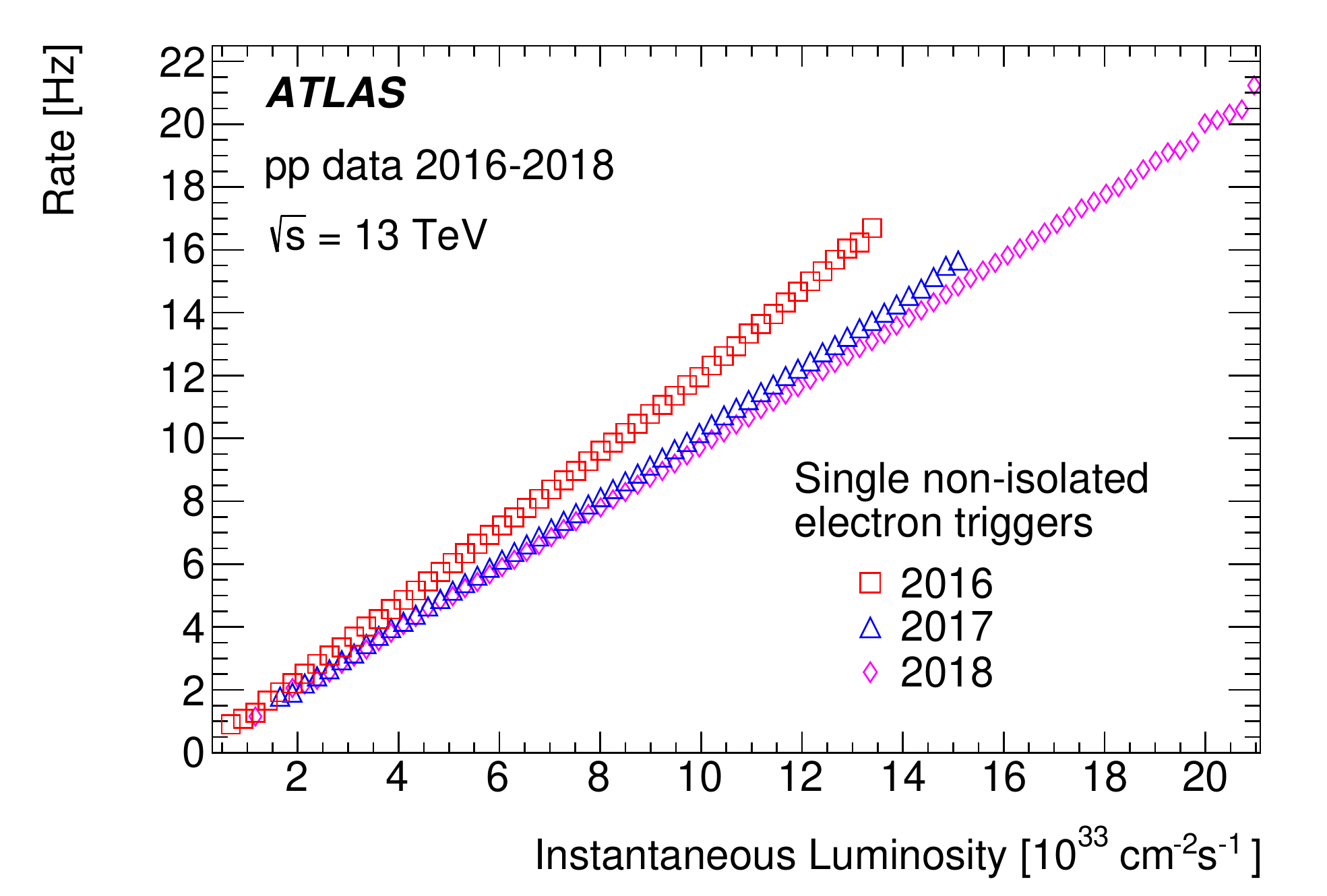}}}
\subfloat[][]{            {\includegraphics[width=0.45\textwidth]{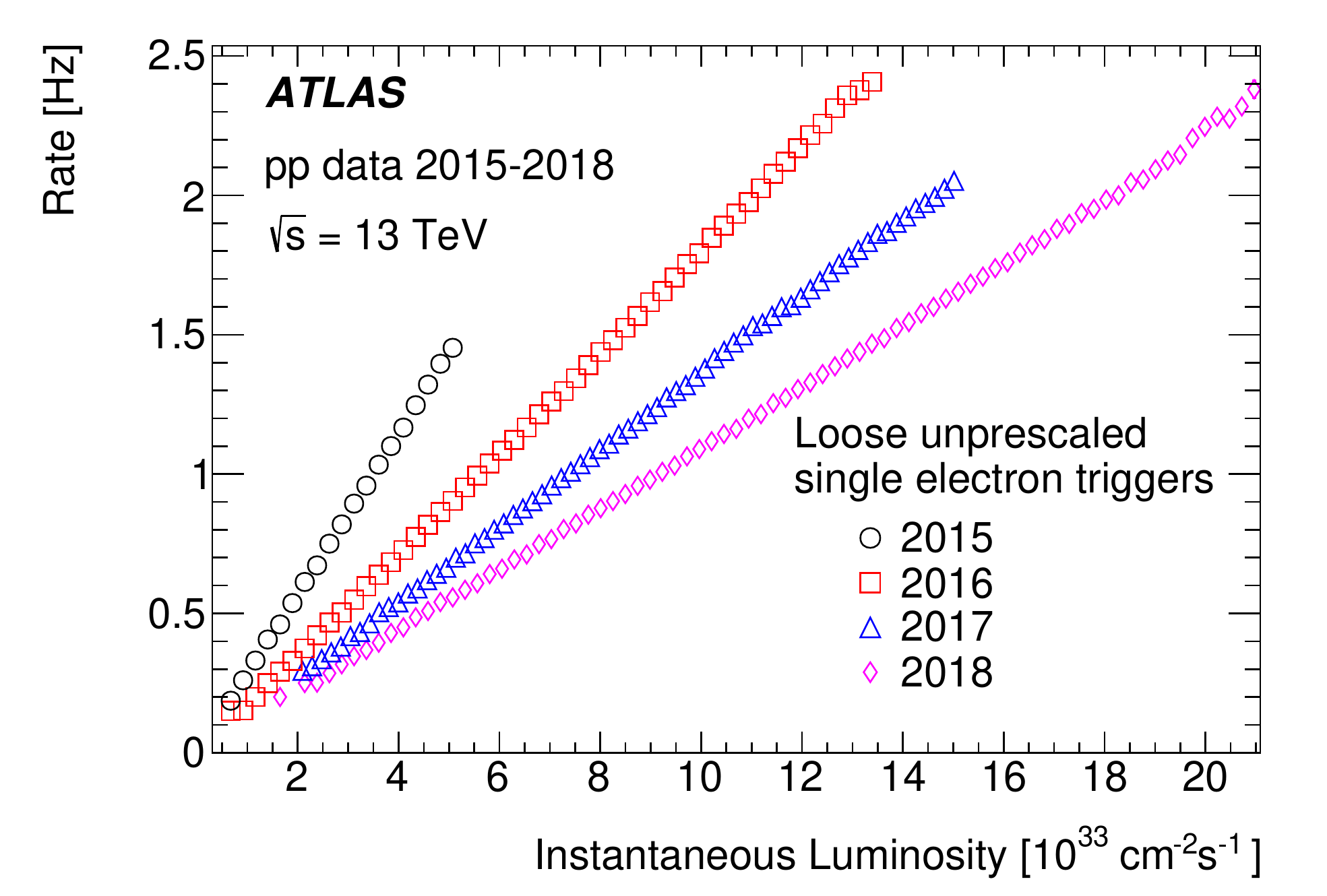}}}
\caption{Dependence of the single-electron-trigger rates on the luminosity for
(a) lowest-threshold unprescaled, non-isolated and (b)
loose unprescaled single-electron triggers in 2015--2018.
The changes between years are detailed in Section~\ref{sec:trig_e_evol}.
}
\label{fig:e-rate2}
\end{figure}
 
At higher \et{}, additional triggers with no isolation requirements
and looser identification are introduced. The rates for the lowest-threshold unprescaled, non-isolated
triggers with a requirement of \et{} above 60~\GeV{} and with `lhmedium' identification are
shown in Figure~\ref{fig:e-rate2}(a). The rates for the loose, unprescaled triggers with the
`lhloose' identification working point and \et{} above 120--140~\GeV{} are shown in Figure~\ref{fig:e-rate2}(b).
These higher-\et{} triggers have rate reductions of one and two orders of magnitude, respectively, compared
with the lowest-threshold unprescaled isolated single-electron trigger. The three single-electron triggers,
the exact configuration for which is detailed in Table~\ref{tab:eltriggers}, are used simultaneously
in a typical analysis selection, allowing an event to be selected if it passes any of them.
This configuration is called the `single-electron trigger combination'. There is also a very high \et{} trigger,
e300\_etcut, running at a rate of up to 5$\,$Hz at $2\cdot10^{34}$~cm$^{-2}$s$^{-1}$.
This trigger allows the collection of an unbiased sample of events with very high energy deposits in the EM calorimeter,
as discussed in Section~\ref{sec:eringer}.
 
\begin{figure}[!ht]
\centering
\subfloat[][]{{\includegraphics[width=0.45\textwidth]{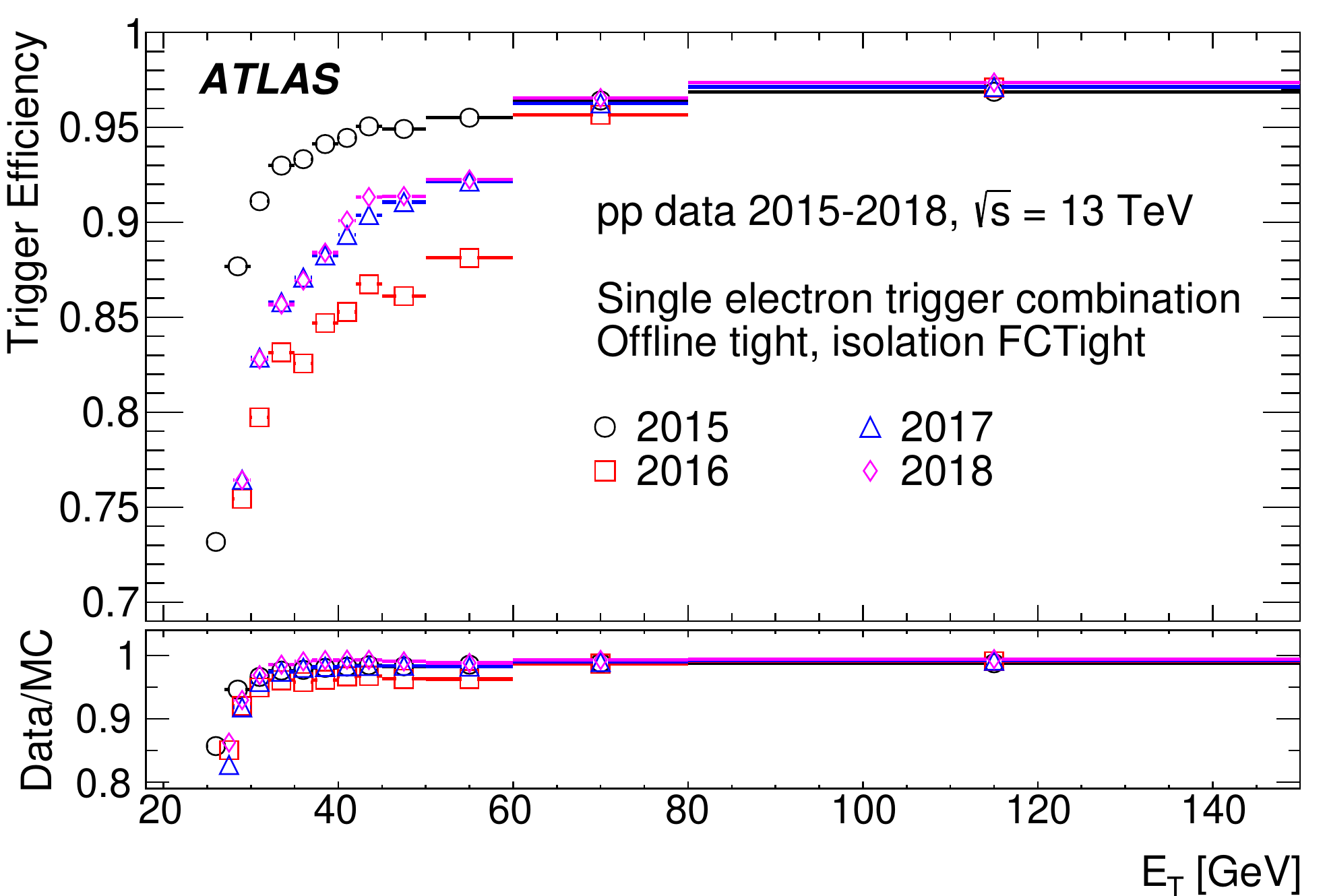}}}
\subfloat[][]{{\includegraphics[width=0.45\textwidth]{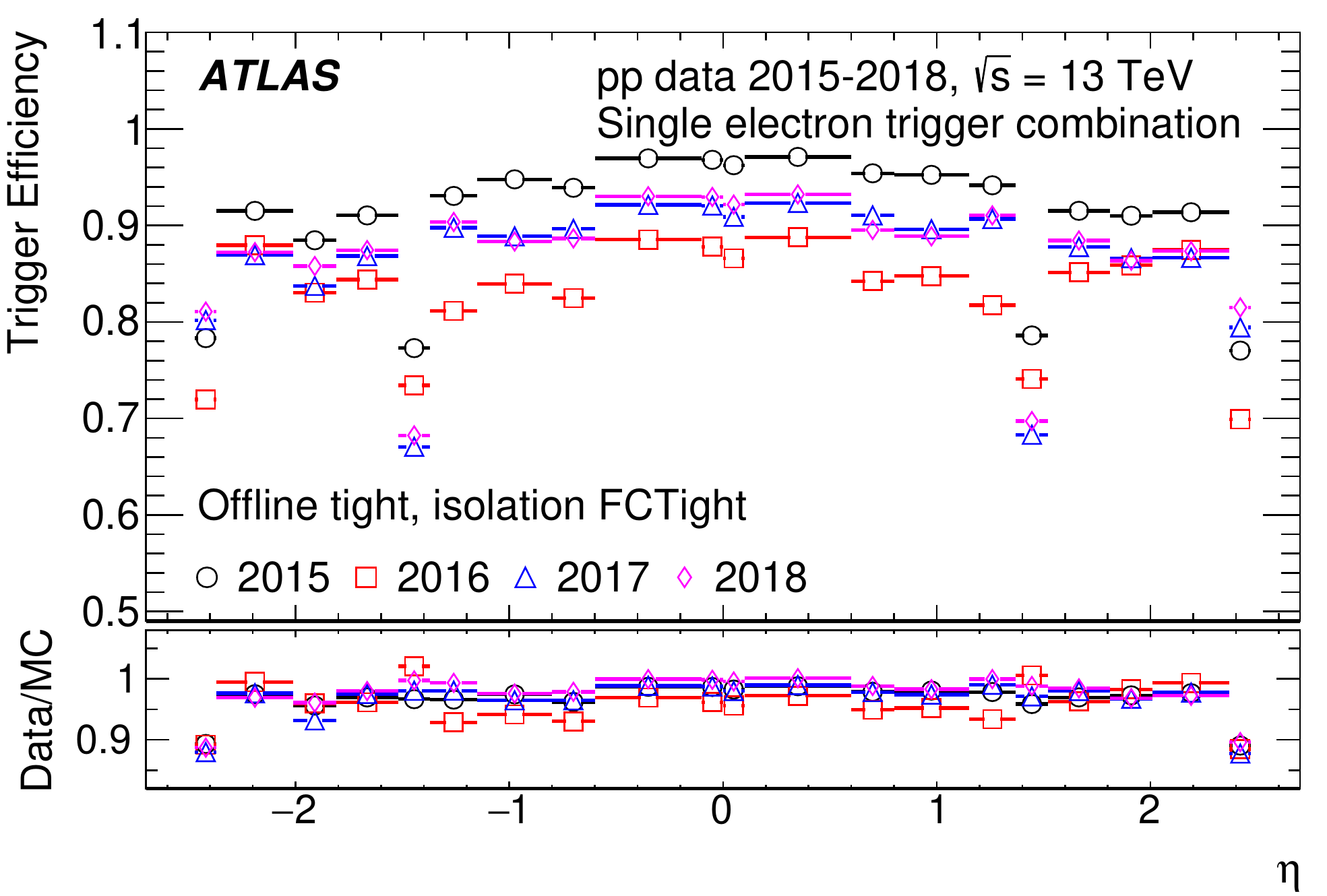}}}
\caption{Evolution of the single-electron trigger combination efficiency as a function of the
offline electron (a) \et{} and (b) $\eta$ during Run~2.
The changes between years are detailed in Section~\ref{sec:trig_e_evol}.
Efficiency is given with respect to offline tight identification and the FCTight isolation
working point. The ratios of data to MC simulation efficiencies are also shown.
The error bars indicate statistical and systematic uncertainties combined in quadrature.
For (b), only offline candidates with \et{} values at least 1~\GeV{} above the
corresponding trigger threshold are used.
}
\label{fig:Sing_all}
\end{figure}
 
The evolution of the single-electron trigger combination efficiency in 2015--2018 is shown in
Figure~\ref{fig:Sing_all}. The offline electron is required to pass the tight identification
and FCTight isolation requirements. The FCTight isolation requirement is chosen because it is
the only one which has a more restrictive isolation configuration than is used online. The sharper efficiency
turn-on as a function of \et{} in 2015 shown in Figure~\ref{fig:Sing_all}(a)
is due to a looser identification requirement (`lhmedium' versus `lhtight' from 2016),
a lower \et{} threshold (24~\GeV{} versus 26~\GeV{} from 2016) and no isolation requirement.
Although similar identification, isolation, and \et{} requirements are imposed in the single-electron triggers in 2016--2018,
some inefficiency at $\et{} < 60\,$\GeV{} is observed in 2016. This is explained by the different
electron trigger configuration used in 2016, in particular the
inefficiency of the calorimeter-only LH selection at the precision step.
In 2015--2016, triggers used simulation-based LH and were optimised relative to a different offline
selection~\cite{PERF-2017-01}, which results in some inefficiency; however, from 2017 a data-driven
likelihood selection and introduction of a
looser fast selection with the Ringer algorithm recover the trigger efficiency at $\et{}<60$~\GeV.
The main remaining sources of inefficiency are the L1 electromagnetic isolation requirements discussed
in Section~\ref{sec:trig_l1}.
As shown in Figure~\ref{fig:Sing_all}(b), the single-electron trigger combination efficiency
is lower in the $1.37<|\eta|<1.52$ and $|\eta|>2.37$ regions, where a significant amount of inactive material is
present. Further, detailed investigation into the sources of the inefficiency
relative to the offline selection is discussed below.
 
The MC simulation efficiency correction factors, defined in Section~\ref{sec:tnp}
and shown in the lower panels of Figure~\ref{fig:Sing_all}, are as large
as 18\% close to the trigger \et{} threshold and at most 4\% above 40~\GeV{}. Their $\eta$-dependence
is fairly smooth for 2015 and 2017--2018,
with typical values of less than 4\% (11\%) outside (inside) the $|\eta|>2.37$ region.
These efficiency correction factors are measured with a typical precision of 0.1\%.
 
\begin{figure}[!ht]
\centering
{\includegraphics[width=0.45\textwidth]{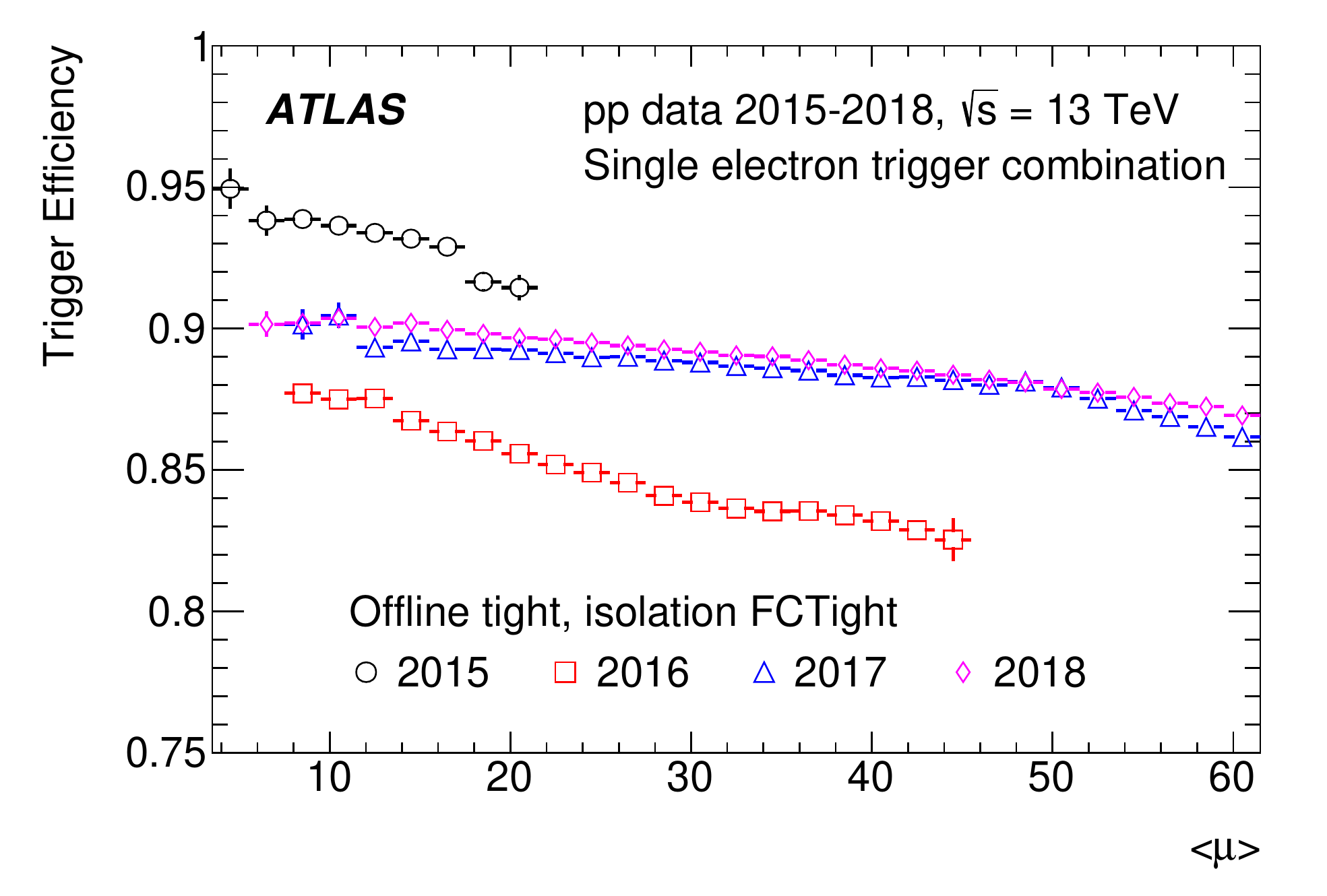}}
\caption{Evolution of the single-electron trigger combination efficiency as a function of
the pile-up during Run~2, showing measurements in data only.
The changes between years are detailed in Section~\ref{sec:trig_e_evol}.
The efficiency is given with respect
to offline tight identification and the FCTight isolation working point. Background subtraction is not
applied, as the effect is expected to be negligible. Poorly populated bins are removed. Only offline
candidates with \et{} values at least 1~\GeV{} above the corresponding trigger threshold are used.
The error bars indicate statistical and systematic uncertainties combined in quadrature. }
\label{fig:Sing_all_mu}
\end{figure}

Figure~\ref{fig:Sing_all_mu} shows the trigger efficiency dependence on pile-up.
This was reduced towards the end of Run~2. The residual dependence is caused by the isolation requirements
both in the HLT and at L1.
 
Sources of inefficiency for the e26\_lhtight\_nod0\_ivarloose and e60\_lhmedium\_nod0 triggers
relative to the offline reconstruction and the corresponding L1 requirements (EM22VHI)
are shown in Table~\ref{tab:ineff}. The sources are broken down for each of the selection steps in the HLT.
A description of the steps is provided in Section~\ref{sec:hlt-electron}. The offline reconstructed
electron is required to have $\et>27$ (61)~\GeV{} and pass the `lhtight' identification.
The inefficiencies are determined by the percentage of candidates that pass the offline
identification, but fail the online identification at the indicated step, measured with a tag-and-probe
method using $Z \rightarrow ee$ decays providing approximately $25\,000$ and $15\,000$ suitable probe electrons,
respectively. The sizes of the contributions of the individual selection steps to the overall
inefficiency depend on the \et{} of the electron and on the tightness of the selection requirements.
The dominant source of inefficiency for the lowest-threshold unprescaled isolated trigger is the electron identification,
while for the lowest-threshold unprescaled non-isolated trigger, the sources of inefficiency are more diverse.
These inefficiencies are driven by differences between the online and offline reconstruction criteria described in
Section~\ref{sec:egselectiontrig}.
 
\begin{table}[h!]
\caption{Sources of inefficiency for the e26\_lhtight\_nod0\_ivarloose and e60\_lhmedium\_nod0
triggers at each selection step in the HLT. Isolation requirements on the precision electron
candidate may be applied: if the candidate fails the `Precision Electron selection' but
passes isolation, `Electron selection only' is filled; if the candidate passes the
precision electron selection buts fails isolation, `Isolation only' is filled; if both fail,
`Electron selection and isolation' is filled. Data collected in October 2017 are used for this study.}
\label{tab:ineff}
\begin{tabular}{@{\extracolsep{\fill}}  l | c | c }
\toprule
&  \multicolumn{2}{c}{Inefficiency [$\%$]} \\
Trigger          &  e26\_lhtight\_nod0\_ivarloose & e60\_lhmedium\_nod0 \\
\midrule
Fast step        & $0.72 \pm 0.02$ & $1.3\pm 0.1$ \\\hline
Precision steps: & & \\
Calorimeter reconstruction and \et{} selection & $0.11 \pm 0.01$ & $1.63 \pm 0.11$ \\
Track reconstruction and track--cluster matching & $0.87 \pm 0.02$ & $0.67 \pm 0.07$ \\
Electron selection only & $6.03 \pm 0.05$ & $2.60 \pm 0.13$ \\
Electron selection and isolation & $0.26 \pm 0.01$ & - \\
Isolation only & $0.75 \pm 0.02$ & - \\
Other & $0.22 \pm 0.01$ & $0.18 \pm 0.04$ \\\hline
Total & $8.9 \pm 0.1$ & $6.38 \pm 0.21$ \\
\bottomrule
\end{tabular}
\end{table}
 
\subsection{Dielectron triggers}
\label{sec:de}
 
\begin{figure}[!ht]
\centering
{\includegraphics[width=0.6\textwidth]{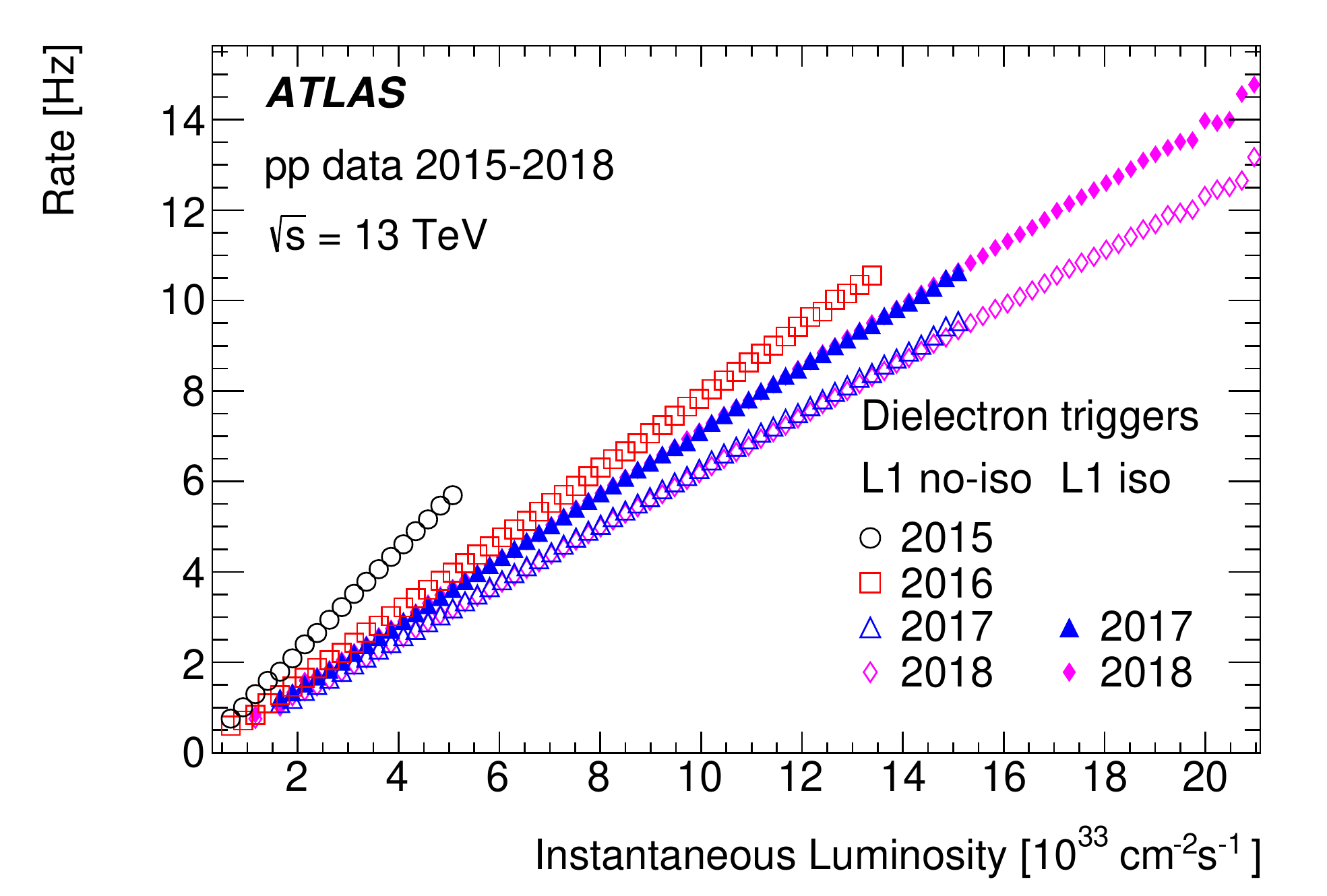}}
\caption{Dependence of the dielectron trigger rates on the luminosity in 2015--2018.
The changes between years are detailed in Section~\ref{sec:trig_e_evol}. Open markers represent L1 triggers with
no EM isolation (L1 no-iso), while filled markers represent EM isolated L1 triggers (L1 iso).
}
\label{fig:de-rate}
\end{figure}

Dielectron triggers allow the use of electron \et{} thresholds at least
9~\gev\ below those of the single-electron triggers and looser identification
and isolation requirements with only a very small increase in HLT rate.
The major constraint for dielectron triggers comes from their L1 \et{} thresholds, while
the corresponding HLT rates, shown in Figure~\ref{fig:de-rate}, are of the order of 10~Hz,
which allow triggers with a very loose selection in the HLT to be kept. The L1\_2EM15VH
threshold had to be increased to L1\_2EM20VH due to rate considerations in 2017.
An additional set of dielectron triggers which start from L1\_2EM15VHI was introduced.
 
\begin{figure}[!ht]
\centering
\subfloat[][]{      {\includegraphics[width=0.45\textwidth]{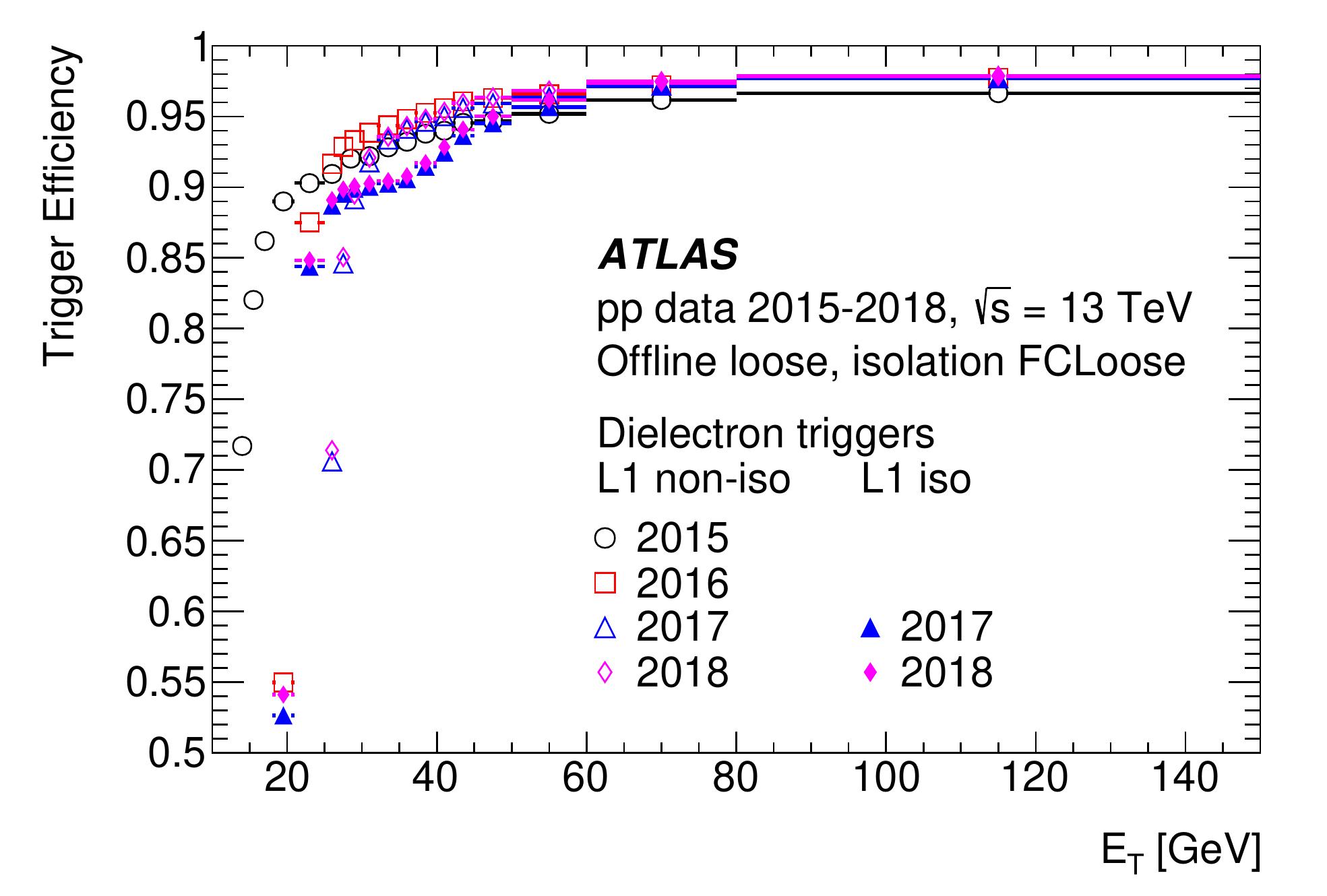}}}
\subfloat[][]{      {\includegraphics[width=0.45\textwidth]{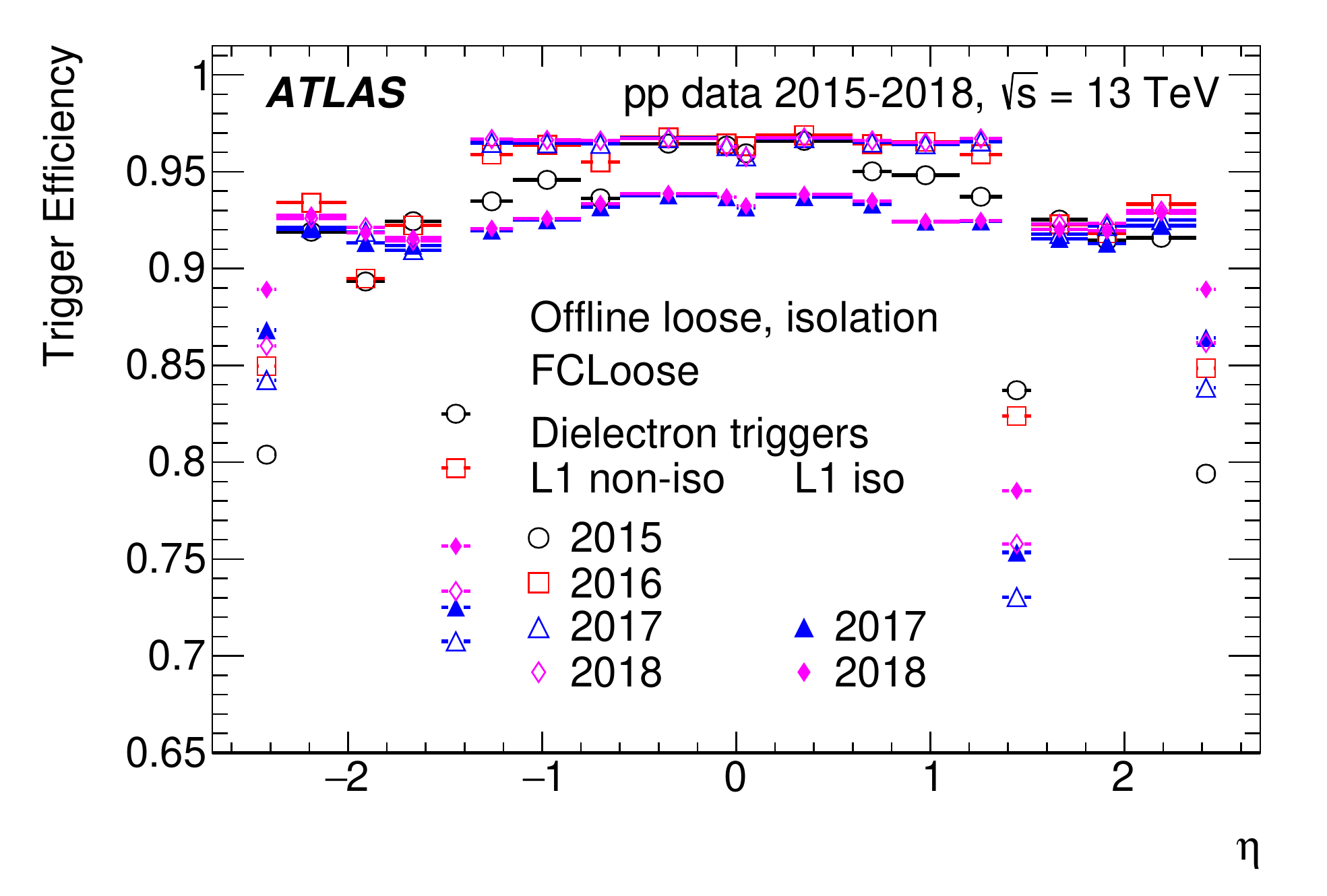}}}
\caption{Evolution of efficiencies for dielectron trigger legs as a function of the
offline electron (a) \et{} and (b) $\eta$
during Run~2, showing measurements in data only.
The changes between years are detailed in Section~\ref{sec:trig_e_evol}.
The efficiency is given with respect
to the loose offline identification and the FCLoose isolation working point. For (b),
only offline candidates with \et{} values 1~\GeV{} above the corresponding trigger threshold are used.
The error bars indicate statistical and systematic uncertainties combined in quadrature.
}
\label{fig:Di_all}
\end{figure}

The efficiencies of the dielectron triggers as a function of the offline electron \et{}
shown in Figure~\ref{fig:Di_all}(a) are calculated for a single electron trigger leg
of the dielectron trigger. Thus for the 2015 dielectron trigger, 2e12\_lhloose,
the efficiency of e12\_lhloose is shown.
The dielectron trigger had a lower \et{} threshold in 2015, and a slightly
tighter identification point (`lhloose' instead of `lhvloose'), which results in a different efficiency curve.
The dielectron triggers with an \et{} threshold of 17~\GeV{} have a lower efficiency in 2017
and 2018 than in previous years for \et{} below 60~\GeV. This is due to the L1 seed,
which has an electromagnetic isolation requirement. To recover the lost efficiency,
a combination of a lower-\et{} trigger (isolated at L1) and a higher-\et{} trigger
(with only the L1 hadronic veto applied) is typically used in ATLAS physics analyses.
The $\eta$-dependence of the efficiencies of the dielectron trigger legs is
shown in Figure~\ref{fig:Di_all}(b). The efficiency shown is
lower in the $1.37<|\eta|<1.52$ and $|\eta|>2.37$ regions, similar to the single-electron triggers.
Outside these regions, efficiencies of dielectron triggers without L1 EM isolation are
about 5\% lower in the endcaps than in the barrel region, while those for triggers
with L1 EM isolation have at most 3\% variations.
 
\begin{figure}[!ht]
\centering
{\includegraphics[width=0.45\textwidth]{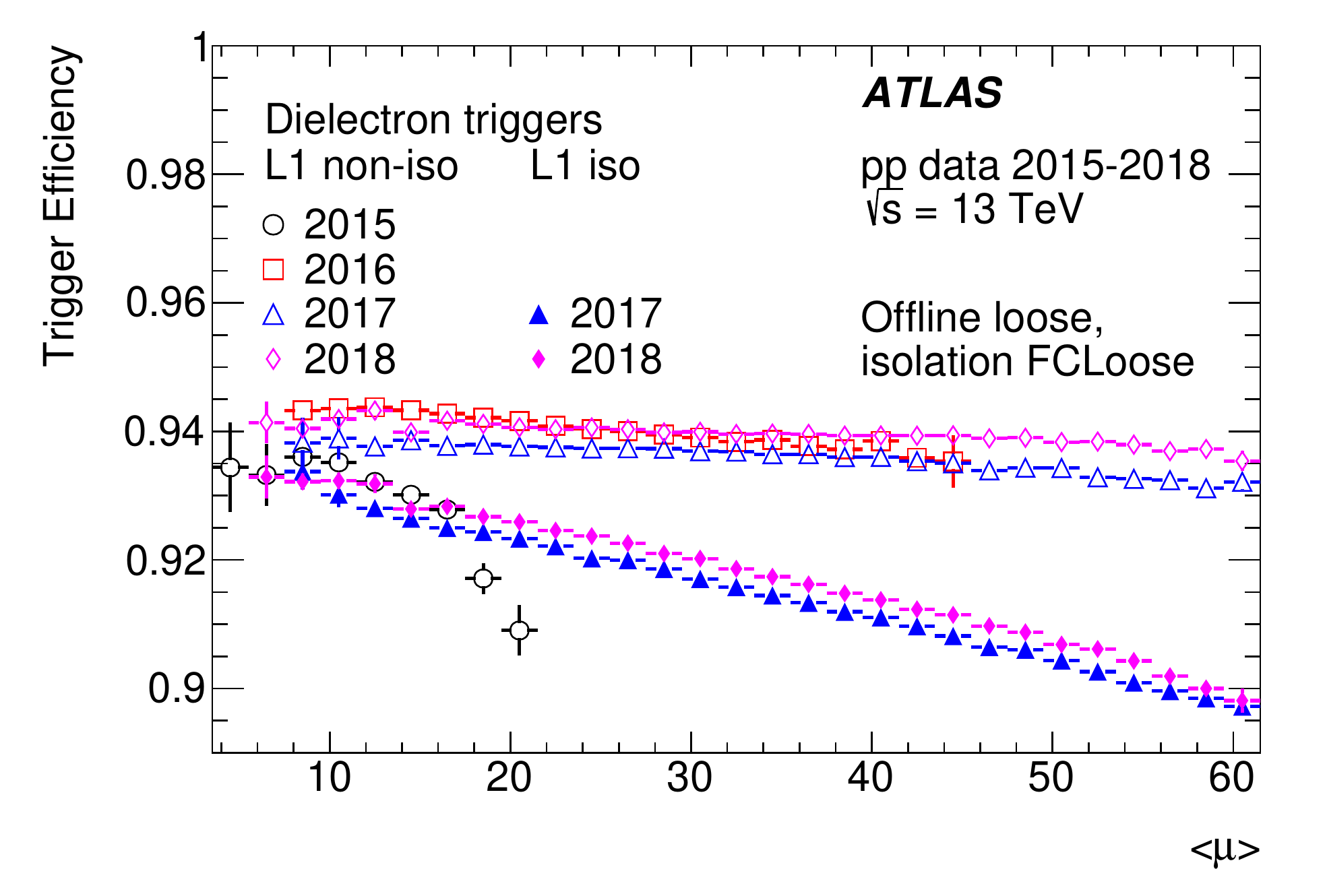}}
\caption{Evolution of efficiencies for the dielectron trigger legs as a function of
pile-up during Run~2, showing measurements in data only.
The changes between years are detailed in Section~\ref{sec:trig_e_evol}.
The efficiency is given with
respect to the loose offline identification and the FCLoose isolation working point.
No background subtraction is applied, as the effect is expected to be
negligible. Poorly populated bins are removed. Only offline
candidates with \et{} values at least 1~\GeV{} above the corresponding trigger threshold are used.
The error bars indicate statistical and systematic uncertainties combined in quadrature.
}
\label{fig:Di_all_mu}
\end{figure}

Figure~\ref{fig:Di_all_mu} shows the dielectron trigger efficiency as a function of pile-up.
It decreases slightly with $\langle\mu\rangle$ for non-isolated L1 triggers,
and has a much stronger $\langle\mu\rangle$ dependence (due to the L1
electromagnetic isolation requirement) for the isolated L1 triggers.
 
\begin{figure}[!ht]
\centering
\subfloat[][]{      {\includegraphics[width=0.45\textwidth]{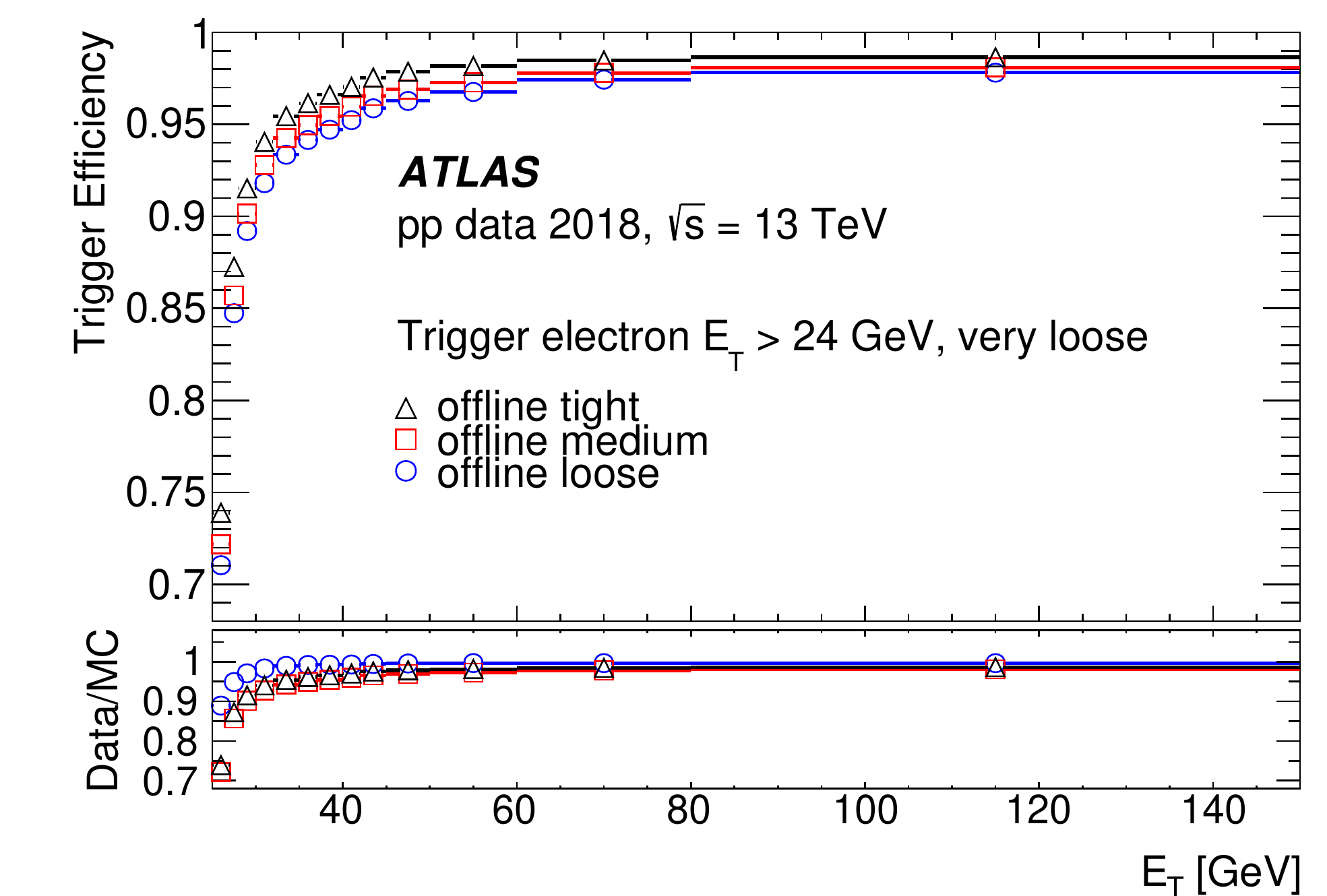}}}
\subfloat[][]{      {\includegraphics[width=0.45\textwidth]{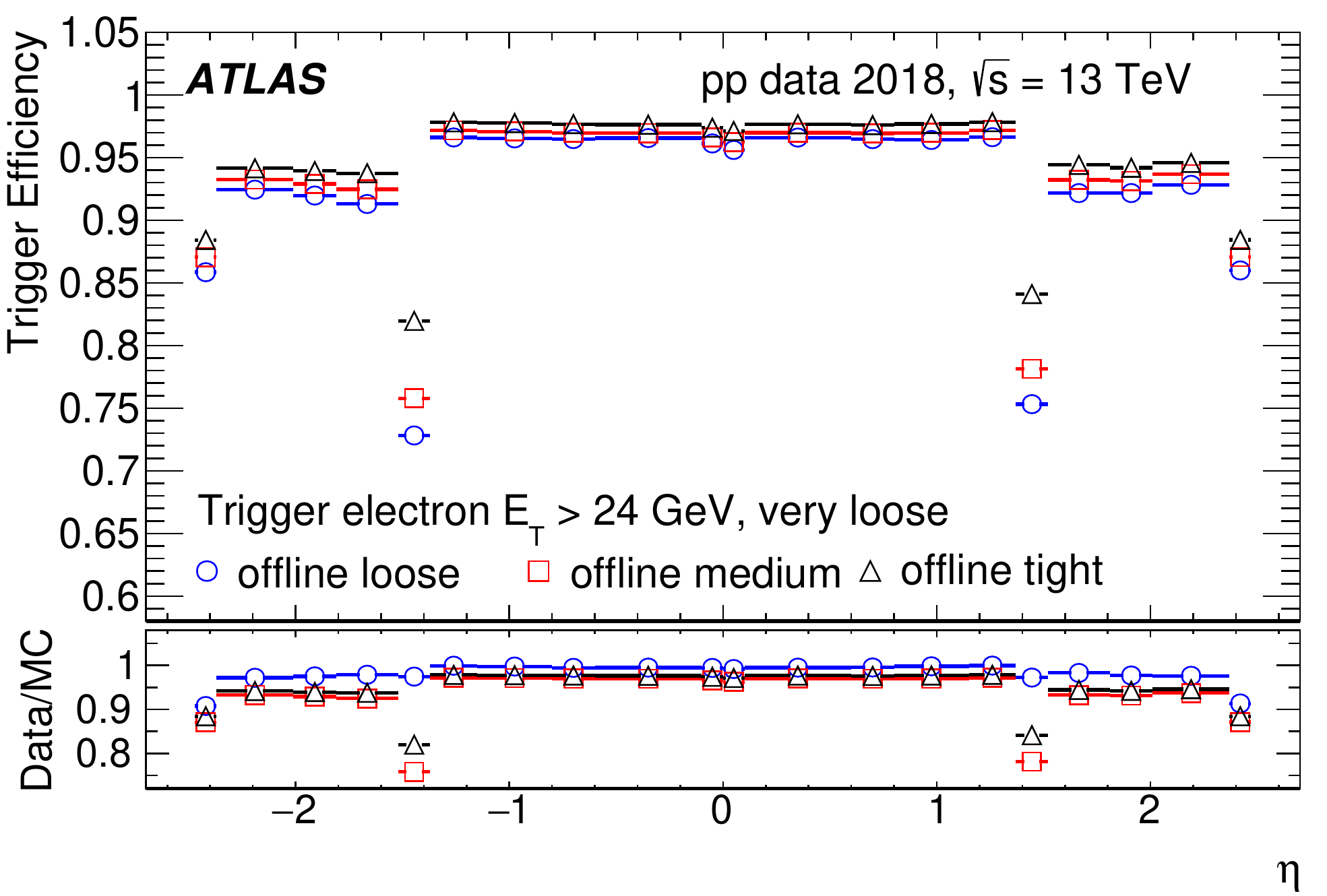}}}
\caption{Efficiencies of the e24\_lhvloose\_nod0 trigger as a function of the
offline electron (a) \et{} and (b) $\eta$ with respect to offline tight, medium, and
loose identification, and no isolation requirements. The efficiencies are measured in
data and shown with corresponding statistical and systematic uncertainties.
The ratios of data to MC simulation are also shown. For (b), only offline candidates
with \et$>$25~\GeV{} are used.
}
\label{fig:Di_datamc}
\end{figure}

The efficiency of the e24\_lhvloose\_nod0 trigger in 2018 with respect to various offline identification WPs is shown in Figure~\ref{fig:Di_datamc}.
The MC efficiency corrections for this trigger with very loose online
selection reach up to 30\% (10\%) at low \et{} relative to loose and medium
(tight) offline selections, but above 40~\gev{} they remain below 5\%. The data--simulation discrepancies are
mostly driven by the performance in the $1.37<|\eta|<1.52$ and $|\eta|>2.37$
regions. The efficiency correction factors are measured with a typical precision of $0.1\%$.
 
\subsection{Other electron and combined triggers for physics}
\label{sec:electron-other}
 
\begin{figure}[!ht]
\centering
\subfloat[][]{      {\includegraphics[width=0.45\textwidth]{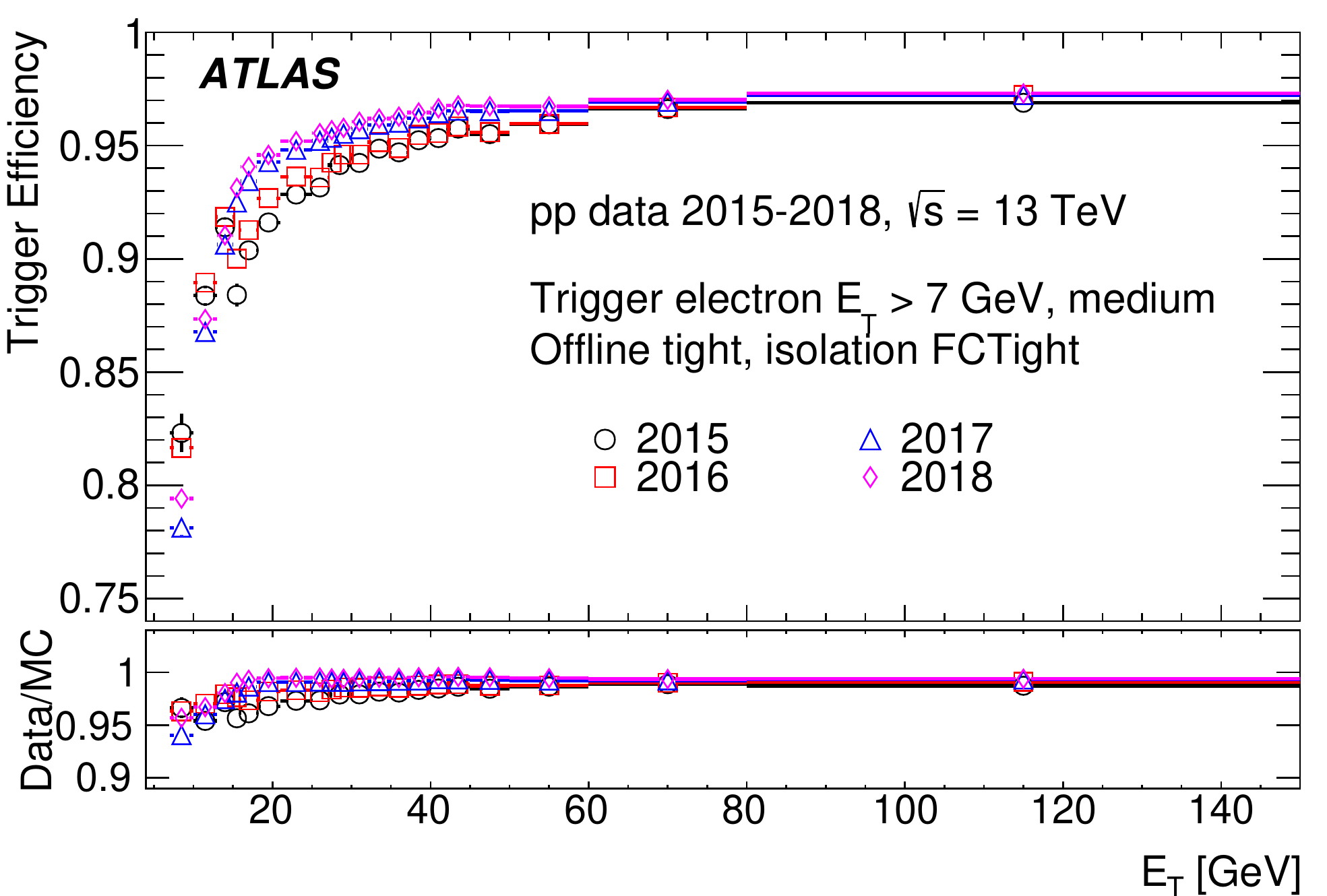}}}
\subfloat[][]{      {\includegraphics[width=0.45\textwidth]{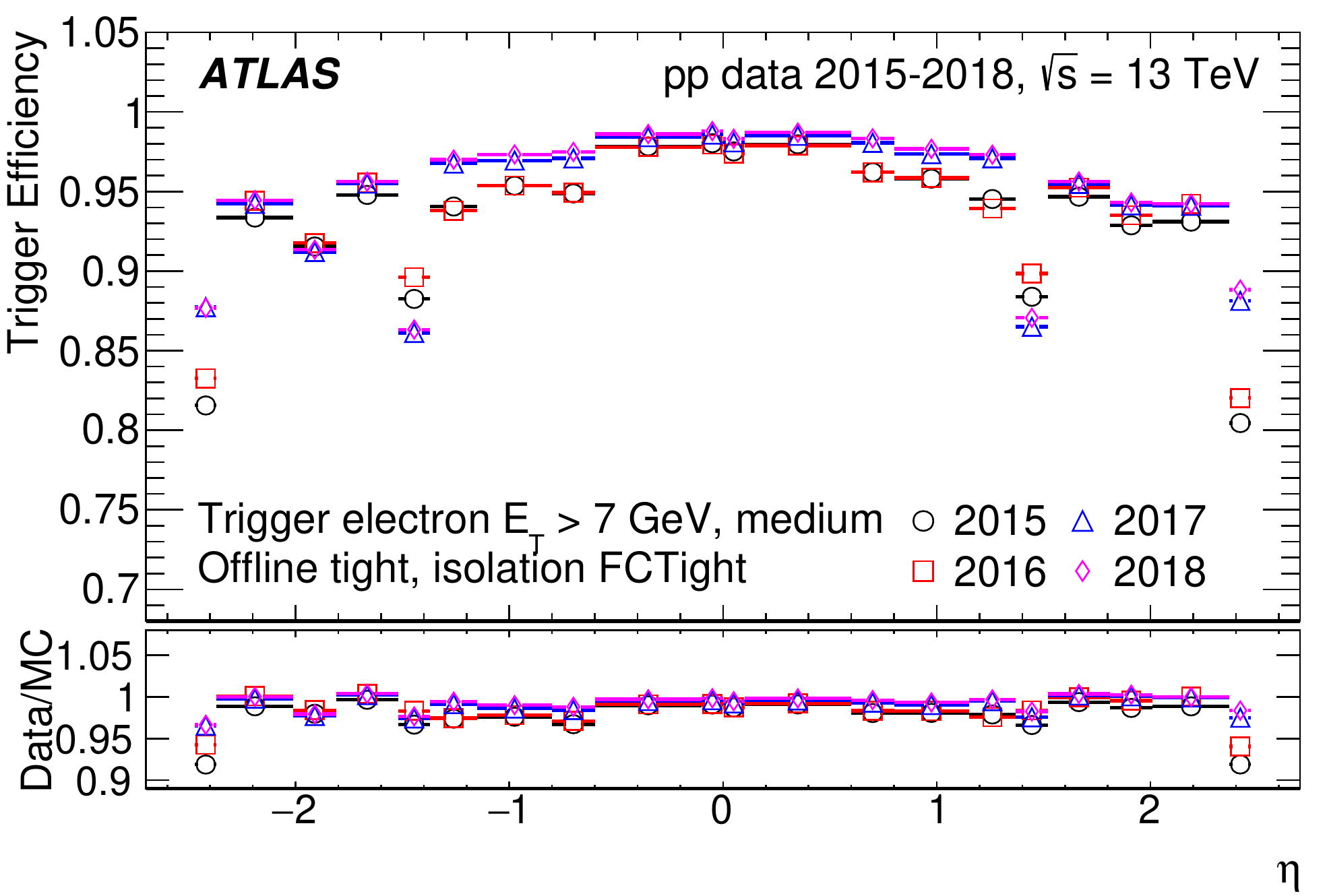}}}
\caption{Evolution of efficiencies of triggers with a 7~\gev{} electron-\et{} threshold and `lhmedium' identification
as a function of the offline electron (a) \et{} and (b) $\eta$ during Run~2.
The changes between years are detailed in Section~\ref{sec:trig_e_evol}. The efficiency is given with respect to the offline tight identification criteria
and the FCTight isolation working point.
The ratios of data to MC simulation efficiencies are also shown.
The error bars indicate statistical and systematic uncertainties combined in quadrature.
For (b), only offline candidates with $\et>8$~\gev{} are used.
}
\label{fig:e7_all}
\end{figure}

Lower thresholds or looser identification criteria than those used in the single-electron and dielectron
triggers described above can be used for `combined' triggers which target specific final states with
other physics objects (photons, muons, $\tau$-leptons, jets, $b$-jets, missing transverse momentum, etc.)
in addition to an electron. The lowest electron-\et{} threshold used in combined triggers is 7~\GeV.
The efficiency of the electron leg of this trigger, shown in Figure~\ref{fig:e7_all}, is similar to
that for the single-electron and dielectron triggers.
 
\begin{figure}[!ht]
\centering
{\includegraphics[width=0.45\textwidth]{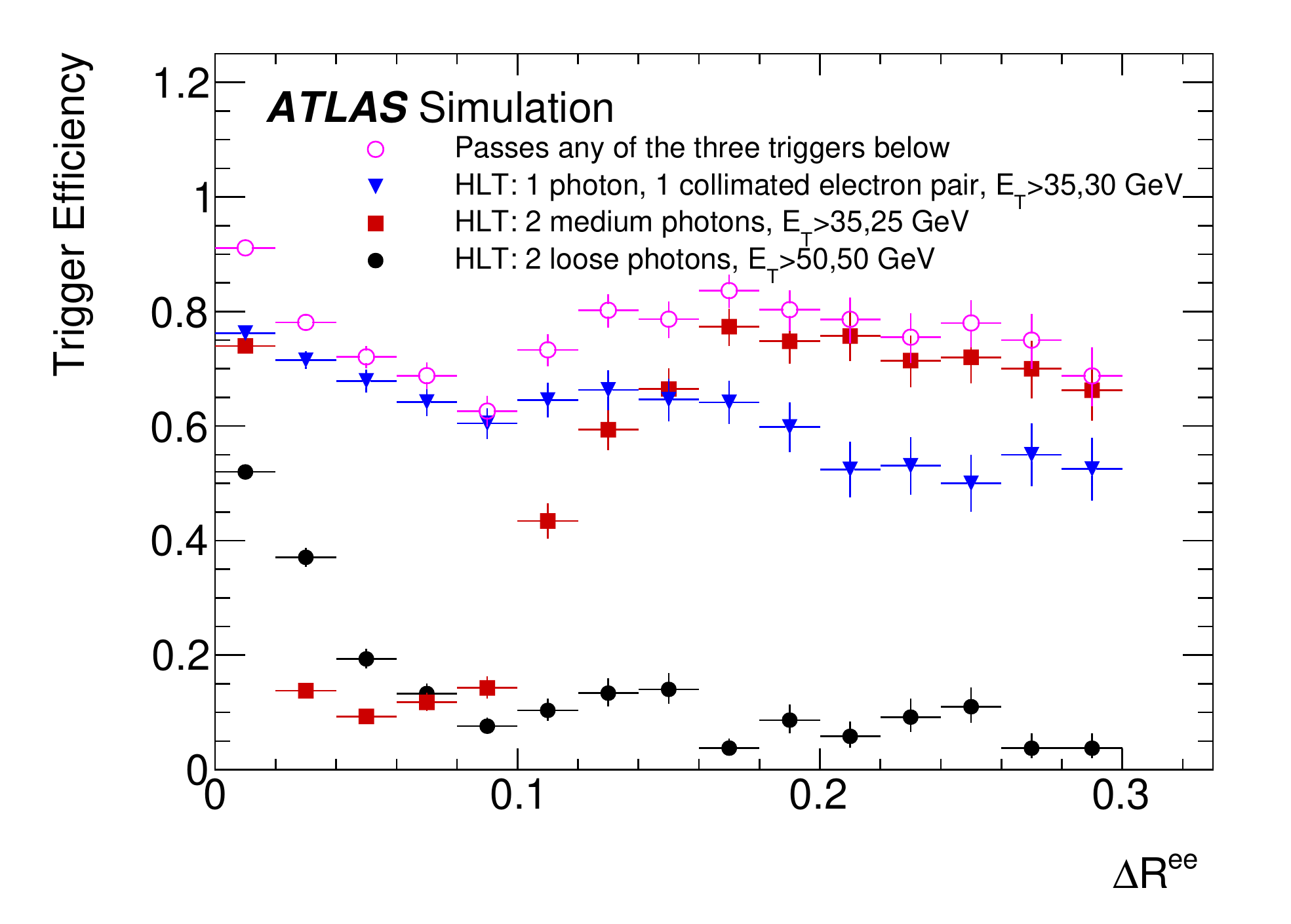}}
\caption{
Trigger efficiency in a sample of simulated $H\rightarrow \gamma^*\gamma \rightarrow ee\gamma$ events.
The specially developed trigger is called `1 photon, 1 collimated electron pair'.
The error bars indicate statistical uncertainties only.
}
\label{fig:mergedel}
\end{figure}

There are also dedicated triggers which allow events with unusual topologies to be collected.
For example, for final states with two closely spaced electrons, standard triggers are not very efficient.
This is shown in Figure~\ref{fig:mergedel} for the Higgs boson Dalitz decay
$H\rightarrow \gamma^*\gamma \rightarrow ee\gamma$.
The efficiency is measured in a sample of simulated events with $m_{ee}<10$~\GeV,
$E_\textrm{T}^\gamma>35$~\GeV{} and both electrons with \et$>15$~\GeV. A dedicated, cut-based electron
identification WP, `mergedtight', was introduced in 2017, allowing events with two collimated electrons
to pass the trigger. As shown in Table~\ref{tab-h}, this new trigger recovers a significant
fraction of events not recorded by the standard diphoton triggers (2g50\_loose and g35\_medium\_g25\_medium),
especially in the regime of $\Delta R(ee) < 0.1$. An additional requirement on the invariant mass of the
photon and dielectron requiring compatibility with the Higgs boson mass is introduced to reduce
the background rate, with negligible impact on the signal efficiency.
 
\begin{table*}[ht!]
\small
\centering
\caption{Trigger efficiency in a sample of simulated $H\rightarrow \gamma^*\gamma \rightarrow ee\gamma$ events.
The specially developed trigger is called `1 photon, 1 collimated electron pair'.
The signal selection efficiency for the combination of three triggers is given in the `Combined efficiency' line.
}
\label{tab-h}
\begin{tabular}{ll}
\toprule
{Trigger} & {Signal efficiency [\%]} \\
\midrule
2 loose photons, \et$>$50~\gev  & $41.3\pm0.5$ \\
2 medium photons, \et$>$35,\,25~\gev & $61.7\pm0.5$ \\
1 photon, 1 collimated electron pair, \et$>$35,\,30~\gev &  $72.3\pm0.4$ \\\hline
Combined efficiency &  $85.8\pm0.3$ \\
\bottomrule
\end{tabular}
\end{table*}
 
\subsection{Support electron triggers}
\label{sec:electron-support}
 
Samples of $J/\psi\rightarrow ee$ events for various performance measurements were collected with prescaled `support' dielectron triggers.
These require electron \et{} thresholds in the range 4--14~\GeV{} with one electron leg passing an `lhtight'
identification requirement and the other an `etcut' requirement.
 
There are also sets of single-electron prescaled triggers with either only an \et{} requirement in the HLT, or with
`lhvloose' identification at various \et{} thresholds. These have a rate of about 1$\,$Hz each
and are used for performance and background studies.
 
Additionally, for each combined trigger electron leg there is a corresponding single electron trigger
which is enabled only in `rerun' mode to allow electron trigger efficiency measurements.
 
\section{Electron and photon trigger performance in HI data-taking}
\label{sec:trig_hi}
 
In the 2015 $PbPb$ run, only cut-based `HI loose' electron triggers were activated.
The number of $Z\rightarrow ee$ candidates in the entire run was limited to about 4000,
which limited the precision of the trigger efficiency evaluation.
 
\begin{figure}[!ht]
\centering
\subfloat[][]{\includegraphics[width=0.48\textwidth]{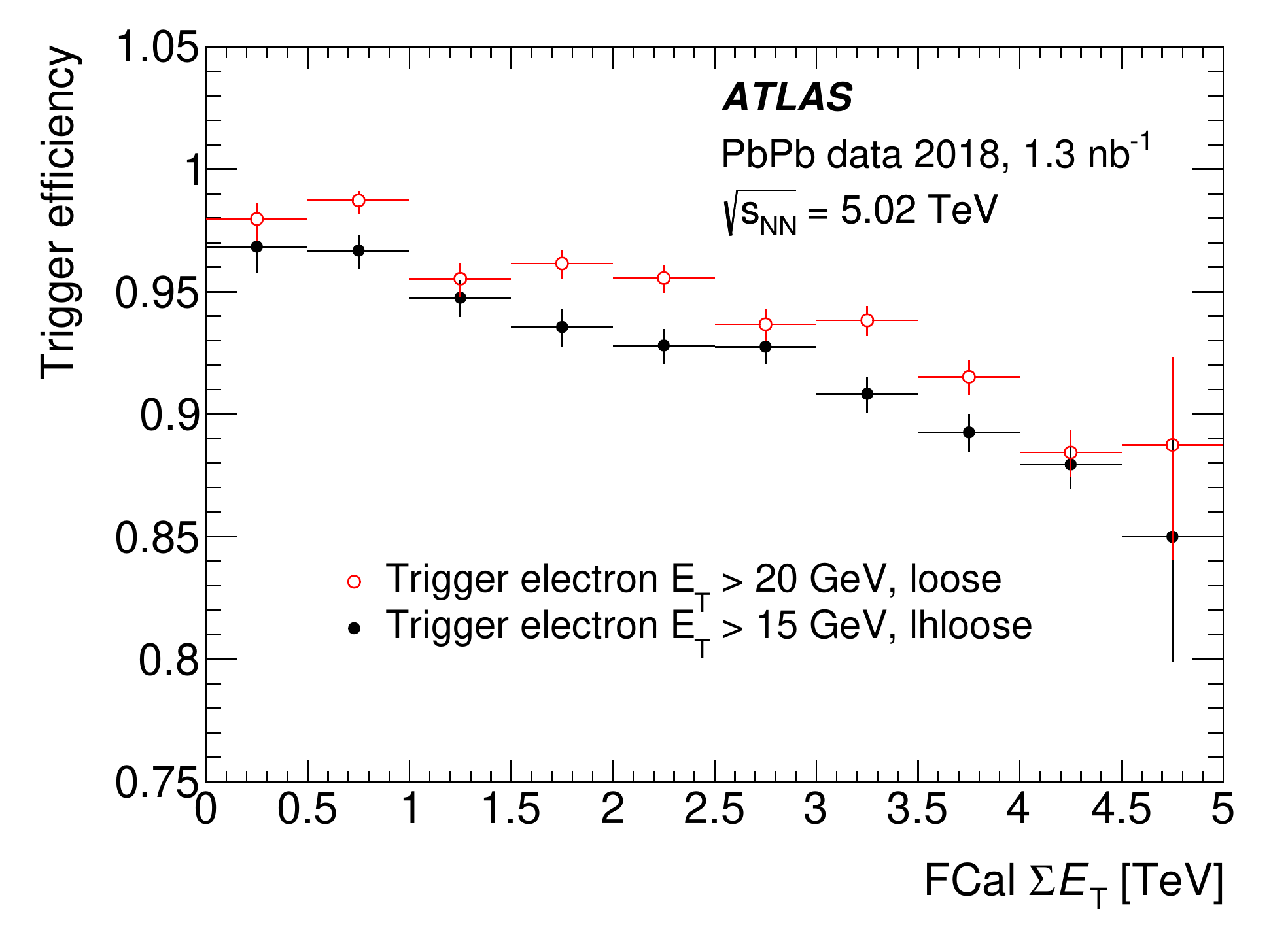}}
\subfloat[][]{\includegraphics[width=0.48\textwidth]{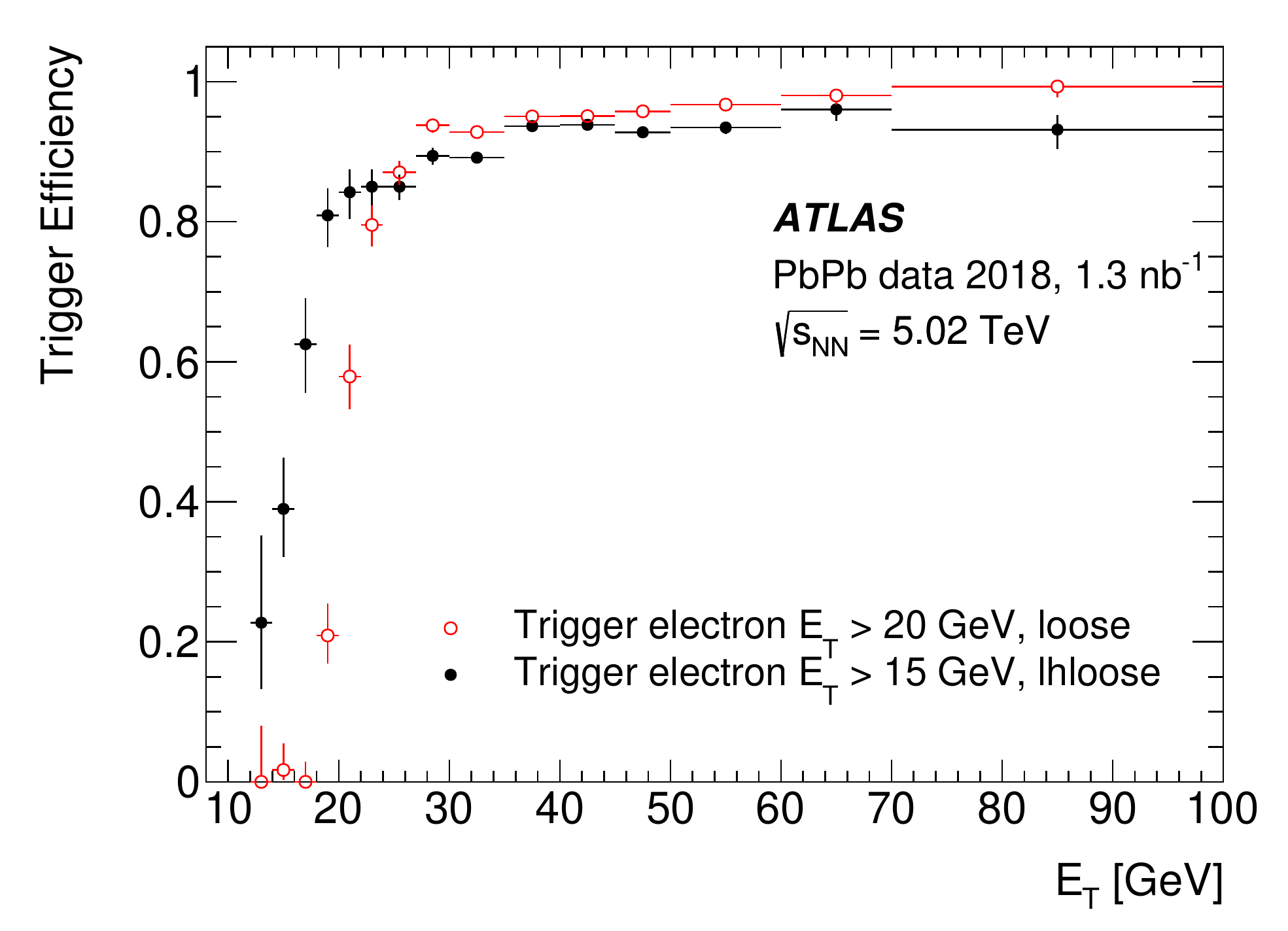}}
\caption{HI electron trigger efficiency as a function of (a) FCal $\Sigma \et$ and (b) the offline electron \et.
The efficiency is calculated with respect to the loose offline identification WP
using offline monitoring tools described in Section~\ref{sec:ops}.
For (a), only offline candidates with \et$>$20~\GeV{} are used.
The error bars indicate statistical uncertainties only.}
\label{fig:hi-eff}
\end{figure}

In the 2018 $PbPb$ run,
a factor of 3.5 more integrated luminosity was provided which resulted in a significant improvement
in the number of $Z\rightarrow ee$ candidates for electron trigger performance studies.
Two electron trigger sequences are activated for data-taking: a cut-based `loose\_ion' trigger
with an $\et=20$~\GeV{} threshold, and an LH-based `lhloose\_ion' trigger with an
$\et>15$~\GeV{} requirement. In the latter, the standard $pp$ pdfs are evaluated using
the UE-corrected variables. An advantage of the LH-based approach is a significant
reduction in the output rate in comparison with the cut-based trigger at the same \et{} threshold.
The LH trigger has significantly better purity at the cost of a moderate loss in trigger efficiency.
Figure~\ref{fig:hi-eff} shows the trigger efficiency as a function of FCal $\Sigma\et$
and offline electron \et{} for the `loose\_ion' and `lhloose\_ion' triggers with 20 and 15$\,$\GeV{}
thresholds, respectively. The trigger efficiencies are evaluated using the tag-and-probe method
on $Z\rightarrow ee$ candidate events.  Probe electrons are required to pass a version of
the loose LH identification optimised for $PbPb$ collisions. The `loose\_ion' trigger is
slightly more efficient in the plateau region, which is reached at around 25~\GeV{}
in both cases. Both trigger sequences have a small (below 12\%) collision-centrality dependence.
 
\begin{figure}[t]
\begin{center}
\subfloat[][]{\includegraphics[width=0.48\textwidth]{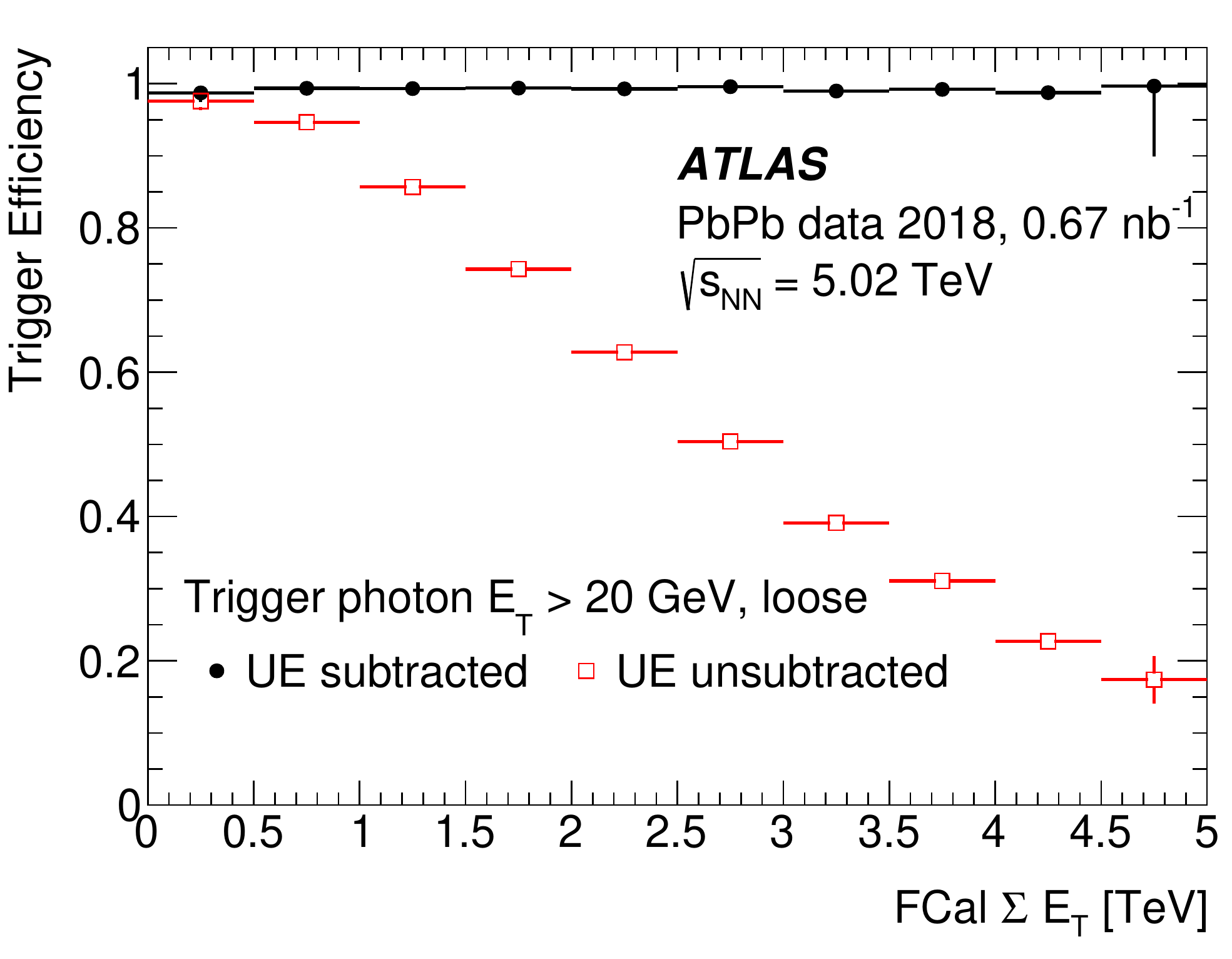}}
\subfloat[][]{\includegraphics[width=0.48\textwidth]{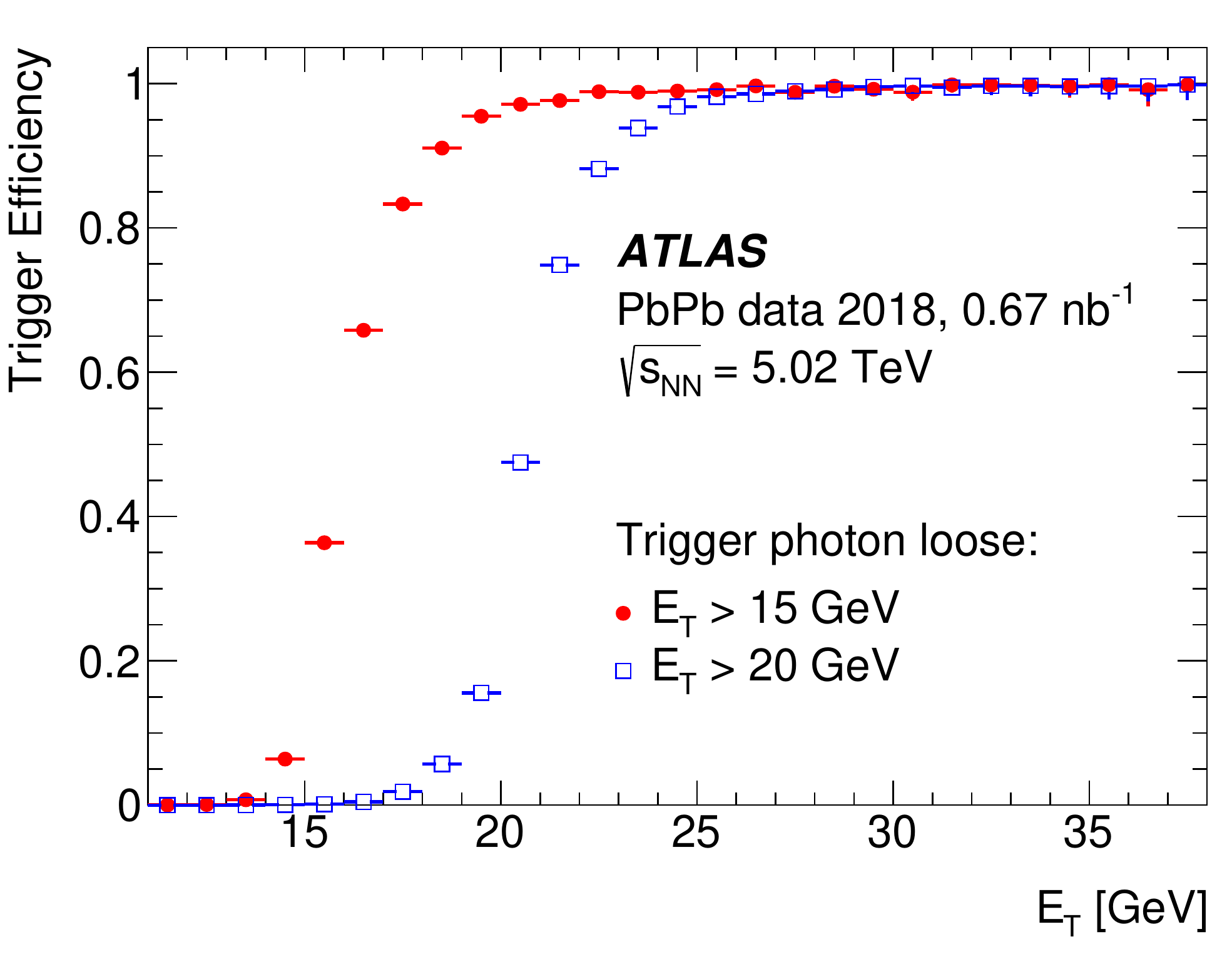}}
\caption{Photon trigger efficiencies as a function of (a) FCal $\Sigma \ET$ and (b) offline photon \et.
In (a), the closed (open) marker indicate data with (without) UE subtraction;
only offline candidates with \et$>$20~\GeV{} are used.
In (b), UE subtraction is applied in both cases, with closed (open) markers indicating
\et{} thresholds of 15 (20)~\GeV{}. The efficiencies are computed with respect to
offline photons satisfying tight identification criteria
using offline monitoring tools described in Section~\ref{sec:ops}.
Offline photon candidates in the calorimeter transition region $1.37 < |\eta| < 1.52$ are not considered.
The error bars indicate statistical uncertainties only.
}
\label{fig:eff}
\end{center}
\end{figure}

The primary unprescaled photon trigger used in 2015 and 2018 $PbPb$ data-taking had a $20~\GeV$ \ET threshold, and
the photon candidate was required to satisfy loose identification criteria.
Figure~\ref{fig:eff} shows the 2018 photon trigger efficiency using the BS method.
The efficiency is shown in Figure~\ref{fig:eff}(a) as a function of FCal $\Sigma\et$,
with and without UE subtraction applied in the online reconstruction. When the reconstruction
is run without UE subtraction, i.e.\ in the same manner as done in $pp$ collision data-taking,
the efficiency shows a strong dependence on collision centrality. This is primarily due to a strong distortion
of the shower shapes and subsequent inefficiency associated with the identification requirements.
When the reconstruction is run with the UE subtraction procedure, the photon trigger efficiency
remains high across the full range of centralities. In Figure~\ref{fig:eff}(b), the (offline, calibrated)
photon-\ET dependence of photon trigger efficiencies using UE subtraction is shown for photon
triggers with 15 and $20~\GeV$ \ET thresholds. The efficiency is determined with respect
to offline reconstructed photons which pass a tighter set of identification cuts,
identical to those used in typical physics analyses. The HI photon triggers become fully efficient
at about $5~\GeV$ above the nominal online trigger threshold, similar to the photon triggers used
for the $pp$ data-taking.
 
\section{Monitoring and data quality}
\label{sec:ops}
 
During data-taking, the performance of the electron and photon triggers is monitored online while data
are being collected, and offline, right after data are recorded.
 
The online monitoring check is performed during the data-taking by shift personnel based in the ATLAS control room at CERN,
located at ground level, 100~m above the ATLAS detector, in the Point 1 of the LHC ring.
Observables related to electron and photon candidates at different stages of the HLT reconstruction are checked.
These observables are defined in the reconstruction and hypothesis-testing algorithms executed online.
Only a set of representative electron and photon triggers are monitored online.
Monitoring of the full set of triggers would require a large fraction of the HLT farm's computing power and is not necessary, as many of the electron and photon triggers share the same algorithms. In the fast reconstruction step, only the coverage in $\eta$--$\phi$ space
and the distributions of \et are monitored for both the electrons and photons.
Ringer algorithm variables, track positions, and distances between tracks and calorimeter clusters are monitored for electron triggers.
In the precision step, in addition to those observables, the value of $\langle\mu\rangle$ at the HLT and some calorimeter shower shapes are monitored. For electrons, the distribution of the value of the likelihood discriminant is also monitored.
 
An `express stream' is defined, containing a fraction of the collected data reconstructed with
high priority for offline monitoring and data-quality purposes. It takes around one day for the ATLAS
computing centre at CERN to calibrate and reconstruct data from the express stream. Once these data are
available, offline monitoring of the performance and data quality for several electron and photon triggers
is performed. The list of triggers that are monitored offline include versions of all primary triggers (single, isolated electron,
single photon, dielectron, diphoton, tight diphoton), and supporting triggers
(e.g.\ dedicated $Z\rightarrow ee$ and $J/\psi\rightarrow ee$ tag-and-probe triggers).
All of the express stream triggers are highly prescaled;
3~Hz of the total 20~Hz express stream rate is reserved for the electron and photon trigger monitoring.
 
Almost all of the variables used by the online algorithms are monitored offline, at L1 and in both the fast and precision HLT reconstruction steps. In addition to distributions of physics observables, the efficiencies of different triggers are computed and monitored. Figure~\ref{fig:l1iso} is an example of an efficiency plot produced with the monitoring tools.
The offline monitoring is performed by an expert among the shift personnel, who compares the distributions of HLT
physics observables and trigger performance with those from a reference set of data, selected in
advance, for which the ATLAS detector is known to have good operational performance.
There were no data-quality issues caused by electron and photon triggers in Run 2.
 
\section{Conclusion}
\label{sec:conclusion}
This paper describes the ATLAS electron and photon triggers and their evolution during Run~2.
To cope with a fourfold increase of peak LHC luminosity in Run 2 (2015--2018),
to 2.1$\times$\highL{}, and a similar increase in the number of interactions
per beam-crossing, trigger algorithms and selections needed
to be optimised to control the trigger rates while retaining a high efficiency for offline analyses.
The main triggers for the proton--proton data-taking were a single-electron trigger with
a transverse energy threshold of 26~\GeV{} and a diphoton trigger with transverse
energy thresholds of 25 and 35~\GeV. 
The single-electron trigger efficiency relative to a single-electron offline selection is at least 75\%
for an offline electron of 31~\GeV{}, and rises to 96\% at 60~\GeV.
The trigger efficiency of a 25~\gev{} leg of the primary diphoton trigger
relative to a tight offline photon selection is more than 96\% for an offline photon of 30~\GeV{}.
Trigger efficiencies are comparable in the heavy-ion runs, in which electron and photon trigger transverse energy thresholds
were in the range 15--20~\GeV.
 
 
\section*{Acknowledgements}
 
 
We thank CERN for the very successful operation of the LHC, as well as the
support staff from our institutions without whom ATLAS could not be
operated efficiently.
 
We acknowledge the support of ANPCyT, Argentina; YerPhI, Armenia; ARC, Australia; BMWFW and FWF, Austria; ANAS, Azerbaijan; SSTC, Belarus; CNPq and FAPESP, Brazil; NSERC, NRC and CFI, Canada; CERN; CONICYT, Chile; CAS, MOST and NSFC, China; COLCIENCIAS, Colombia; MSMT CR, MPO CR and VSC CR, Czech Republic; DNRF and DNSRC, Denmark; IN2P3-CNRS, CEA-DRF/IRFU, France; SRNSFG, Georgia; BMBF, HGF, and MPG, Germany; GSRT, Greece; RGC, Hong Kong SAR, China; ISF and Benoziyo Center, Israel; INFN, Italy; MEXT and JSPS, Japan; CNRST, Morocco; NWO, Netherlands; RCN, Norway; MNiSW and NCN, Poland; FCT, Portugal; MNE/IFA, Romania; MES of Russia and NRC KI, Russian Federation; JINR; MESTD, Serbia; MSSR, Slovakia; ARRS and MIZ\v{S}, Slovenia; DST/NRF, South Africa; MINECO, Spain; SRC and Wallenberg Foundation, Sweden; SERI, SNSF and Cantons of Bern and Geneva, Switzerland; MOST, Taiwan; TAEK, Turkey; STFC, United Kingdom; DOE and NSF, United States of America. In addition, individual groups and members have received support from BCKDF, CANARIE, CRC and Compute Canada, Canada; COST, ERC, ERDF, Horizon 2020, and Marie Sk{\l}odowska-Curie Actions, European Union; Investissements d' Avenir Labex and Idex, ANR, France; DFG and AvH Foundation, Germany; Herakleitos, Thales and Aristeia programmes co-financed by EU-ESF and the Greek NSRF, Greece; BSF-NSF and GIF, Israel; CERCA Programme Generalitat de Catalunya, Spain; The Royal Society and Leverhulme Trust, United Kingdom.
 
The crucial computing support from all WLCG partners is acknowledged gratefully, in particular from CERN, the ATLAS Tier-1 facilities at TRIUMF (Canada), NDGF (Denmark, Norway, Sweden), CC-IN2P3 (France), KIT/GridKA (Germany), INFN-CNAF (Italy), NL-T1 (Netherlands), PIC (Spain), ASGC (Taiwan), RAL (UK) and BNL (USA), the Tier-2 facilities worldwide and large non-WLCG resource providers. Major contributors of computing resources are listed in Ref.~\cite{ATL-GEN-PUB-2016-002}.
 

\printbibliography

\clearpage 
 
\begin{flushleft}
{\Large The ATLAS Collaboration}

\bigskip

G.~Aad$^\textrm{\scriptsize 102}$,    
B.~Abbott$^\textrm{\scriptsize 129}$,    
D.C.~Abbott$^\textrm{\scriptsize 103}$,    
A.~Abed~Abud$^\textrm{\scriptsize 71a,71b}$,    
K.~Abeling$^\textrm{\scriptsize 53}$,    
D.K.~Abhayasinghe$^\textrm{\scriptsize 94}$,    
S.H.~Abidi$^\textrm{\scriptsize 167}$,    
O.S.~AbouZeid$^\textrm{\scriptsize 40}$,    
N.L.~Abraham$^\textrm{\scriptsize 156}$,    
H.~Abramowicz$^\textrm{\scriptsize 161}$
% ...
\bigskip
\\

% ...
$^{40}$Niels Bohr Institute, University of Copenhagen, Copenhagen; Denmark.\\
% ...
$^{53}$II. Physikalisches Institut, Georg-August-Universit\"{a}t G\"ottingen, G\"ottingen; Germany.\\
% ...
$^{71}$$^{(a)}$INFN Sezione di Pavia;$^{(b)}$Dipartimento di Fisica, Universit\`a di Pavia, Pavia; Italy.\\
% ...
$^{94}$Department of Physics, Royal Holloway University of London, Egham; United Kingdom.\\
% ...
$^{102}$CPPM, Aix-Marseille Universit\'e, CNRS/IN2P3, Marseille; France.\\
$^{103}$Department of Physics, University of Massachusetts, Amherst MA; United States of America.\\
% ...
$^{129}$Homer L. Dodge Department of Physics and Astronomy, University of Oklahoma, Norman OK; United States of America.\\
% ...
$^{156}$Department of Physics and Astronomy, University of Sussex, Brighton; United Kingdom.\\
% ...
$^{161}$Raymond and Beverly Sackler School of Physics and Astronomy, Tel Aviv University, Tel Aviv; Israel.\\
% ...
$^{167}$Department of Physics, University of Toronto, Toronto ON; Canada.\\
% ...
$^{a}$ Also at Borough of Manhattan Community College, City University of New York, New York NY; United States of America.\\
$^{b}$ Also at CERN, Geneva; Switzerland.\\
% ...

\end{flushleft}


\end{document}